\documentclass[aps,prd,nofootinbib,showpacs,preprintnumbers,longbibliography,amssymb,11pt]{revtex4-1} 
\pdfoutput=1
\usepackage[sort&compress]{natbib}

\usepackage{latexsym}
\usepackage{amsthm}
\usepackage{float}
\usepackage{amsmath,amssymb}
\usepackage{graphicx,subfigure,grffile}
\usepackage{fancybox,feynmp}
\usepackage{indentfirst}
\usepackage{epsfig}
\usepackage{epsfig,subfigure}
\usepackage{dcolumn}
\usepackage{bm}
\usepackage{slashed}
\usepackage{cancel}
\usepackage{color}
\usepackage{tabularx}
\usepackage[titletoc]{appendix}

\usepackage{hyperref} 
\usepackage[noabbrev]{cleveref} 
\graphicspath{{./Figures/}}

\usepackage{epstopdf}
\usepackage{cleveref}

\usepackage{epstopdf}

\usepackage{booktabs, tabularx}

\usepackage{tikz}
\usepackage[utf8]{inputenc}
\usepackage{array,booktabs,calc}
\usepackage{multirow}

\newcolumntype{N}{>{\centering\arraybackslash}m{4cm}}
\newcolumntype{G}{>{\bfseries\centering\arraybackslash}m{3cm+6\tabcolsep}}
\newcolumntype{M}[1]{>{\centering\arraybackslash}m{#1}}

\newcommand{\eqnref}[1]{Eq.~(\ref{eqn:#1})}

\newcommand{\secref}[1]{Sec.~\ref{sec:#1}}

\newcommand{\subsecref}[1]{Subsec.~\ref{subsec:#1}}

\newcommand{\appref}[1]{Appendix~\ref{sec:#1}}
\newcommand{\figref}[1]{Fig.~\ref{fig:#1}}

\newcommand{\tableref}[1]{Table~\ref{table:#1}}

\begin{document}

\preprint{MITP/17-005}

\title{A Tale of Two Portals: Testing Light, Hidden New Physics at
  Future $e^+ e^-$ Colliders}

\author{Jia Liu}
\email{liuj@uni-mainz.de}
\affiliation{PRISMA Cluster of Excellence \& Mainz Institute for
  Theoretical Physics, Johannes Gutenberg University, 55099 Mainz,
  Germany}

\author{Xiao-Ping Wang}
\email{xiaowang@uni-mainz.de}
\affiliation{PRISMA Cluster of Excellence \& Mainz Institute for
  Theoretical Physics, Johannes Gutenberg University, 55099 Mainz,
  Germany}

\author{Felix Yu}
\email{yu001@uni-mainz.de}
\affiliation{PRISMA Cluster of Excellence \& Mainz Institute for
  Theoretical Physics, Johannes Gutenberg University, 55099 Mainz,
  Germany}

\date{\today}

\begin{abstract}
We investigate the prospects for producing new, light, hidden states
at a future $e^+ e^-$ collider in a Higgsed dark $U(1)_D$ model, which
we call the Double Dark Portal model.  The simultaneous presence of
both vector and scalar portal couplings immediately modifies the
Standard Model Higgsstrahlung channel, $e^+ e^- \to Zh$, at leading
order in each coupling.  In addition, each portal leads to
complementary signals which can be probed at direct and indirect
detection dark matter experiments.  After accounting for current
constraints from LEP and LHC, we demonstrate that a future $e^+ e^-$
Higgs factory will have unique and leading sensitivity to the two
portal couplings by studying a host of new production, decay, and
radiative return processes.  Besides the possibility of exotic Higgs
decays, we highlight the importance of direct dark vector and dark
scalar production at $e^+ e^-$ machines, whose invisible decays can be
tagged from the recoil mass method.
\end{abstract}


\maketitle
\tableofcontents

\section{Introduction}
\label{sec:Introduction}

Searches for new, light, hidden states are strongly motivated from the
overriding question of determining the particle nature of dark matter.
The possible couplings to such light states, however, remain highly
model-dependent.  Because higher dimension operators are expected to
be suppressed in scattering processes at low energies, the most
promising couplings give marginal Lagrangian operators at dimension
four.  Along these lines, two well-studied couplings are a new kinetic
mixing term $\epsilon$ between a new, light, hidden photon and the
hypercharge gauge boson and a new quartic Higgs portal coupling
$\lambda_{HP}$ between a hidden charged scalar field and the Standard
Model Higgs field.

In this work, we argue and demonstrate that both marginal couplings
can be simultaneously probed in future measurements of a high energy
$e^+ e^-$ collider.  Such a collider is, of course, very strongly
motivated by a rich and diverse set of possible Higgs measurements,
with leading sensitivity to the total Higgs width, Higgs couplings to
Standard Model (SM) particles, exotic Higgs decays, and additional
precision measurements of the top quark mass and exotic $Z$ boson
decays if additional running conditions are
afforded~\cite{Dawson:2013bba, Djouadi:2007ik, Gomez-Ceballos:2013zzn,
  CEPC-SPPCStudyGroup:2015csa}.  We highlight that such a machine also
has leading sensitivity to new, weakly coupled, hidden sectors, which
can be probed via both radiative return processes and exotic invisible
and semi-visible Higgs decays.  We will show that these measurements
are enabled because of the expected high precision photon resolution
in the electromagnetic calorimeter, the exquisite reconstruction of
charged leptons, and clean discrimination of exotic signals from SM
background processes.

Both of these marginal operators have been studied autonomously at
electron colliders in the hidden photon context~\cite{Yin:2009mc,
  Li:2009wz, Lees:2014xha, Curtin:2014cca, Babusci:2015zda,
  Karliner:2015tga, Prasad:2015mxa, Hochberg:2015vrg, Won:2016pjz,
  Kaneta:2016uyt, He:2017ord, Araki:2017wyg, Biswas:2017lyg} and the
hidden scalar context~\cite{Acciarri:1996um,
  Acciarri:1997tr,Abreu:1999vu, Searches:2001ab, Schael:2006cr,
  Flacke:2016szy}.  Some works study both operators in
tandem~\cite{TheBelle:2015mwa, Biswas:2015sha, Angelescu:2017jyj} or
adopt an effective operator approach~\cite{Dreiner:2012xm}.  The
current status of light, sub-GeV hidden photon searches and future
prospects is summarized in Ref.~\cite{Alexander:2016aln}.  In contrast
with previous studies, we focus on higher mass hidden photons beyond
the reach of $B$-physics experiments and beam-dump experiments.  For
illustrative purposes, we show our projections to dark photons as
light as $1$~GeV to demonstrate the complementarity with recent
results from $B$-physics experiments such as
BaBar~\cite{Lees:2017lec}.  In addition, we will emphasize the unique
capability of $e^+ e^-$ machines to reconstruct invisible decays,
which is a marked improvement over the reconstruction prospects at
hadron colliders.

The lack of evidence for weakly interacting massive particles (WIMPs)
in direct detection (DD) experiments~\cite{Agnese:2015nto,
  Angloher:2015ewa, Tan:2016zwf, Akerib:2016vxi}, increasingly strong
constraints on thermal WIMPs from indirect detection (ID)
experiments~\cite{Ade:2015xua, Ackermann:2015zua, Ahnen:2016qkx,
  Massari:2015xea, Aguilar:2014mma}, and non-observation of beyond the
Standard Model (BSM) missing transverse energy signatures at the
LHC~\cite{Aaboud:2016tnv, Sirunyan:2017hci}, combine to an increasing
unease with the standard WIMP miracle paradigm.  On the other hand,
dark matter coupled to kinetically mixed hidden photons suffers from
strong direct detection constraints (see, {\it
  e.g.},~\cite{Liu:2014cma}).  A consistent dark matter model must
hence simultaneously address the relic density mechanism and
non-observation in the current experimental probes, and thus minimal
models either require nonthermal dark matter production in the early
universe, coannihilation channels~\cite{Griest:1990kh, Baker:2015qna,
  Buschmann:2016hkc}, or resonant dark matter annihilation in order to
divorce the early universe dynamics from collider processes 
(see, {\it e.g.}, \cite{Alves:2016cqf}).
Moreover, while the nuclear recoil energy spectrum at direct detection
experiments requires the dark matter mass as input, colliders instead
probe mediator masses if they are on-shell, which shows the
complementarity between both approaches.  In our work, we will further
demonstrate these complementary aspects between dark matter
experiments and hadron and lepton colliders in the context of our dark
matter model.

In~\secref{review}, we review the theoretical framework for the Double
Dark Portal model, which unifies the kinetic mixing portal and the
scalar Higgs portal into a minimal setup with dark matter.
In~\secref{DDIDpheno}, we detail the phenomenology of the dark matter
for direct detection and indirect detection experiments.  We discuss
the extensive collider phenomenology of the model and review the
current constraints from experiments at the Large Electron-Positron
(LEP) collider and the Large Hadron Collider (LHC)
in~\secref{ColliderLEPLHC}.  We then present the prospects for
exploring new, light hidden states at a future $e^+ e^-$ machine
in~\secref{futureprospects} and conclude in~\secref{conclusions}.
In~\appref{limits}, we offer some detailed discussion of limiting
cases in our Double Dark Portal model for pedagogical clarity, and we
discuss a cancellation effect in scattering processes via kinetic
mixing in~\appref{cancellation}.  We also present the dark matter
annihilation cross sections for charged SM final states
in~\appref{annXS}.

\section{Overview of the Double Dark Portal Model: Simultaneous Kinetic Mixing and Scalar Portal Couplings}
\label{sec:review}

We begin with the Lagrangian of the Double Dark Portal Model,
\begin{equation}
\begin{array}{ccl}
\mathcal{L} &\supset& - \frac{1}{4} B_{\mu \nu} B^{\mu \nu} 
- \frac{1}{4} W_{\mu \nu}^i W^{i \, \mu \nu} - \frac{1}{4} K_{\mu \nu} K^{\mu \nu} 
+ \frac{\epsilon}{2 \cos \theta_W} B_{\mu \nu} K^{\mu \nu} \\ 
&+& |D_\mu H|^2 + |D_\mu \Phi|^2 + \mu_H^2 |H|^2 - \lambda_H |H|^4 +
\mu_D^2 |\Phi|^2 - \lambda_D |\Phi|^4 - \lambda_{HP} |H|^2 |\Phi|^2 \\
&+& \bar{\chi} (i \slashed{D} - m_\chi) \chi \ ,
\end{array}
\label{eqn:DDP}
\end{equation}
where $K_{\mu \nu}$ is the field strength tensor for the $U(1)_D$
gauge boson, $\Phi$ is a dark Higgs scalar field with charge $+1$
under $U(1)_D$, and $\chi$ is the dark matter and a SM gauge singlet
fermion with charge $+1$ under $U(1)_D$.  We take $\mu_H^2 > 0$ and
$\mu_D^2 > 0$, which trigger spontaneous symmetry breaking of the SM
electroweak symmetry and the $U(1)_D$ dark gauge symmetry,
respectively.  The $\theta_W$ parameter is the tree-level SM weak
mixing angle, $\theta_W = \tan^{-1} (g' / g)$. The nonzero Higgs
portal coupling, $\lambda_{HP}$, induces mass mixing between the $h$
and $\phi$ scalars, which results in mass eigenstates $H_0$ and $S$.
Simultaneously, the kinetic mixing $\epsilon$ will result in an
effective mass mixing between the SM $Z$ gauge boson and the $K$ dark
gauge boson, which results in the mass eigenstates $\tilde{Z}$ and
$\tilde{K}$.  The two marginal couplings, $\epsilon$ and
$\lambda_{HP}$, are commonly referred to as vector and scalar portals,
respectively~\cite{Essig:2013lka}.  Because the phenomenology of such
portal couplings changes significantly when a light dark matter
particle is added, we call the Lagrangian in~\eqnref{DDP} the Double
Dark Portal (DDP) model.

We solve the Lagrangian in the broken phase after the Higgs and the
dark Higgs obtain their vacuum expectation values (vevs),
\begin{align}
\Phi &= \frac{1}{\sqrt{2}} (v_D + \phi) \ , 
\label{eqn:phivev} \\
H &= \frac{1}{\sqrt{2}} (v_H + h) \ ,
\label{eqn:Hvev}
\end{align}
by diagonalizing and canonically normalizing the kinetic terms for the
electrically neutral gauge bosons and diagonalizing their mass matrix.
We can rewrite the Lagrangian using matrix notation, with mass terms
acting on the gauge basis vector $(\begin{array}{ccc} W_\mu^3 & B_\mu
  & K_\mu \end{array})^T$ as
\begin{align}
\mathcal{L} \supset \dfrac{1}{2} \left(
\begin{array}{ccc}
W^{\mu\, 3} & B^\mu & K^\mu \\
\end{array} \right) \left(
\begin{array}{ccc}
g^2 \dfrac{v_H^2}{4} & -g' g \dfrac{v_H^2}{4} & 0 \\
-g' g \dfrac{v_H^2}{4} & g'^2 \dfrac{v_H^2}{4} & 0 \\
0 & 0 & g_D^2 v_D^2
\end{array}
\right)
\left(
\begin{array}{c}
W_\mu^3 \\ B_\mu \\ K_\mu
\end{array} \right) \ .
\label{eqn:WBKmasses}
\end{align}
In this breaking of $SU(2)_L \times U(1)_Y \times U(1)_D \to
U(1)_{\text{em}}$, the resulting field strength tensors of the individual
neutral vectors corresponding to the gauge eigenstates $W^3$, $B$, and
$K$ all have Abelian field strengths, while non-Abelian vector
interactions are inherited from the $SU(2)_L$ gauge boson field
strength tensor.  We will not explicitly write the non-Abelian vector
interactions in the following, but instead understand that they are
correspondingly modified when we perform the rescaling needed to
canonically normalize the Abelian field strengths of the neutral
vectors.\footnote{We remark that the Stueckelberg
  mechanism~\cite{Ruegg:2003ps, Kors:2005uz} provides an alternative
  mass generation for $\tilde{K}$, which we do not employ here.  The
  collider phenomenology of a dark neutral gauge boson with mass
  arising from the Stueckelberg mechanism is presented in
  Ref.~\cite{Kors:2005uz}.}

\subsection{Neutral vector boson mixing}
\label{subsec:neutralvectors}
To simplify the Lagrangian in the broken phase, we first rotate by the
tree-level SM weak mixing angle,
which reduces the mass matrix to rank 2 and correspondingly modifies
the kinetic mixing between the Abelian field strengths.  Explicitly,
we sandwich $R_{\theta W} R_{\theta W}^T$ twice in~\eqnref{WBKmasses}, with
\begin{align}
R_{\theta W} = \left( \begin{array}{ccc}
c_W & s_W & 0 \\
-s_W & c_W & 0 \\
0 & 0 & 1 
\end{array} \right) \ ,
\end{align}
$c_W = \cos \theta_W$ and $s_W = \sin \theta_W$, which gives
\begin{align}
\mathcal{L} &\supset
\dfrac{-1}{4} \left(
\begin{array}{ccc}
Z_{\text{SM}}^{\mu \nu} & A_{\text{SM}}^{\mu \nu} & K^{\mu \nu} \\
\end{array} \right) \left(
\begin{array}{ccc}
1 & 0 & \epsilon t_W \\
0 & 1 & -\epsilon \\
\epsilon t_W & -\epsilon & 1
\end{array}
\right) \left( \begin{array}{c}
Z_{\mu \nu, \text{ SM}} \\
A_{\mu \nu, \text{ SM}} \\
K_{\mu \nu}
\end{array}
\right) \nonumber \\
&+ \dfrac{1}{2} \left(
\begin{array}{ccc}
Z_{\text{SM}}^\mu & A_{\text{SM}}^\mu & K^\mu \\
\end{array} \right) \left(
\begin{array}{ccc}
m_{Z, \text{ SM}}^2 & 0 & 0 \\
0 & 0 & 0 \\
0 & 0 & m_K^2 
\end{array}
\right) \left(
\begin{array}{c}
Z_{\mu, \text{ SM}} \\
A_{\mu, \text{ SM}} \\
K_\mu
\end{array}
\right) \ ,
\label{eqn:firststep}
\end{align}
where $t_W = \tan \theta_W$, $m_{Z, \text{ SM}}^2 = (g^2 + g'^2) v_H^2
/ 4$ is the tree-level SM $Z$-boson mass, and $m_K^2 = g_D^2 v_D^2 $
is the tree-level $U(1)_D$ gauge boson mass.  To canonically normalize
the kinetic terms for the neutral gauge bosons, we use the successive
transformations
\begin{align}
U_1 = \left( \begin{array}{ccc}
1 & 0 & 0 \\
-\epsilon^2 t_W & 1 & \epsilon \\
-\epsilon t_W & 0 & 1 
\end{array}
\right) \ , \quad
U_2 = \left( \begin{array}{ccc}
\sqrt{\dfrac{1 - \epsilon^2}{1 - \epsilon^2 c_W^{-2}}} & 0 & 0 \\
0 & 1 & 0 \\
\dfrac{-\epsilon^3 t_W}{\sqrt{ (1 - \epsilon^2) 
(1 - \epsilon^2 c_W^{-2})}} & 0 & \dfrac{1}{\sqrt{1 - \epsilon^2}} 
\end{array}
\right) \ ,
\label{eqn:U1U2def}
\end{align}
which give
\begin{align}
\mathcal{L} &\supset
 \dfrac{-1}{4} \left(
\begin{array}{ccc}
Z_{\text{SM}}^{\mu \nu} & A_{\text{SM}}^{\mu \nu} & K^{\mu \nu} \\
\end{array} \right) (U_1^T)^{-1} (U_2^T)^{-1} \mathbb{I}_3 U_2^{-1} U_1^{-1}
\left( \begin{array}{c}
Z_{\mu \nu, \text{ SM}} \\
A_{\mu \nu, \text{ SM}} \\
K_{\mu \nu}
\end{array}
\right) \nonumber \\
\vspace{0.2cm}
&+ \dfrac{1}{2} \left(
\begin{array}{ccc}
Z_{\text{SM}}^\mu & A_{\text{SM}}^\mu & K^\mu \\
\end{array} \right) (U_1^T)^{-1} (U_2^T)^{-1}
\left( \begin{array}{ccc}
\dfrac{m_{Z, \text{ SM}}^2 (1 - \epsilon^2)^2 + m_K^2 \epsilon^2 t_W^2}{(1 - \epsilon^2)(1 - \epsilon^2 c_W^{-2})} & 0 & \dfrac{-m_K^2 \epsilon t_W}{(1 - \epsilon^2) \sqrt{1 - \epsilon^2 c_W^{-2}}} \\
0 & 0 & 0 \\
\dfrac{-m_K^2 \epsilon t_W}{(1 - \epsilon^2) \sqrt{1 - \epsilon^2 c_W^{-2}}} & 0 & \dfrac{m_K^2}{1 - \epsilon^2}
\end{array} \right) \nonumber \\ 
&\times U_2^{-1} U_1^{-1} \left(
\begin{array}{c}
Z_{\mu, \text{ SM}} \\
A_{\mu, \text{ SM}} \\
K_\mu
\end{array}
\right) \ ,
\label{eqn:intermediateKZmasses}
\end{align}
where the kinetic terms are now canonically normalized and only one
further unitary rotation is needed to diagonalize the mass matrix.  We
remark that $|\epsilon| < c_W$ is required to ensure the kinetic
mixing matrix in~\eqnref{firststep} has a positive definite
determinant, which allows $U_2$ to remain non-singular.  The final
Jacobi rotation required is
\begin{align}
R_M = \left( \begin{array}{ccc}
c_M & 0 & s_M \\
0 & 1 & 0 \\
-s_M & 0 & c_M 
\end{array} \right)
\label{eqn:RMdef}
\end{align}
for $c_M = \cos \theta_M$ and $s_M = \sin \theta_M$ and $\theta_M$
defined by $\tan \theta_M = \dfrac{1}{\beta \pm \sqrt{\beta^2 + 1}}$
for
\begin{align}
\beta \equiv \dfrac{ m_{Z, \text{ SM}}^2 (1 - \epsilon^2)^2 - m_K^2 
(1 - \epsilon^2 c_W^{-2} (1 + s_W^2))}{2 m_K^2 \epsilon t_W 
\sqrt{1 - \epsilon^2 c_W^{-2}}} \ ,
\end{align}
and the upper (lower) sign in $\tan \theta_M$ corresponds to $m_{Z,
  \text{ SM}} > m_K$ ($m_{Z, \text{ SM}} < m_K$).  The resulting
non-zero mass eigenvalues are
\begin{align}
m_{\tilde{K}}^2, \, m_{\tilde{Z}}^2 = \dfrac{m_{Z, \text{ SM}}^2 (1 - \epsilon^2) 
+ m_K^2 \pm \sqrt{ (-m_{Z, \text{ SM}}^2 (1 - \epsilon^2) + m_K^2)^2 
+ 4 m_{Z, \text{ SM}}^2 m_K^2 
\epsilon^2 t_W^2}}{2 (1 - \epsilon^2 c_W^{-2})} \ ,
\label{eqn:KZmasses}
\end{align}
and the corresponding neutral vector basis is $R_M^T U_2^{-1} U_1^{-1}
(\begin{array}{ccc} Z_{\mu, \text{ SM}} & A_{\mu, \text{ SM}} &
  K_\mu \end{array})^T$.  We remark that these are exact expressions
valid for arbitrary $\epsilon$.

For $\epsilon \ll 1$, we provide compact expressions for the masses
and the corresponding gauge fields in the mass basis.  To
$\mathcal{O}(\epsilon^3)$,
\begin{align}
m_{\tilde{K}}^2 = m_K^2 + \dfrac{m_K^2 c_W^{-2} \epsilon^2 (m_{Z, \text{ SM}}^2 c_W^2 - m_K^2)}{m_{Z, \text{ SM}}^2 - m_K^2} \ , \quad m_{\tilde{Z}}^2 = m_{Z, \text{ SM}}^2 + \dfrac{m_{Z, \text{ SM}}^4 t_W^2 \epsilon^2}{m_{Z, \text{ SM}}^2 - m_K^2}
\end{align}
and
\begin{align}
& \left( \begin{array}{c}
\tilde{Z}_{\mu} \\
\tilde{A}_{\mu} \\
\tilde{K}_{\mu} 
\end{array} \right) = 
R_M^T U_2^{-1} U_1^{-1} \left( \begin{array}{c} 
Z_{\mu, \text{ SM}} \\ 
A_{\mu, \text{ SM}} \\
K_\mu  
\end{array} \right) = \\
&\left( \begin{array}{c}
Z_{\mu, \text{ SM}} - \dfrac{ t_W m_K^2}{m_{Z, \text{ SM}}^2 - m_K^2} \epsilon K_\mu 
- \dfrac{m_{Z, \text{ SM}}^4 t_W^2}{2(m_{Z, \text{ SM}}^2 - m_K^2)^2} \epsilon^2 
Z_{\mu, \text{ SM}} \\
A_{\mu, \text{ SM}} - \epsilon K_\mu \\
K_\mu + \dfrac{t_W m_{Z, \text{ SM}}^2}{m_{Z, \text{ SM}}^2 - m_K^2} \epsilon 
Z_{\mu, \text{ SM}} - \left( \dfrac{1}{2} 
+ \dfrac{m_K^4 t_W^2}{2(m_{Z, \text{ SM}}^2 - m_K^2)^2} \right) \epsilon^2 K_\mu
\end{array}
\right) \ .
\end{align}
We note that this expansion for $\epsilon \ll 1$ is insufficient for
$m_K \to 0$ or $m_K \to m_{Z, \text{ SM}}$.  These two limits are
discussed in~\appref{limits}.  Given that $\epsilon$ is small, the
masses of $\tilde{K}$ and $\tilde{Z}$ are altered only at the
$\epsilon^2$ level.

With the $\mathcal{O}(\epsilon^3)$ expressions for the mass eigenstate
vectors with canonically normalized kinetic terms, we can now write
down the corresponding currents associated with the mass eigenstate
vectors:
\begin{align}
\mathcal{L} &\supset
 g Z_{\mu, \text{ SM}} J_Z^{\mu} + e A_{\mu, \text{ SM}} J_{\text{em}}^\mu 
+ g_D K_\mu J_D^\mu \nonumber \\
&= \tilde{Z}_\mu \left( g J_Z^\mu - 
g_D \dfrac{m_{Z, \text{ SM}}^2 t_W}{m_{Z, \text{ SM}}^2 - m_K^2} \epsilon J_D^\mu 
+ g \dfrac{m_{Z, \text{ SM}}^2 
(m_{Z, \text{ SM}}^2 - 2 m_K^2) t_W^2}{2(m_K^2 - m_{Z, \text{ SM}}^2)^2} 
\epsilon^2 J_Z^\mu - e \dfrac{m_{Z, \text{ SM}}^2 t_W}{m_{Z, \text{ SM}}^2 - m_K^2} 
\epsilon^2 J_{\text{em}}^\mu \right) \nonumber \\
&+ \tilde{K}_\mu \left( g_D J_D^\mu + 
g \dfrac{m_K^2 t_W}{m_{Z, \text{ SM}}^2 - m_K^2} \epsilon J_Z^\mu 
+ e \epsilon J_{\text{em}}^\mu + g_D \dfrac{(m_{Z, \text{ SM}}^4 c_W^2 
- 2 m_K^2 m_{Z, \text{ SM}}^2 + m_K^4) c_W^{-2}}{2(m_{Z, \text{ SM}}^2 - m_K^2)^2} 
\epsilon^2 J_D^\mu \right) \nonumber \\
&+ \tilde{A}_\mu e J_{\text{em}}^\mu \,.
\label{eqn:currents}
\end{align}
Again, the situation for $m_K \to 0$ or $m_K \to m_{Z, \text{ SM}}$ is
discussed in~\appref{limits}.  From these expressions, we see
explicitly that SM fermions, encoded via $J_{\text{em}}^\mu$ and
$J_Z^\mu$, obtain an $\mathcal{O}(\epsilon)$ electric charge and an
$\mathcal{O}(\epsilon)$ neutral weak charge mediated by
$\tilde{K}_\mu$.  Matter charged in the $U(1)_D$ sector
correspondingly receives an $\mathcal{O}(\epsilon)$ dark charge
mediated by $\tilde{Z}_\mu$.

\subsection{Scalar boson mixing}
\label{subsec:scalarmixing}

The analysis of the scalar sector is simpler and follows previous
discussions of scalar Higgs portals in the literature (see, {\it
  e.g.}~\cite{Kumar:2012ww}).  From~\eqnref{phivev} and~\eqnref{Hvev},
we have
\begin{align}
\mu_D^2 &= \lambda_D v_D^2 + \frac{1}{2} \lambda_{HP} v_H^2 \ ,
\label{eq:muD2} \\
\mu_H^2 &= \lambda_H v_H^2 + \frac{1}{2} \lambda_{HP} v_D^2 \ .
\label{eq:muH2}
\end{align}
The scalar mass eigenstates are then
\begin{align}
\left(
\begin{array}{c} S \\ H_0 \end{array}
\right) = \left(
\begin{array}{cc} 
\cos \alpha & \sin \alpha \\ 
-\sin\alpha & \cos\alpha 
\end{array}
\right) \left( 
\begin{array}{c} \phi \\ h \end{array}
\right) ,
\end{align}
where
\begin{align}
\tan 2\alpha = \frac{\lambda_{HP} v_H v_D}{\lambda_D v_D^2 - \lambda_H v_H^2}
\label{eq:alpha} 
\end{align}
is the scalar mixing angle.  The scalar masses are
\begin{align}
m_{S,\, H_0}^2 = \lambda_H v_H^2 + \lambda_D v_D^2 \pm
\sqrt{(\lambda_H v_H^2 - \lambda_D v_D^2)^2 + \lambda_{HP} v_H^2
  v_D^2} \ .
\end{align}

We can thus reparametrize the scalar Lagrangian couplings $\mu_D$,
$\mu_H$, $\lambda_D$, $\lambda_H$, $\lambda_{HP}$ as $m_S$, $m_{H_0}$,
$v_D$, $v_H$, and $\alpha$.  The reparametrizations for $\mu_D^2$,
$\mu_H^2$ and $\lambda_{HP}$ are given above, while the
reparametrization for $\lambda_D$ and $\lambda_H$ are
\begin{align}
\lambda_H=&\frac{1}{4 v_H^2} \left(m_{H_0}^2 + m_S^2 +(m_{H_0}^2 - m_S^2)\cos 2\alpha \right) \ , \\
\lambda_D=&\frac{1}{4 v_D^2} \left(m_{H_0}^2 + m_S^2 -(m_{H_0}^2 - m_S^2)\cos 2\alpha \right) \ .
\end{align}

We also calculate the scalar interactions in the mass eigenstate basis
$H_0$ and $S$.  The cubic scalar interactions are
\begin{align}
\mathcal{L} \supset & - S^3 m_S^2  \frac{ v_H \cos^3\alpha + v_D \sin^3\alpha }{2 v_D v_H}
- H_0^3 m_{H_0}^2  \frac{ v_D \cos^3\alpha - v_H \sin^3\alpha }{2 v_D v_H}  \nonumber \\
& + H_0 S^2 \frac{m_{H_0}^2 + 2 m_{S}^2}{4 v_D v_H} (v_H \cos\alpha - v_D \sin\alpha)\sin 2 \alpha \nonumber \\
& - H_0^2 S \frac{2 m_{H_0}^2 +  m_{S}^2}{4 v_D v_H} (v_D \cos\alpha + v_H \sin\alpha)\sin 2 \alpha \ .
\label{eqn:SSSinteractions}
\end{align}
We have, of course, $m_{H_0} = 125$~GeV and $v_H = 246$~GeV, but the
other observables are free parameters.  We will restrict $\lambda_{HP}
> 0$ in our analysis, recognizing that $\lambda_{HP} < 0$ and
$|\lambda_{HP}| > \sqrt{\lambda_H \lambda_D}$ can cause tree-level
destabilization of the electroweak vacuum.

Lastly, the scalar-vector-vector interactions of $\tilde{K}$,
$\tilde{Z}$, $S$ and $H_0$ in the mass basis to
$\mathcal{O}(\epsilon^2)$ are
\begin{align}
\mathcal{L} &\supset m^2_{Z,\text{SM}}  \left(\frac{\cos\alpha}{v_H} - \epsilon^2  t_W^2 \frac{m_K^2 m^2_{Z,\text{SM}}  }{(m_K^2- m^2_{Z,\text{SM}})^2} \frac{\sin{\alpha}}{v_D}\right) \tilde{Z}_\mu \tilde{Z}^\mu H_0 \nonumber \\
&+ 2 \epsilon t_W \frac{ m_K^2 m_{Z, \text{ SM}}^2  }{(m_{Z, \text{ SM}}^2 - m_K^2)} \left( \frac{\cos\alpha}{v_H} +\frac{\sin\alpha}{v_D} \right)  \tilde{Z}_\mu\tilde{K}^\mu H_0 \nonumber \\
&+ m_K^2 \left( - \frac{\sin \alpha}{v_D}  + \epsilon^2 t_W^2 \frac{m_K^2 m_{Z, \text{ SM}}^2  }{(m_K^2 - m_{Z, \text{ SM}}^2)^2 } \frac{\cos\alpha}{v_H}\right) \tilde{K}_\mu \tilde{K}^\mu H_0 \nonumber \\
&+ m^2_{Z,\text{SM}}  \left(\frac{\sin \alpha }{v_H} + \epsilon^2 t_W^2  \frac{m_K^2 m_{Z, \text{ SM}}^2  }{(m_K^2 - m_{Z, \text{ SM}}^2)^2 } \frac{\cos\alpha}{v_D} \right) \tilde{Z}_\mu \tilde{Z}^\mu S  \nonumber \\
&+  2 \epsilon t_W \frac{ m_K^2 m_{Z, \text{ SM}}^2  }{(m_{Z, \text{ SM}}^2 - m_K^2)} \left( - \frac{\cos\alpha}{v_D} + \frac{\sin\alpha}{v_H} \right) \tilde{Z}_\mu \tilde{K}^\mu S \nonumber \\
&+ m_K^2 \left( \frac{\cos \alpha}{v_D}  + \epsilon^2 t_W^2 \frac{m_K^2 m_{Z, \text{ SM}}^2  }{(m_K^2 - m_{Z, \text{ SM}}^2)^2 } \frac{\sin\alpha}{v_H}\right)  \tilde{K}_\mu \tilde{K}^\mu S \ .
\label{eqn:SVVinteractions}
\end{align}
We reiterate that both $\alpha$ and $\epsilon$ are theoretical
parameters that must be constrained by data, and hence a particular
hierarchy between $\alpha$ and $\epsilon$ would reflect
model-dependent assumptions.  As a result,~\eqnref{SVVinteractions}
forms a consistent basis for determining the sensitivity to $\alpha$
and $\epsilon$ simultaneously.

We can characterize the changes in the phenomenology of the Higgs-like
$H_0$ state as a combination of modified SM-like production and decay
modes and the opening of new exotic production and decay channels.
One main effect of $\alpha$ is to suppress all of the SM fermion
couplings of the $H_0$ state by $\cos \alpha$, while the $S$ state
acquires Higgs-like couplings to SM fermions proportional to $\sin
\alpha$.  This feature also applies to the loop-induced couplings to
gluons and photons for $H_0$ and $S$.  On the other hand, the coupling
between $H_0$ to $\tilde{Z}$ bosons is changed not only by $\cos
\alpha$ but also by $\epsilon^2 \sin \alpha$, while the $S$ state
acquires a $\tilde{Z}$ coupling proportional to $\sin \alpha$ and also
$\epsilon^2 \cos \alpha$.

In addition, if kinematically open, the $H_0$ state can decay to pairs
of $S$ or pairs of $\tilde{K}$, with $S \to \tilde{K} \tilde{K} \to 4
\chi$ and $\tilde{K} \to \bar{\chi} \chi$ as possible subsequent
decays.  These Higgs invisible decays are also mimicked by the exotic
$H_0 \to \tilde{Z} \tilde{K}$ decay, when $\tilde{Z} \to \bar{\nu}
\nu$.  As a result, the total invisible width of $H_0$ is sensitive to
a combination of different couplings and masses
in~\eqnref{SSSinteractions} and~\eqnref{SVVinteractions}, further
demonstrating the viability of the Double Dark Portal model as a
self-consistent theoretical framework for constraining Higgs
observables.  We remark that we have not added a direct Yukawa
coupling between $\Phi$ and $\chi$, {\it e.g.} if $\Phi$ has charge
$+2$ under $U(1)_D$, we would introduce a direct decay from $S$ to
dark matter and also split the Dirac dark matter into Majorana
fermions~\cite{TuckerSmith:2001hy}.

\subsection{Dark matter interactions}
\label{subsec:DMints}
Finally, we will consider the DM interactions with the mass
eigenstates of the gauge bosons and scalars.  The main observations
can be obtained by recognizing that DM inherits its couplings to SM
particles via the $J_D$ current shown in~\eqnref{currents}.
Explicitly, the dark matter particle Lagrangian reads
\begin{align}
\mathcal{L} &\supset i\bar{\chi} \slashed{\partial}_\mu \chi + g_D
\bar{\chi} \left(\tilde K_\mu + \frac{t_W m_{Z, \text{ SM}}^2}{m_K^2 -
  m_{Z, \text{ SM}}^2} \epsilon \tilde{Z}_\mu \right) \gamma^\mu \chi
- m_{\chi} \bar{\chi} \chi \ .
\end{align}

\section{Direct detection and indirect detection phenomenology and constraints}
\label{sec:DDIDpheno}

The Double Dark Portal model presented in~\eqnref{DDP} offers many
phenomenological opportunities, including dark matter signals at
direct detection, indirect detection, and collider experiments and
modifications of electroweak precision and Higgs physics at colliders.
We remark that aside from the vacuum stability requirement on
$\lambda_{HP}$ and upper bound on $|\epsilon|$, the theory parameter
space of the Double Dark Portal model is wide open and subject only to
experimental constraints.  This vast parameter space has been
extremely useful in motivating searches for light, hidden mediators at
high intensity, beam-dump experiments, as reviewed in
Refs.~\cite{Essig:2013lka, Alexander:2016aln}.

Our focus, however, is the $\mathcal{O}(10-100~\text{GeV})$ scale for
the $\tilde{K}$ vector mediator and its accompanying Higgs partner
$S$, which will both dominantly decay to the dark matter particle
$\chi$.  This is readily motivated by considering $m_{\chi} <
m_{\tilde{K}}/2$ and $g_D \gg \epsilon$, so that $\tilde{K}$ has an
on-shell two-body decay to $\bar{\chi} \chi$ and an $\epsilon^2 /
g_D^2$ suppressed branching ratio to SM charged particles.  Moreover,
for $m_{\tilde{K}} < m_S /2$ and $\sin \alpha \ll g_D$, the SM gauge
singlet scalar $S$ dominantly decays to pairs of $\tilde{K}$ and only
have $\sin^2 \alpha / g_D^2$ suppressed rates to SM pairs.  All of
these choices, however, can be reversed to give markedly different
phenomenology.  If $m_{\chi} > m_{\tilde{K}} / 2$, for example, then
the total width of $m_{\tilde{K}}$ scales as
$\epsilon^2$~\cite{Karliner:2015tga} and $\tilde{K}$ decays to pairs
of SM charged fermions, as long as it is heavier than $2 m_e$.  For
very small $\epsilon$, however, the $\tilde{K}$ lifetime can be long,
leading to either displaced vertex signatures or missing energy
signatures.  The lifetime and decay length of $\tilde{K}$ can be
estimated to be
\begin{align}
\tau = \dfrac{1}{\Gamma} = 0.9 \times 10^{-2} \text{ ps} 
\left( \dfrac{1 \text{ GeV}}{m_{\tilde{K}}} \right) 
\left( \dfrac{10^{-4}}{\epsilon} \right)^2 \dfrac{1}{\tilde{N}_F} \ , \\
L = \gamma c \tau \approx
150 \times 10^{-6} \text{m} 
\left( \dfrac{1 \text{ GeV}}{m_{\tilde{K}}} \right) 
\left( \dfrac{10^{-4}}{\epsilon} \right)^2 \dfrac{1}{\tilde{N}_F} \ ,
\end{align}
for $\gamma = 60$ (as from a Higgs two-body decay) and $\tilde{N}_F$
is the effective factor for kinematically open charge-weighted
two-body SM final states.  If $m_S$, $m_{\tilde{K}} > m_{H_0} / 2$,
then any possible exotic decay of the SM-like Higgs will be strongly
suppressed by multi-body phase space and a combination of $g_D$, $\sin
\alpha$, or $\lambda_{HP}$.  We remark that choosing $m_S < m_{H_0} /
2$ already gives an exotic Higgs decay, $H_0 \to S S$, which is
sensitive directly to $\lambda_{HP}$.

Given our mass hierarchy, the dominant collider signature from
production of either $\tilde{K}$ or $S$ is missing energy from
escaping $\chi$ particles, while the relic density of $\chi$ in our
local dark matter halo can be probed via nuclear recoils in
terrestrial direct detection experiments or through their annihilation
products in satellite indirect detection experiments.  We will discuss
the direct and indirect constraints from dark matter searches in the
remainder of this section and focus on the collider signatures for
vector and scalar mediator production in~\secref{ColliderLEPLHC}
and~\secref{futureprospects}.

\subsection{Direct detection and relic abundance}
\label{subsec:DDpheno}

Dark matter direct detection experiments search for anomalous nuclear
recoil events consistent with the scattering of the dark matter halo
surrounding Earth.  Direct detection scattering occurs via $t$-channel
exchange of $\tilde{Z}$ and $\tilde{K}$, as evident from the
$J_D^{\mu}$ and $J_{\text{em}}^\mu$ interactions shown
in~\eqnref{currents}.  Because of the relative sign between the
$\tilde{K}$ and $\tilde{Z}$ terms, dark matter scattering proportional
to $g^2 g_D^2 \epsilon^2$ is naturally suppressed by extra
$\epsilon^2$ or $Q^2/m_K^2$ factors, where $Q$ is the momentum
transfer scale, and the leading contribution is hence proportional to
$e^2 g_D^2 \epsilon^2$.  This cancellation between $\tilde{K}$ and
$\tilde{Z}$ mediators is generic, and we outline the details
in~\appref{cancellation}.  As a result, the dominant DM-nucleon
interaction for direct detection is mainly from DM-proton scattering.
With the SM and DM currents from~\eqnref{currents}, the DM-proton
scattering cross-section is
\begin{align}
\sigma_p \simeq \frac{\epsilon^2 g_D^2 e^2}{\pi} 
\frac{\mu_{\chi p}^2}{m_{\tilde{K}}^4} 
\approx 10^{-44} \text{ cm}^2 
\left( \frac{g_D}{e} \right)^2 \left( \frac{\epsilon}{10^{-5}} \right)^2
\left( \frac{10 \text{ GeV}}{m_{\tilde{K}}} \right)^2 \ ,
\label{eqn:DDxsec}
\end{align}
where $\mu_{\chi p}$ is the reduced mass of the dark matter $\chi$ and
the proton and $e = \sqrt{4 \pi/137}$.  The cross-section $\sigma_p$
is calculated at leading order in $\epsilon$ and $v_{\text{in}}$, the
incoming DM velocity, and agrees with previous results when DM only
interacts via $t$-channel $\tilde{K}$ exchange with strength
proportional to the SM electromagnetic current~\cite{Liu:2014cma}.

Given that the momentum transfer in the propagator is smaller than
gauge boson masses $m_{\tilde{K}}$ and $m_{\tilde{Z}}$, then the
scattering amplitude is $\mathcal{O}(Q^2/m_V^2)$ suppressed after
summing all the vector boson contributions, where $m_V$ is the smaller
of either gauge boson mass.  For DM direct detection, the momentum
transfer is about $Q^2 \sim (m_\chi v_{\text{in}})^2 \ll m_{\tilde{K},
  \tilde{Z}}^2$, hence the contribution induced by the $J_Z$ current
cancels and we arrive at the same result in Ref.~\cite{Liu:2014cma}.

We can also motivate particular contours in the $\epsilon$
vs.~$m_{\tilde{K}}$ plane by considering first the requirement that
the DM obtains the correct relic density and second the constraints on
the possible rates for DM annihilation to SM particles coming from
indirect detection experiments.  We first calculate the annihilation
cross sections for $\chi \bar{\chi} \to f \bar{f}$, $W^+ W^-$, where
$f$ denotes a SM fermion.  We focus on the region $m_\chi <
m_{\tilde{K}}$, since the annihilation channel $\chi \bar{\chi} \to
\tilde{K} \tilde{K}$ opens up otherwise and the dominant
self-annihilation cross section is insensitive to $\epsilon$.

In this setup, the annihilation cross section will be proportional to
$g_D^2 \epsilon^2$.  We calculate the annihilation in center of mass
frame, and give the annihilation cross sections before thermal
averaging in~\appref{annXS}. We perform the thermal averaging of the
annihilation cross section numerically according to
Ref.~\cite{Gondolo:1990dk}.  The annihilation cross section generally
has three physical resonances, $m_{\tilde{K}} = 2 m_\chi$,
$m_{\tilde{Z}} = 2 m_\chi$ and $m_{\tilde{K}} = m_{\tilde{Z}}$.  The
first two resonances are from the $s$-channel resonant exchange of
$\tilde{K}$ and $\tilde{Z}$, while the last one is due to maximal
mixing between $\tilde{K}$ and $\tilde{Z}$ when $m_K$ is close to
$m_{\text{Z, SM}}$, as discussed in~\appref{limits}.

\begin{figure*}[tb!]
\includegraphics[width=0.48\textwidth]{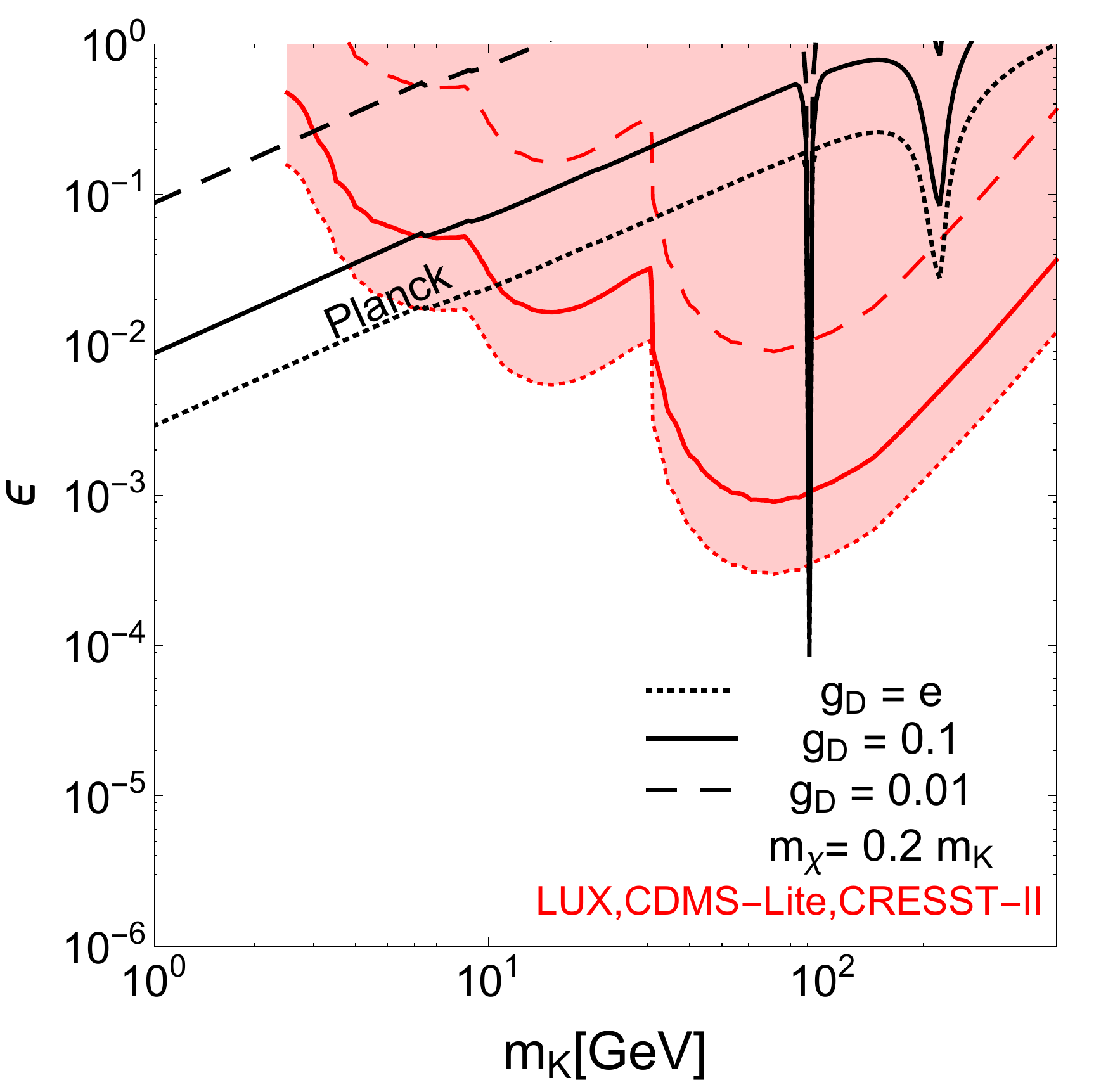}
\includegraphics[width=0.48\textwidth]{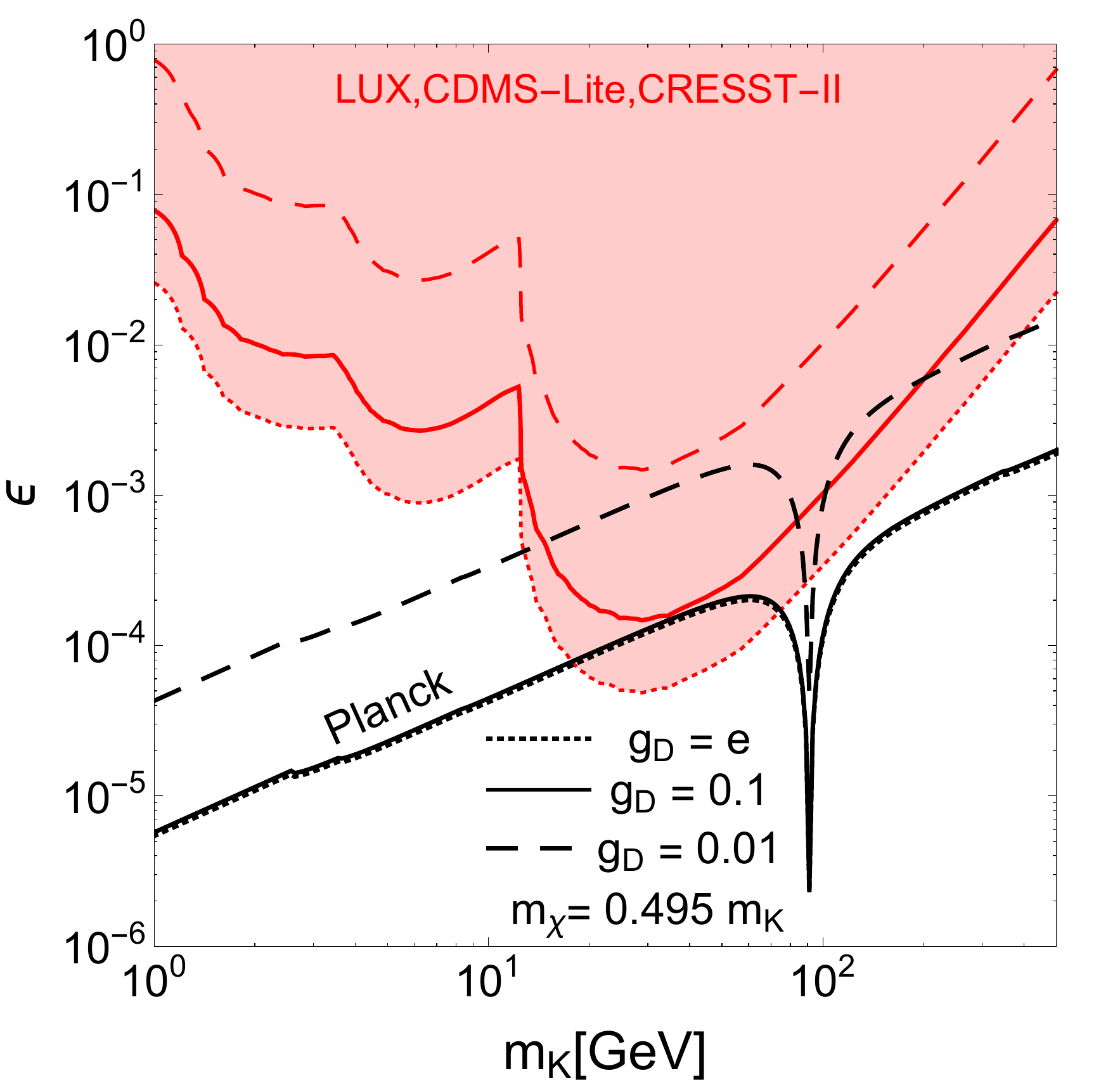}
\\
\includegraphics[width=0.48\textwidth]{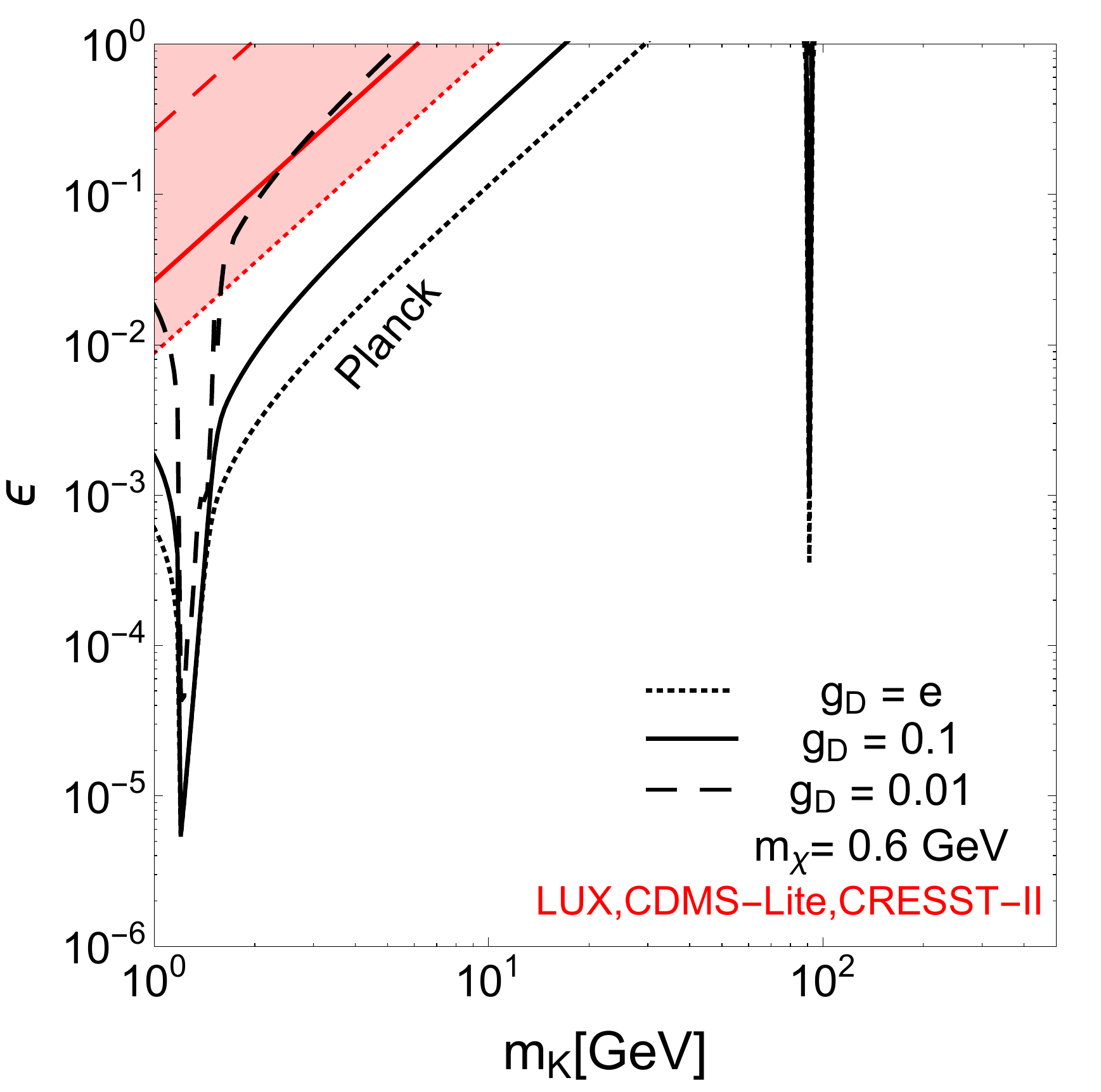}
\includegraphics[width=0.48\textwidth]{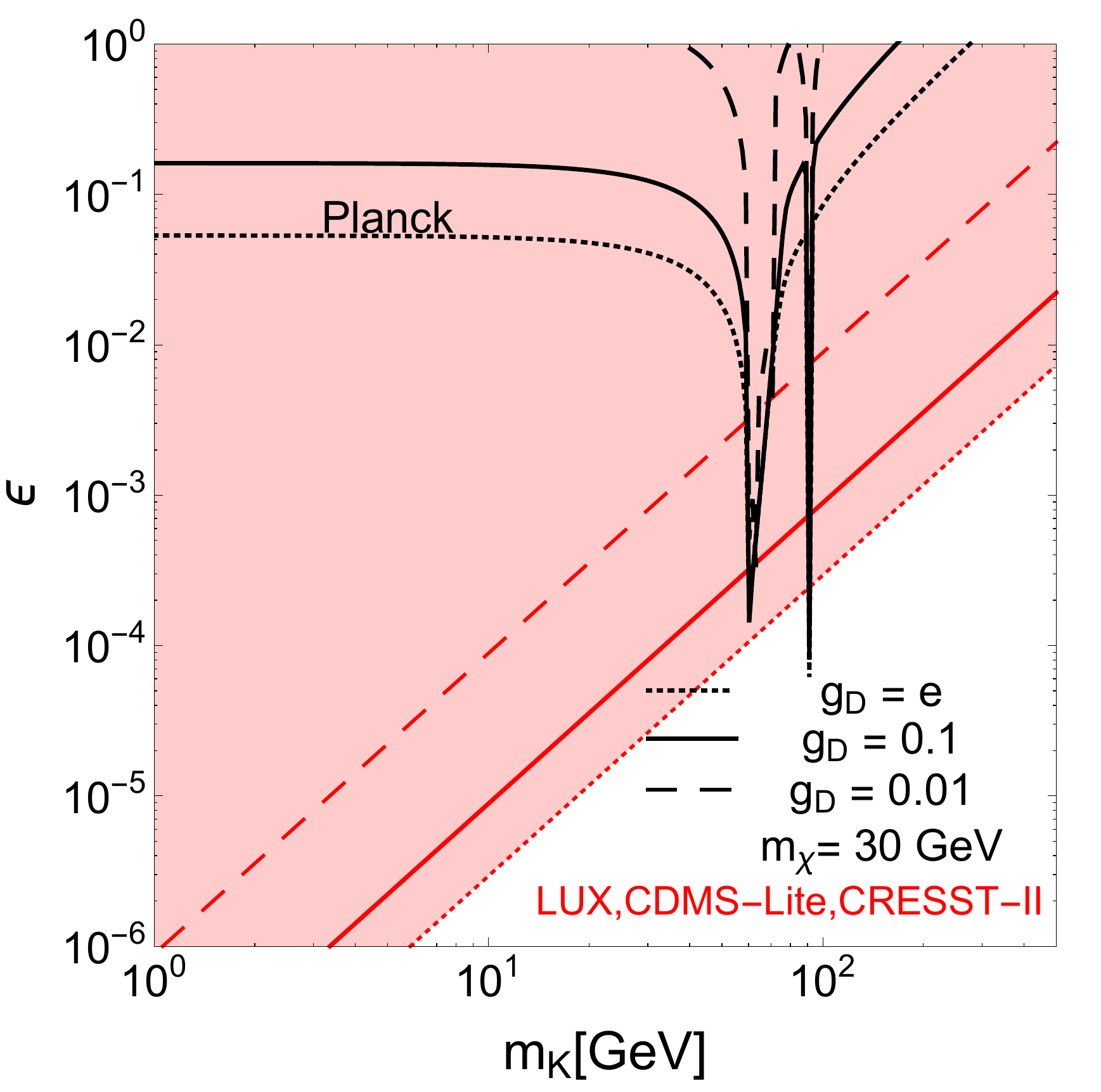}
\caption{The leading direct detection constraints from
 LUX~\cite{Akerib:2016vxi}, PANDAX-II~\cite{Tan:2016zwf}, and
  CRESST-II~\cite{Angloher:2015ewa}, as well as
  CDMSlite~\cite{Agnese:2015nto}, shown in the $\epsilon$ vs.~$m_K$
  plane for various choices of $g_D$ and $m_{\chi}$.  We fix $m_{\chi}
  = 0.2 m_K$ (top left), $m_{\chi} = 0.495 m_K$ (top right), $m_{\chi}
  = 0.6$~GeV (bottom left), and $m_{\chi} = 30$~GeV (bottom right), to
  demonstrate the dependence on the dark matter mass.  In each panel,
  the shaded regions show the exclusions from direct detection
  experiments for $g_D = e$ (dotted), $g_D =
  0.1$ (solid), and $g_D = 0.01$ (dashed), and we overlay black
  contours to mark the relic density requirement from the Planck
  collaboration~\cite{Ade:2015xua}.  Note that $m_K$ is approximately
  the $m_{\tilde{K}}$ mass eigenvalue according to~\eqnref{KZmasses}.}
\label{fig:DMdirectdetection}
\end{figure*}

In~\figref{DMdirectdetection}, we show the direct detection
constraints in the $\epsilon$ vs.~$m_K$ plane from experiments
LUX~\cite{Akerib:2016vxi} with data from 2013 to 2016,
PANDAX-II~\cite{Tan:2016zwf}, and CRESST-II~\cite{Angloher:2015ewa} as
well as CDMSlite~\cite{Agnese:2015nto} for low mass DM.  Each panel
shows choices of $g_D = e$, $0.1$, and $0.01$, and the dark matter
mass fixed to $0.2 m_K$, $0.495 m_K$, $0.6$~GeV, or $30$~GeV.

The thermal relic abundance limit on $\epsilon$ is given
in~\figref{DMdirectdetection} using the $\Omega h^2 = 0.12$
requirement from the Planck collaboration~\cite{Ade:2015xua}.  The dip
around $m_{K} \sim m_{\text{Z, SM}}$ reflects increasing mixing
between $\tilde{K}$ and $\tilde{Z}$.  While for $m_{K} \sim 2 m_\chi$,
the annihilation cross section is enhanced by the $s$-channel
$\tilde{K}$ resonance, thus the required $\epsilon$ is very small.
When $2 m_{\chi} - m_{\tilde{K}} < T_f$, where $T_f \approx m_{\chi} /
25$ is the DM freeze-out temperature, the annihilation cross section
is enhanced by $(m_{\tilde{K}}/T_f)^2$ due to thermal averaging.  When
$2 m_{\chi} - m_{\tilde{K}} > 0 $, the thermal average will not
benefit the resonance effect anymore.

In recasting the direct detection limits, we recognize that the
experiments assume that the local DM density is fixed to
$0.3~\text{GeV} / \text{cm}^3$.  Hence, the respective constraints are
identically meaningful only when the DDP model parameters give this
assumed local relic density.  For other parameter space points, in
particular for fixed $m_K$ and varying $\epsilon \neq
\epsilon_{\text{relic}}$, with $\epsilon_{\text{relic}}$ corresponding
to $\langle \sigma v \rangle = 0.3~\text{GeV} / \text{cm}^3$, the
predicted rate of direct detection scattering events will be
independent of $\epsilon$.  This is because the predicted local DM
relic density will scale with $(\epsilon / \epsilon_{\text{relic}})^2$
while the scattering cross section will scale with
$(\epsilon_{\text{relic}} / \epsilon)^2$, leaving the product, and
thus the predicted direct detection rate, insensitive to $\epsilon$.

In our recasting, however, we keep the local DM relic density fixed to
$0.3~\text{GeV} / \text{cm}^3$ regardless of $\epsilon$, in order to
determine the sensitivity to the direct detection cross section.  For
large $\epsilon$, when the local DM relic density predicted in the DDP
model is generally underabundant, extra dark matter particles beyond
the DDP model are needed, while for small $\epsilon$, the DM relic
density is generally overabundant and extra annihilation channels are
typically needed.  

Hence, the direct detection exclusion contours in each panel simply
illustrate the fractional $\chi$ relic density, relative to $\Omega
h^2 = 0.12$, that is excluded by the direct detection constraint.
When the DD contours are weaker than the relic density contours, the
model only minimally requires extra inert dark matter to make up the
absent relic abundance.  When the DD contours are stronger than the
relic density contours, an extra contribution to the thermal relic
annihilation cross section for $\chi$ is required to satisfy the
$0.3~\text{GeV} / \text{cm}^3$ assumption, and $\epsilon$ is excluded
by DD experiments as shown in the red shaded region.  In particular,
for fixed $m_K$, the strengthening to the annihilation cross section
can be parametrized by the squared ratio of $\epsilon$ from the blue
contour to $\epsilon$ at the red contour.

We see that light DM masses are much less constrained, because of the
$\mu_{\chi p}^2 / m_{\tilde{K}}^2$ suppression in~\eqnref{DDxsec}.
For $m_{\chi} \propto m_{\tilde{K}}$, the sensitivity on $\epsilon$
generally follows the experimental constraint on $\sigma_p$.  We see
that for heavy $\tilde{K}$ and light $\chi$, the direct detection
sensitivity can be weak, leaving significant parameter space to be
probed by colliders.  Interesting parameter space also exists for
$m_\chi \lesssim m_{\tilde{K}}/2$, which will be discussed further in
the next section.

\subsection{Indirect constraints from CMB, Gamma-ray and $e^{\pm}$ measurements}
\label{subsec:indirect}

\begin{figure*}[tb!]
\includegraphics[width=0.48\textwidth]{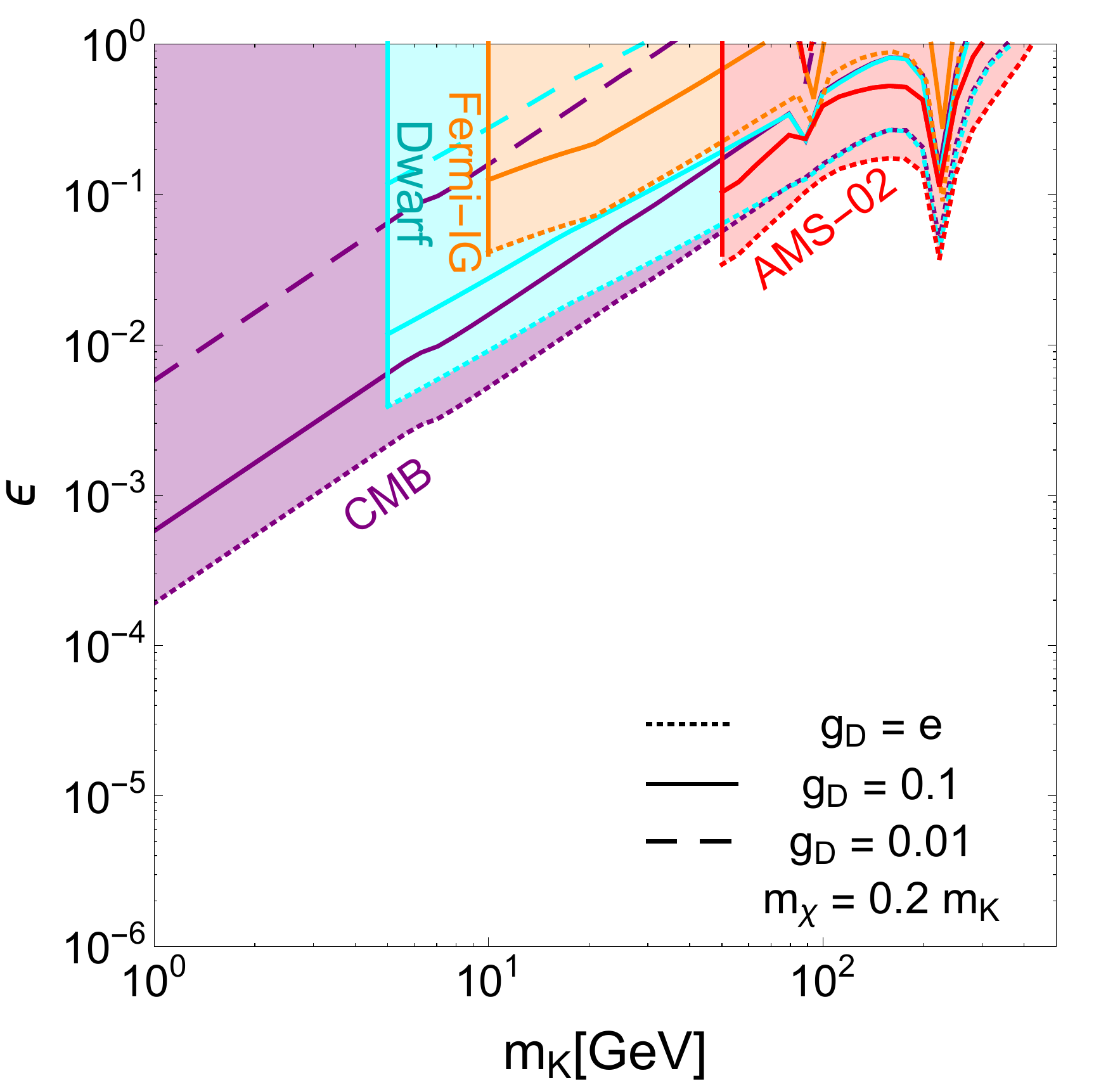}
\includegraphics[width=0.48\textwidth]{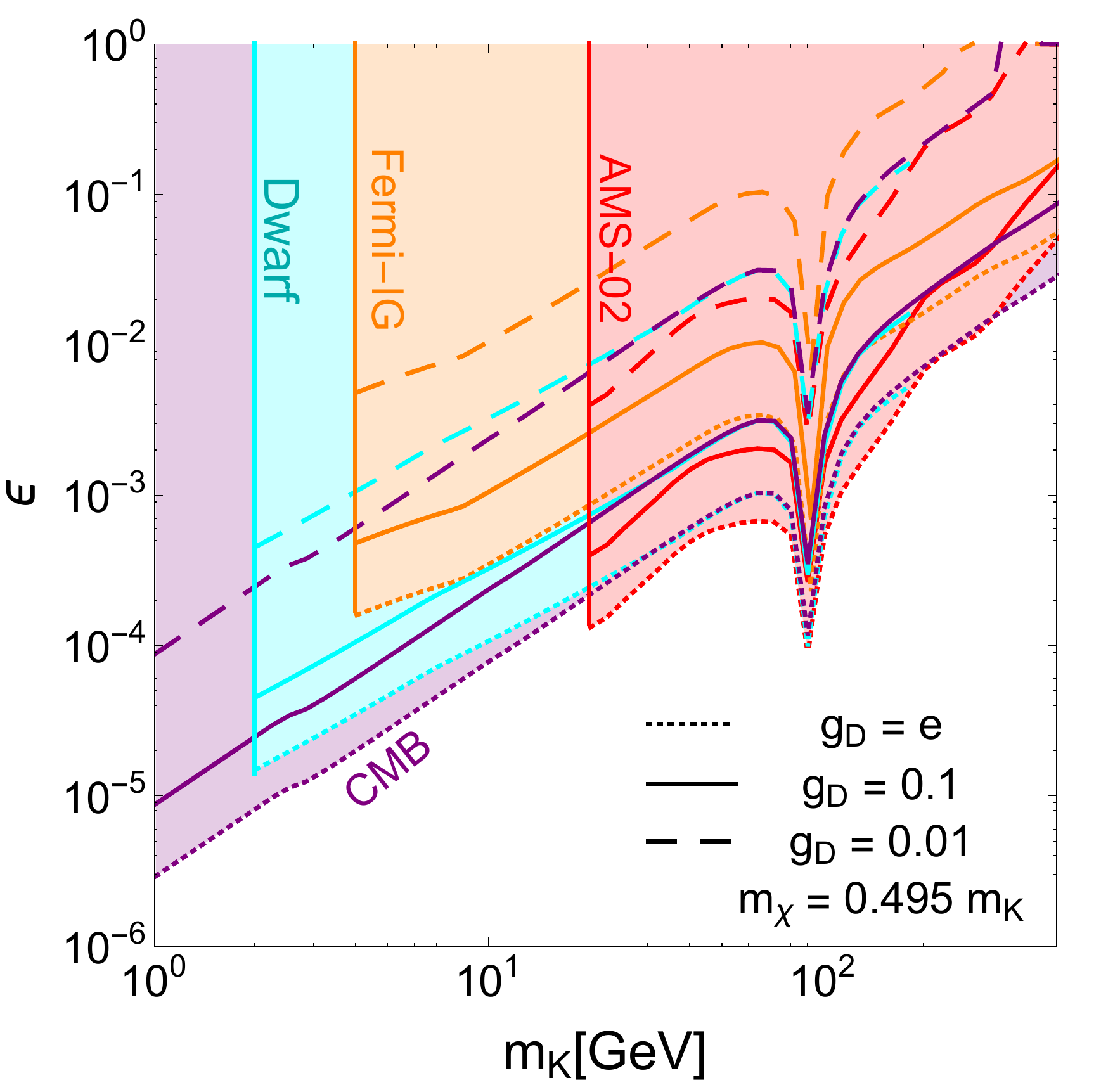} \\
\includegraphics[width=0.48\textwidth]{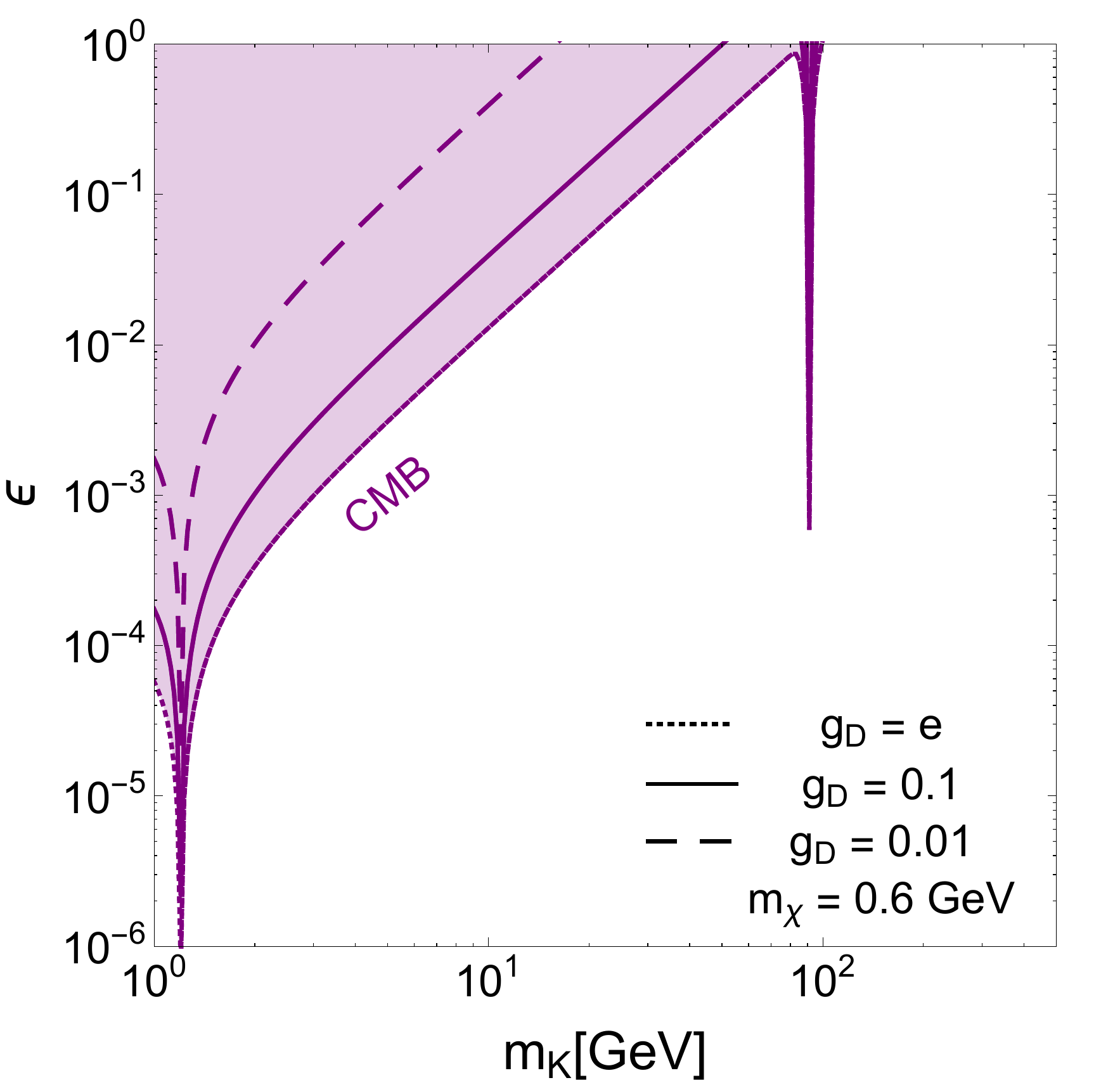}
\includegraphics[width=0.48\textwidth]{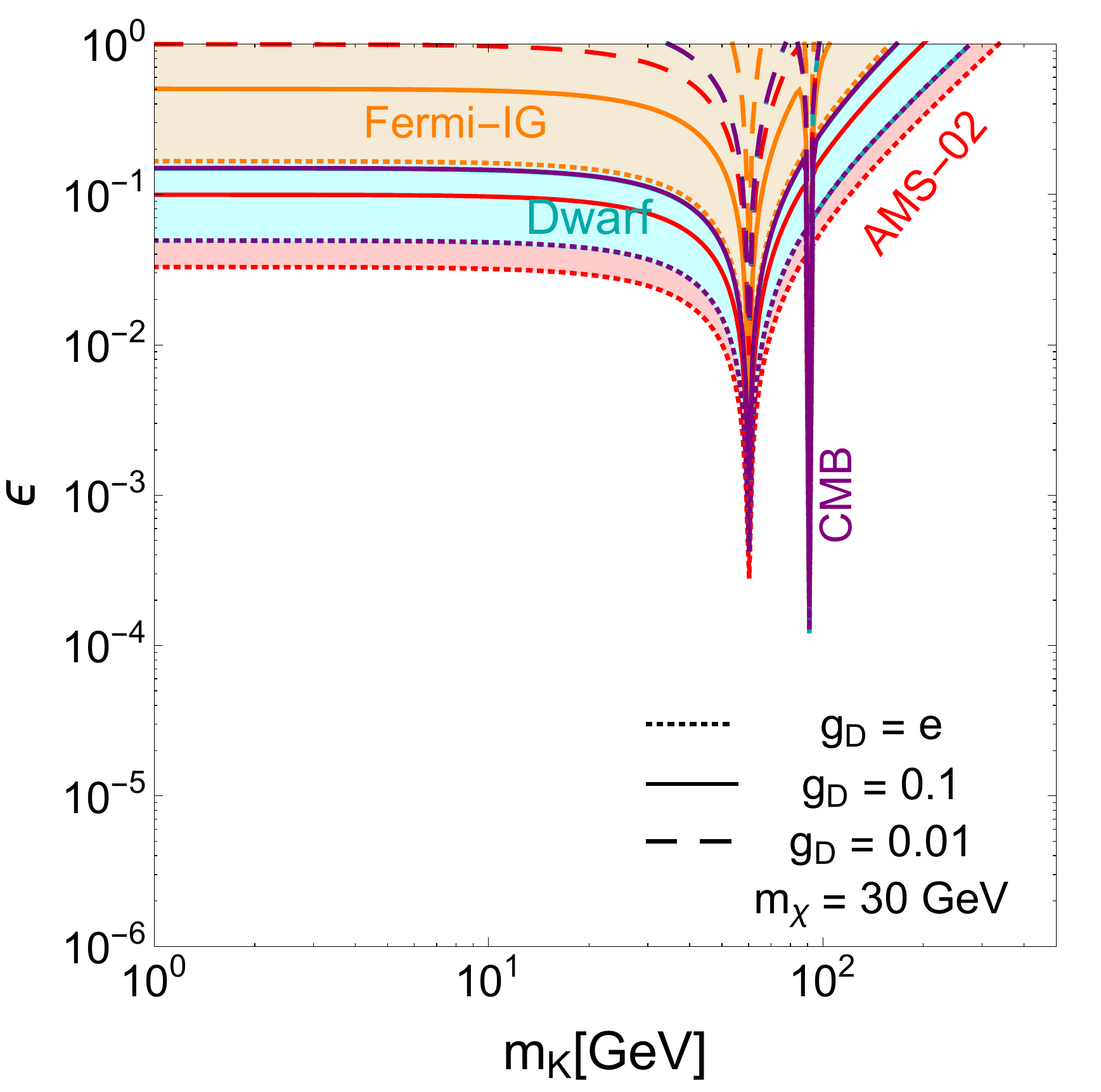}
\caption{The indirect detection constraints from CMB
  measurements~\cite{Ade:2015xua}, gamma-ray measurements from dwarf
  galaxies~\cite{Ackermann:2015zua, Ahnen:2016qkx} and the inner
  galactic region~\cite{Massari:2015xea} and $e^+$ flux measurement
  from AMS-02~\cite{Aguilar:2014mma}. The constraints are shown in the
  $\epsilon$ vs.~$m_K$ plane for $m_{\chi} = 0.2~m_K$ (top left),
  $0.495~m_K$ (top right), $0.6$~GeV (bottom left), and $30$~GeV
  (bottom right), with $g_D = e$ (dotted), $0.1$ (solid), and $0.01$
  (dashed).  Note that $m_K$ is approximately the $m_{\tilde{K}}$ mass
  eigenvalue according to~\eqnref{KZmasses}.}
\label{fig:DMindirectdetection}
\end{figure*}

After the relic abundance constraint, we next consider the constraints
from cosmic microwave background (CMB) observations.  Measurements of
the CMB generally give constraints on DM annihilation or decay
processes, which inject extra energy into the CMB and thus delay
recombination~\cite{Adams:1998nr, Padmanabhan:2005es, Galli:2009zc,
  Slatyer:2009yq, Finkbeiner:2011dx, Madhavacheril:2013cna}.  The
constraint is calculated using the energy deposition yield,
$f_{\text{eff}}^i$, where $i$ denotes a particular annihilation or
decay channel and $f_{\text{eff}}$ describes the efficiency of energy
absorption by the CMB from the energy released by DM in particular
channel.  The constraint is expressed as
\begin{align}
p_\text{ann} =\frac{1}{m_\chi} \sum_i f_{\text{eff}}^i
\left\langle {\sigma v} \right\rangle_i \ ,
\end{align}
where the Planck experiment has constrained $p_\text{ann} < 4.1 \times
10^{-28} \text{ cm}^3 \text{ s}^{-1} \text{
  GeV}^{-1}$~\cite{Ade:2015xua}, and we sum all the SM fermion pair
$\bar{f} f$ and $W^+ W^-$ channels in annihilation. The excluded
parameters are plotted in~\figref{DMindirectdetection} as shaded
purple regions.

The next constraints we consider are the gamma ray observations from
Fermi-LAT and MAGIC in dwarf galaxies~\cite{Ackermann:2015zua,
  Ahnen:2016qkx}.  In Ref.~\cite{Ackermann:2015zua}, Fermi-LAT gives
constraints on $e^+ e^-$, $\mu^+\mu^-$, $\tau^+\tau^-$, $\bar{u}u$,
$\bar{b}b$ and $W^+W^-$ final states, while in
Ref.~\cite{Ahnen:2016qkx}, MAGIC has made a combined analysis with
Fermi-LAT and presented constraints on $\mu^+\mu^-$, $\tau^+\tau^-$,
$\bar{b}b$ and $W^+W^-$.  The computation of constraints on our model
is straightforward, since we can calculate each individual limit on
$\epsilon$ for each channel and take the most stringent constraint for
each $m_\chi$ mass.  The excluded parameter region is shaded by cyan
in~\figref{DMindirectdetection}.  Similarly, we consider the gamma ray
constraints from the inner Milky way~\cite{Massari:2015xea}.  This
analysis sets conservative constraints on various SM final states by
using the inclusive photon spectrum observed by the Fermi-LAT
satellite.  We apply their results by calculating the most stringent
annihilation profile, assuming the Navarro-Frenk-White profile for the
DM density distribution in the galactic center~\cite{Navarro:1995iw}.
We can see in~\figref{DMindirectdetection} that the constraint from
galactic center region, shaded in orange, is much weaker than that
from dwarf galaxies.

The last indirect detection constraint is based on $e^+$ and $e^-$
data from the AMS-02 satellite~\cite{Aguilar:2014mma}.  We use the
constraints from Ref.~\cite{Elor:2015bho} to set bounds on various SM
final states, which mainly derive from the observed positron flux.  We
again adopt the limits from the strongest channel to constrain
$\epsilon$ for each mass parameter choice. We can see the constraint
from AMS-02 is the strongest at the largest $m_{\chi}$ masses
in~\figref{DMindirectdetection}.

To summarize, in~\figref{DMindirectdetection}, we see that the CMB
constraint is strongest at small $m_{\chi}$, while AMS-02 is strongest
at higher $m_{\chi}$.  The constraint from gamma ray observations in
dwarf galaxies is very close to the CMB constraint.  Meanwhile, the
dips in~\figref{DMindirectdetection} nicely show the two $s$-channel
resonances of $\tilde{K}$ and $\tilde{Z}$ as well as the maximal
mixing peak between $\tilde{K}$ and $\tilde{Z}$.

\section{Collider phenomenology of the Double Dark Portal model and current constraints from LEP and LHC}
\label{sec:ColliderLEPLHC}

In this section, we give an overview of the possible probes of the
Double Dark Portal Model at both lepton and hadron colliders.  While
many separate searches have been performed at LEP and LHC experiments
in the context of either kinetic mixing or Higgs mixing scenarios, we
highlight the fact that a future $e^+ e^-$ machine must synthesize
both effects in any given search.  Hence, the Double Dark Portal model
is a natural framework to study light, hidden physics at a future $e^+
e^-$ machine.

The Double Dark Portal model motivates observable deviations in
measurements of both the SM-like $H_0$ and the $\tilde{Z}$ bosons,
which test the scalar mixing angle $\alpha$ as well as the kinetic
mixing parameter $\epsilon$.  Notably, the primary SM Higgsstrahlung
workhorse process at an $e^+ e^-$ Higgs factory, $e^+ e^- \to Z h$,
can deviate significantly from the SM expectation for nonzero
$\epsilon$ or $\alpha$.  For instance, nonzero $\alpha$ causes a
well-known $\cos \alpha$ suppression of the $H_0 \tilde{Z}_\mu
\tilde{Z}^\mu$ vertex, but nonzero $\epsilon$ gives an additional
diagram with intermediate $\tilde{K}$, which becomes on-shell when
$m_{\tilde{K}} > m_{\tilde{Z}} + m_{H_0}$.  We remark that these
effects are not generically captured by a simple $\cos \alpha$
rescaling of the $H_0 \tilde{Z}_\mu \tilde{Z}^\mu$ vertex.

\begin{figure*}[tb!]
 \includegraphics[width=0.75\textwidth]{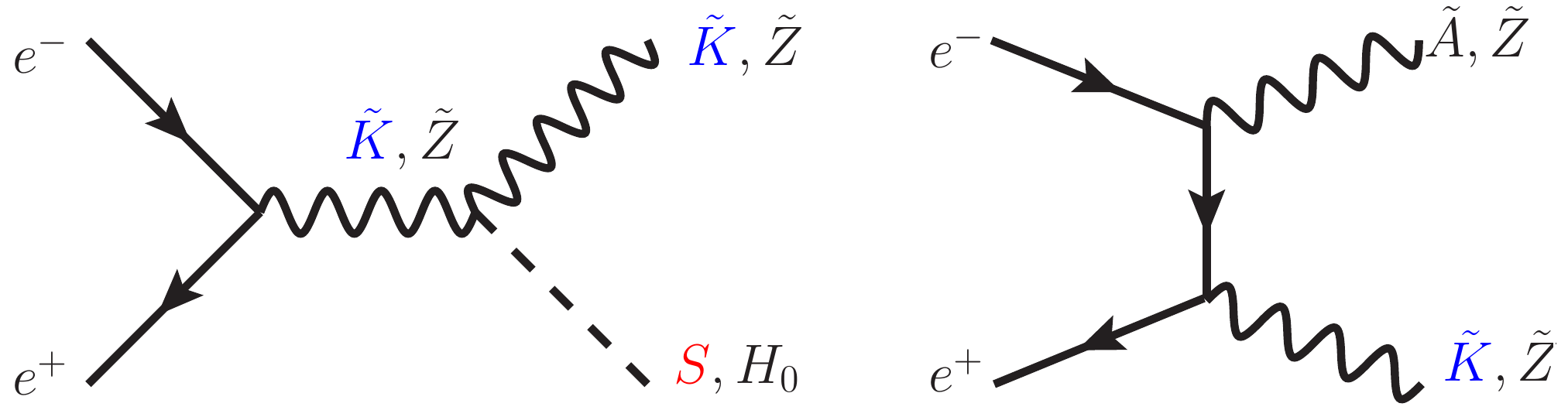}
 \includegraphics[width=0.95\textwidth]{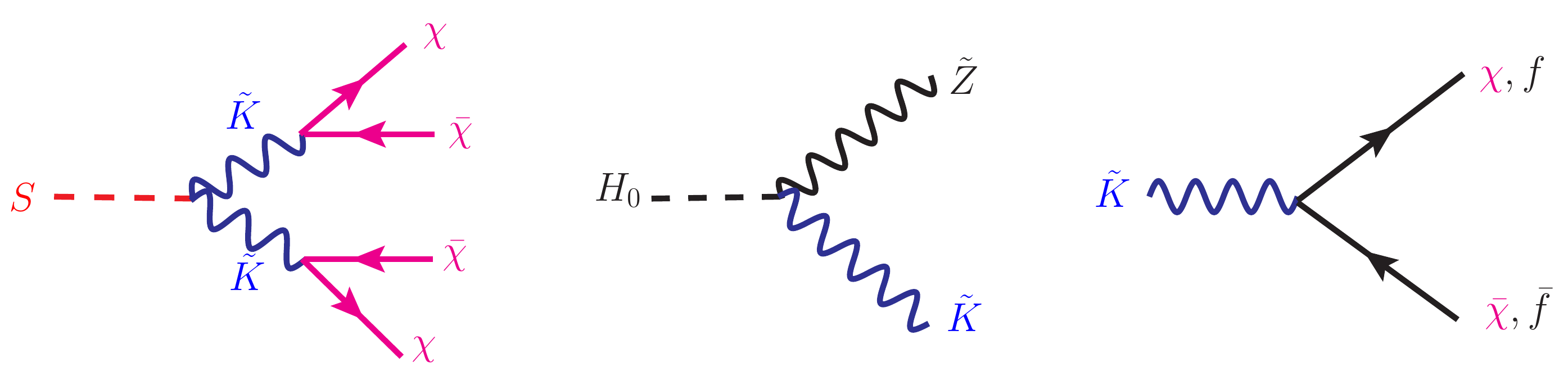}
 \caption{Feynman diagrams for (top left) vector + scalar production,
   (top right) vector + vector production and (bottom row) example new
   decay processes in the Double Dark Portal model sensitive to the
   kinetic mixing $\epsilon$ and scalar Higgs $\lambda_{HP}$ portal
   couplings.  Note $\tilde{Z}$, $\tilde{A}$, and $H_0$ are the mass
   eigenstates corresponding to the SM-like $Z$, photon, and Higgs
   bosons, respectively.}
  \label{fig:diagrams}
\end{figure*}

In~\figref{diagrams}, we show the new possibilities for SM-like and
dark scalar Higgsstrahlung from the intermediate massive vector bosons
$\tilde{K}$ and $\tilde{Z}$.  We also show the radiative return
process for $e^+ e^- \to \tilde{A} \tilde{K}$ or $\tilde{A}
\tilde{Z}$, and the diboson process $e^+ e^- \to \tilde{K} \tilde{Z}$.
All of these processes give different signals at a future $e^+ e^-$
machine.  If we also consider the possibility of $Z$-pole measurements
and Drell-Yan processes probing~\eqnref{currents}, then we can
categorize the collider phenomenology of the Double Dark Portal model
into four groups: electroweak precision and $Z$-pole observables,
Higgs measurements, Drell-Yan measurements, and radiative return
processes.  We point out, however, that $e^+ e^-$ machines offer
unique opportunities for probing new, light, hidden particles by
virtue of the recoil mass method, which we discuss first.

\subsection{Recoil mass method for probing new, light, hidden states}
\label{subsec:recoil}
As long as they are kinematically accessible, both $S$ and $\tilde{K}$
can be produced in $e^+ e^-$ collisions in association with SM
particles.  Hence, even if they decay invisibly, the recoil mass
method can be used to probe the couplings $\sin \alpha$ and
$\epsilon$, according to the interactions from~\eqnref{currents}
and~\eqnref{SVVinteractions}.  This is familiar from the leading $e^+
e^- \to \tilde{Z} H_0$ Higgsstrahlung production process, where the
reconstruction of the $\tilde{Z} \to \ell^+ \ell^-$ decay consistent
with a 125~GeV recoil mass gives a rate dependent only on the $H_0
\tilde{Z}_\mu \tilde{Z}^\mu$ coupling.  We emphasize 
(see also Ref.~\cite{Jia:2016pbe}) that this generalizes to any
scattering process at an $e^+ e^-$ machine if visible SM states are
produced in association with a new, light, hidden particle.  Moreover,
sensitivity to the hidden states $S$ and $\tilde{K}$ can be improved
by scanning over $\sqrt{s}$, where the various production modes of
$\tilde{Z} S$, $\gamma \tilde{K}$, and $\tilde{Z} \tilde{K}$ can be
optimized for the different $S$ and $\tilde{K}$ masses.  This
$\sqrt{s}$ adjustment would be immediately motivated, for example, by
a new physics signal in the recoil mass distribution.

The recoil mass method uses the knowledge that the center-of-mass
frame for the $e^+ e^-$ collision is fixed to be $(\sqrt{s}, 0, 0, 0)$
in the lab frame, where $\sqrt{s}$ is the energy of the collider.
Hence, for an invisibly decaying final state particle $X$ produced in
association with a SM state $Y$, four-momentum conservation requires
\begin{align}
  E_Y = \frac{\sqrt{s}}{2} + \frac{m_Y^2 - m_X^2}{2 \sqrt{s}} \ ,
\label{eqn:visibleE}
\end{align}
or equivalently,
\begin{align}
m_X = \sqrt{s + m_Y^2 - 2 E_Y \sqrt{s}} \ .
\label{eqn:recoilmass}
\end{align}
If there are multiple visible states $Y_i$, this generalizes to
\begin{align}
\sum E_{Y_i} = \frac{\sqrt{s}}{2} + \frac{ (\sum p_i)^2 - m_X^2}{2
  \sqrt{s}} \ , \quad
m_X  = \sqrt{s + (\sum p_i)^2 - 2 (\sum E_{Y_i}) \sqrt{s}} \ ,
\end{align}
where $(\sum p_i)^2$ is the total invariant mass of the $Y_i$ system.
We see that studying the differential distribution of $E_Y$ will show
a characteristic excess at a given $E_Y$ when $X$ is produced.
Identifying this monochromatic peak is formally equivalent to finding
a peak in the recoil mass distribution, but we emphasize that these
two distributions reconstructed differently at $e^+ e^-$ colliders.
Specifically, the recoil mass distribution uses both the energy and
total four-momentum of each detected SM particle, which the
differential energy distribution only requires calorimeter
information.

In particular, for the $\tilde{Z} H_0$, $\tilde{Z} \to \ell^+ \ell^-$
Higgsstrahlung process, the recoil mass method requires measurements
of each individual lepton four-momentum and the event-by-event
invariant mass $m_{\ell \ell}$.  The resulting differential
distribution also includes off-shell contributions and interference,
giving a smeared peak in the recoil mass distribution whose width is
dominated by experimental resolution and not the intrinsic Higgs
width.  On the other hand, in radiative return processes, both the
recoil mass distribution and the photon energy spectrum are only
limited by the possible width of the recoiling new physics particle
and the photon energy resolution (see also~\cite{Chakrabarty:2014pja,
Greco:2016izi}).

For our studies, we assume both $S$ and $\tilde{K}$ have dominant
decay widths to the dark matter $\chi$, which does not leave tracks or
calorimeter energy deposits as it escapes.  The recoil mass technique,
however, also readily probes both the $\tilde{K}$ and $S$ masses in
numerous production modes, when we produce $\tilde{K}$ or $S$ in
association with a visible SM final state.  For example, while the
SM-like $\tilde{Z}$ boson is a canonical choice to study $\tilde{Z}
H_0$ events, we can use the recoil mass technique in the radiative
return process for $\tilde{A} \tilde{K}$ production to identify the
invisible decay of $\tilde{K}$.  An even more striking possibility is
to use the SM-like Higgs boson, $H_0$, as the recoil mass particle to
probe $\tilde{K} H_0$ production.

\subsection{Modifications to electroweak precision}
\label{subsec:EWPOchanges}

We now consider the four categories of collider processes in turn.
The first set of observables we consider are those from electroweak
precision tests.  In the Double Dark Portal model, $Z$-pole
observables will show deviations according to the new decay channel
$\tilde{Z} \to \bar{\chi} \chi$ or $\tilde{Z} \to S \tilde{K}$,
sensitive to $\epsilon$, shifts in the $\tilde{Z}$ mass from the
mixing with $\tilde{K}$, and deviations in the weak mixing angle from
the mixing between $\tilde{K}$, $\tilde{A}$, and $\tilde{Z}$.  In
particular, identifying the $\tilde{Z}$ mass eigenstate of the DDP as
the $91.2$~GeV $Z$ boson studied by LEP, measurements of the $Z$ mass,
total width, and the invisible decay to SM neutrinos give strong
constraints on $\epsilon$ and the possibility of exotic decays.  For
$m_{\tilde{K}} < 10$~GeV, both the visible and invisible channels can
be constrained by various experiments, as reviewed in
Ref.~\cite{Alexander:2016aln}.  We thus focus on the status and
prospects for $m_{\tilde{K}} > 10$~GeV.

\subsubsection{LEP-I and LEP-II constraints}
\label{subsubsec:LEP}
At LEP-II, contact operators $(4\pi / \Lambda^2) \bar{e}\gamma^\mu e
\bar{f} \gamma_\mu f$ were used to test for new physics, analogous to
angular distributions in dijet studies at the LHC.  In the $e^+e^- \to
\ell^+ \ell^-$ channel, the constraint on $\Lambda$ is $\gtrsim 20$
TeV~\cite{Alcaraz:2006mx}.  Since the majority of the dataset was
taken at a fixed $\sqrt{s} = 200$~GeV, we can place a constraint on
$\epsilon$ by matching the coefficient of the contact operator at
tree-level to an intermediate $\tilde{K}$ mediator, $g_{\tilde{K}
  \ell\ell}^2 / (\sqrt{(s - m^2_{\tilde{K}})^2 + m^2_{\tilde{K}}
  \Gamma^2_{\tilde{K}}})$.  The corresponding bound on $\epsilon$ is
around $\mathcal{O}(0.1)$.  There is sharper sensitivity
for $m_K \approx m_Z$ because of maximal mixing, and again around
$\sqrt{s} \sim 200$~GeV from resonant production.  Because the
$\tilde{K}$ decay width is dominated by $\tilde{K} \to \bar{\chi}
\chi$, the sensitivity at resonance is suppressed the branching ratio
of $\tilde{K} \to \ell^+\ell^-$.

As mentioned above, the mixing between $Z_{\text{SM}}$ and $K$ leads
to shifts in the $\tilde{Z}$ mass and couplings to SM fermions,
leading to a constraint of $\epsilon < 0.03$ for $m_{\tilde{K}} < m_Z$
using a combination of electroweak precision
observables~\cite{Hook:2010tw}.  The constraint is weakened for
$m_{\tilde{K}} > m_Z$ where the limit on $\epsilon$ is about $0.1$ at
$m_{\tilde{K}} = 200$ GeV~\cite{Hook:2010tw}, and is shown
in~\figref{combined-epsilon-constraint} and~\figref{combined-epsilon-constraint2} 
as ``LEP-EWPT.''

Recently, the BaBar collaboration has published constraints on dark
photons decaying invisibly, $e^+e^- \to \gamma \tilde{K}$,
$\tilde{K}\to \bar{\chi}\chi$~\cite{Lees:2017lec}.  Their constraints
directly map to our $\epsilon$ vs.~$m_K$ parameter space and place
strong limits on $\epsilon \lesssim 0.001$ for masses between 1~GeV to
10~GeV.  These are reproduced in~\figref{combined-epsilon-constraint}
and~\figref{combined-epsilon-constraint2}, and labeled as ``BaBar.''

Although the canonical SM Higgs production channel $e^+ e^- \to Z h$
was ineffectual at LEP-II, the scalar mixing angle $\sin \alpha$ can
still be probed by the $e^+ e^- \to \tilde{Z} S$ production mode when
$S$ is kinematically accessible by LEP-II.  For $m_S < 114$~GeV, the
non-observation of Higgs-like scalar decays constrains $\sin^2 \alpha
< \mathcal{O}(0.01-0.1)$~\cite{Schael:2006cr}, as long as the $S \to
\tilde{K} \tilde{K}$ decay is turned off.  

The LEP experiments have also searched for a low mass Higgs in the
exotic $Z \to H Z^\star$ decay, with $Z^\star \to \ell^+ \ell^-$ and
$H$ decaying invisibly, which excludes $m_H < 66.7$~GeV if the
invisible branching fraction is 100\%~\cite{Acciarri:1996um}.  The
$ZH$ Higgsstrahlung process is also used to push the mass exclusion to
$114.4$~GeV~\cite{Acciarri:1997tr, Abreu:1999vu, Searches:2001ab},
although the intermediate mass range between these two limits are not
comprehensively covered.  In our model, $S \to \tilde{K} \tilde{K}$ is
the dominant decay when $g_D \gg \sin\alpha$, and the decay branching
fraction $\tilde{Z} \to S \tilde{Z}^\star$ and the production cross
section $\sigma(\tilde{Z} S)$ are hence $\sin^2\alpha$ suppressed
compared to the SM rate.  Therefore, these limits apply to $S$ as
bounds on $\sin \alpha$ and $m_S$, which we will show
in~\figref{sig-XS-ZS} in~\secref{futureprospects}.  Note the
constraint from the exotic $\tilde{Z} \to S \tilde{Z}^\star$ decay is
much stronger than $\tilde{Z} H_0$ Higgsstrahlung process
in~\figref{sig-XS-ZS} due to the high statistics of $Z$ decays, and in
the calculation we accounted for the subsequent decay branching
fractions of $\text{BR}(S \to \tilde{K} \tilde{K})$ and
$\text{BR}(\tilde{K} \to \bar{\chi} \chi)$.

\subsection{Modifications to Higgs physics and LHC constraints}
\label{subsec:Higgschanges}

With the era of precision Higgs characterization underway after the
discovery of a Higgs-like boson~\cite{Aad:2012tfa,
  Chatrchyan:2012xdj}, the ATLAS and CMS collaborations have provided
the strongest constraints on the possible mixing of the SM Higgs boson
with a new gauge singlet $\phi$.  In addition, searches for an
invisible decay of the 125 GeV Higgs boson, sensitive to $H_0 \to SS$
or $H_0 \to \tilde{K} \tilde{K}$ decays, have also given constraints
on $\lambda_{HP}$, $\sin \alpha$, and $\epsilon$.  The growing Higgs
dataset at the LHC continues to show no significant deviations from
the SM expectation, but the current sensitivity of the LHC experiments
to our proposed signals is limited.

The most important constraint comes from the search for an invisible
decay of the 125~GeV Higgs, where the Run 1 combination of ATLAS and
CMS data constrains BR$(h \to \text{ inv}) \leq
0.23$~\cite{Aad:2015pla, Khachatryan:2016whc}.  We highlight, however,
that this limit requires that the Higgs is produced in the $\tilde{Z}
H_0$ and vector boson fusion processes at SM rates, which is violated
in the DDP model.  Moreover, in the DDP model, there are two possible
direct invisible decays, $H_0 \to S S \to 4 \tilde{K} \to 8\chi$ and
$H_0 \to 2 \tilde{K} \to 4\chi$, in addition to the possible exotic
decay $H_0 \to \tilde{Z} \tilde{K} \to \bar{\nu} \nu \bar{\chi} \chi$,
which is often semi-visible.  The $H_0 \to S S$, $H_0 \to \tilde{K}
\tilde{K}$ and $H_0 \to \tilde{K} \tilde{Z}$ decay widths at leading
order in $\epsilon$ and $\alpha$ are
\begin{align}
\Gamma (H_0 \to S S) &= g_D^2 \sin^2 \alpha \frac{m_{H_0}}{32 \pi}
\sqrt{1- \frac{4 m_S^2}{m_{H_0}^2}} 
\frac{(m_{H_0}^2 + 2 m_S^2)^2}{m_{H_0}^2 m_K^2} \ , 
\label{eqn:HSSwidth} \\
\Gamma (H_0 \to \tilde{K} \tilde{K}) &=  
g_D^2 \sin^2 \alpha \frac{m_{H_0}}{32 \pi}
\sqrt{1- \frac{4 m_{\tilde{K}}^2}{m_{H_0}^2}} 
\frac{m_{H_0}^4 - 4 m_{H_0}^2 m_{\tilde{K}}^2 +12 m_{\tilde{K}}^4}{m_{H_0}^2 m_{\tilde{K}}^2} 
\frac{m_K^2}{m_{\tilde{K}}^2} \ ,
\label{eqn:HKKwidth} \\
\Gamma (H_0 \to \tilde{K} \tilde{Z}) & = \frac{\epsilon^2 t_W^2 
\left(\frac{\cos\alpha}{v_H}  + \frac{\sin\alpha}{v_D} \right)^2 }{16 \pi m^3_{H_0} \left(m_K^2 - m_{Z,\text{ SM}}^2 \right)^2} \frac{m_K^4 m_{Z,\text{ SM}}^4}{m_{\tilde{K}}^2 m_{\tilde{Z}}^2} 
\sqrt{m^4_{H_0} + \left(m_{\tilde{K}}^2 - m_{\tilde{Z}}^2 \right)^2 - 2 m^2_{H_0} 
\left(m_{\tilde{K}}^2 + m_{\tilde{Z}}^2 \right)} \nonumber \\
& \times \left( (m^2_{H_0} - m_{\tilde{K}}^2 - m_{\tilde{Z}}^2)^2 + 8 m_{\tilde{K}}^2 m_{\tilde{Z}}^2   \right)
\ .
\label{eqn:HKZwidth}
\end{align}
The first two decay widths are proportional to $m_K^{-2}$ while the
last one is proportional to $m_K^2$, therefore the last one is usually
much smaller comparing with the first two when $m_K$ is light. In the
Higgs invisible studies, the experiments will constrain the rate for
Higgs invisible decays in the DDP model,
\begin{align}
\text{BR}^{\text{eff}}_{\text{inv}} = & \dfrac{\sigma(\tilde{Z} H_0)}{\sigma_{\text{SM}}(Zh)} \times \text{ BR} (H_0 \to \text{inv}) = 
\cos^2 \alpha \dfrac{1}{\Gamma_{H_0, \text{ tot}}} \left(
\Gamma (H_0 \to S S) \text{BR}^2 (S\to \tilde{K}\tilde{K}) \text{BR}^4(\tilde{K}\to \bar{\chi}\chi) \right. 
\nonumber \\
& + \left. \Gamma (H_0 \to \tilde{K} \tilde{K}) \text{BR}^2(\tilde{K}\to \bar{\chi}\chi)
+ \Gamma (H_0 \to \tilde{K} \tilde{Z}) \text{BR}(\tilde{K}\to \bar{\chi}\chi)
\text{BR}(\tilde{Z}\to \bar{\nu}\nu)
\right) \ ,
\end{align}
where $\Gamma_{H_0, \text{ tot}} = \cos^2 \alpha \Gamma_{h, \text{
    SM}} + \Gamma(H_0 \to SS) + \Gamma(H_0 \to \tilde{K} \tilde{K}) +
\Gamma(H_0 \to \tilde{K} \tilde{Z})$.  The decay widths $\tilde{K} \to
\bar{f}f$, $W^+W^-$ can be found in the appendix of
Ref.~\cite{Kopp:2016yji}, while the decay width $\tilde{K} \to
\bar{\chi} \chi$ is
\begin{align}
\Gamma(\tilde{K} \to \bar{\chi} \chi )= \frac{g_D^2}{12 \pi}
\sqrt{m_{\tilde{K}}^2 - 4 m_\chi^2} 
\left(1 +   \frac{2 m_\chi^2}{m_{\tilde{K}}^2}\right) \,.
\end{align}

\begin{figure*}[tb!]
\includegraphics[width=0.32\textwidth]{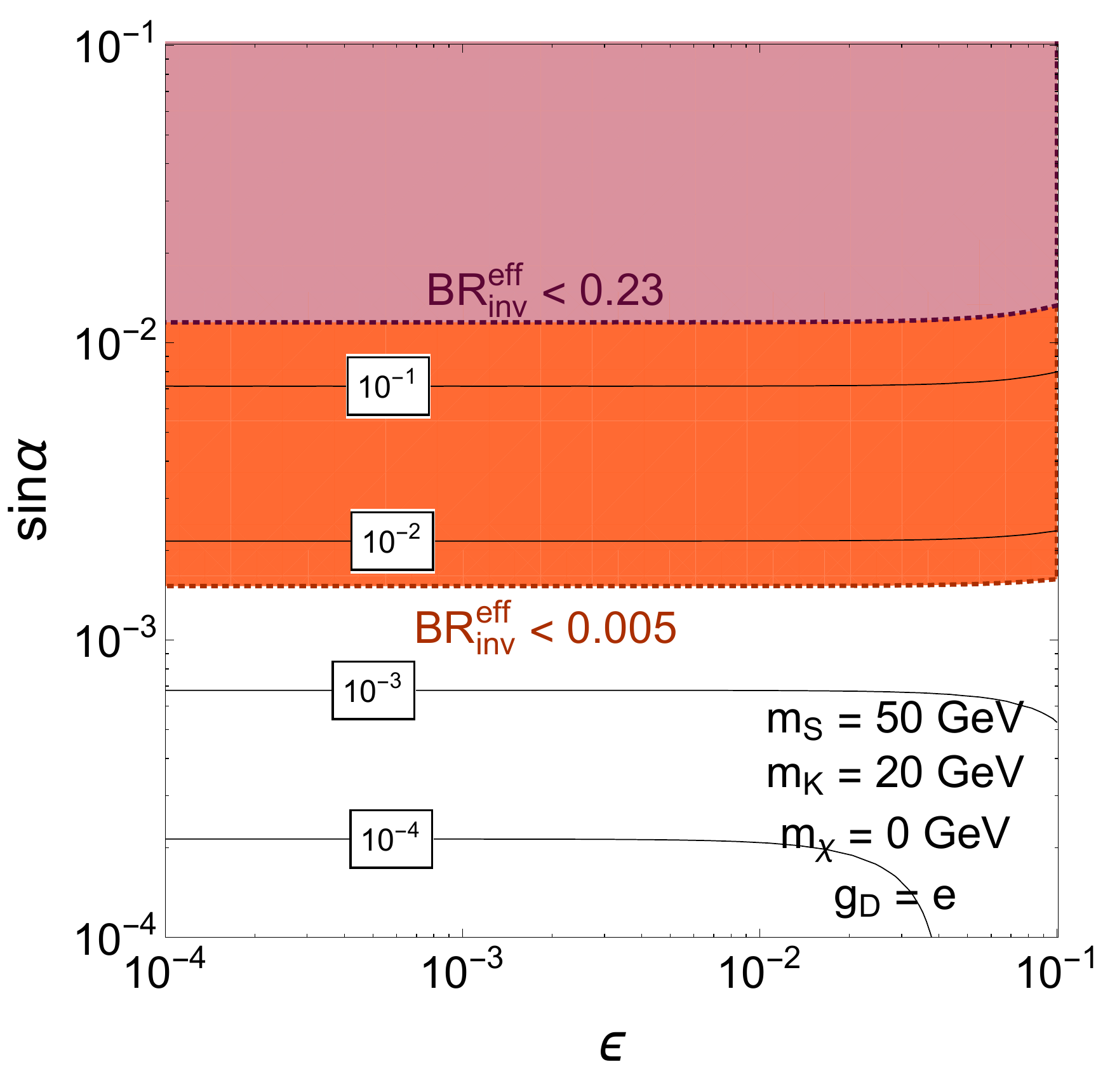}
\includegraphics[width=0.32\textwidth]{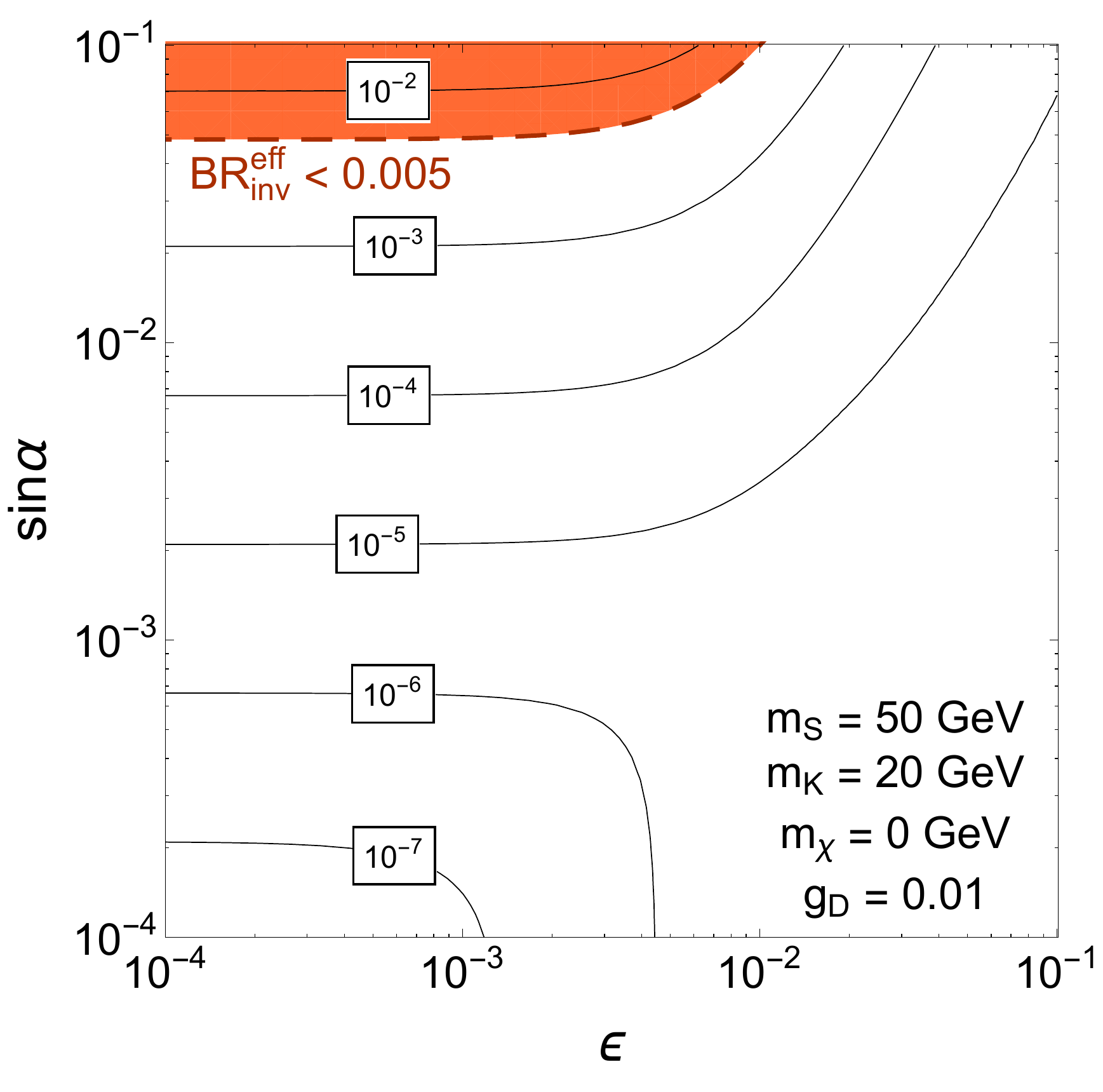} 
\includegraphics[width=0.32\textwidth]{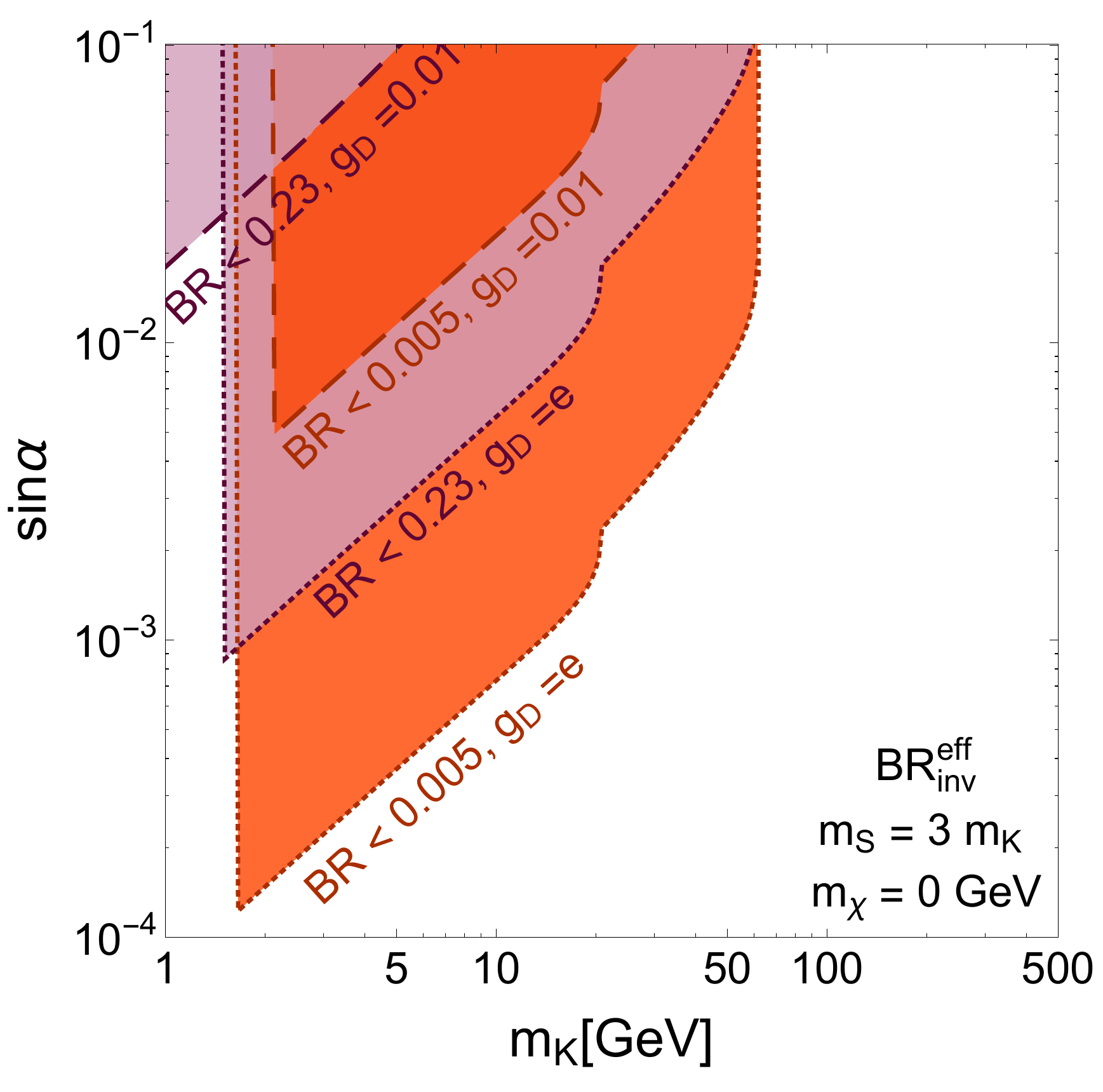}
\caption{(Left and middle panel) Rates for the invisible branching
  fraction of the 125~GeV Higgs in the $\sin \alpha$ vs.~$\epsilon$
  plane, setting $m_S = 50$~GeV, $m_K = 20$~GeV, and $g_D = e$ (left)
  and $0.01$ (middle).  (Right panel) Exclusion regions in the $\sin
  \alpha$ vs.~$m_K$ plane from the search for an invisible decay of
  the 125~GeV Higgs by ATLAS and CMS giving BR$_{\text{inv}} <
  0.23$~\cite{Aad:2015pla, Khachatryan:2016whc}, and projected reach
  from a future $e^+ e^-$ machine giving BR$_{\text{inv}} <
  0.005$~\cite{Dawson:2013bba, Djouadi:2007ik, Gomez-Ceballos:2013zzn,
    CEPC-SPPCStudyGroup:2015csa}.}
\label{fig:Hinv-sinalpha-mk}
\end{figure*}

The prediction for the invisible decay branching fraction of the
125~GeV Higgs is shown in the left and middle panels
of~\figref{Hinv-sinalpha-mk} in the $\sin \alpha$ vs.~$\epsilon$ plane
for $m_S = 50$~GeV, $m_K = 20$~GeV, and $g_D = e$, $0.01$.  The
current constraint of BR$_{\text{inv}} < 0.23$ is adopted from
Refs.~\cite{Aad:2015pla, Khachatryan:2016whc}, while the prospective
sensitivity of BR$_{\text{inv}} < 0.005$ is adopted from the estimate
using 10 ab$^{-1}$ of $\sqrt{s} = 240$~GeV data using unpolarized
beams in Ref.~\cite{Gomez-Ceballos:2013zzn}.  This prospective limit
can be lowered in combined fits, with more luminosity, or with other
assumptions about detector performance to the $\mathcal{O}(0.001)$
level~\cite{Dawson:2013bba, Djouadi:2007ik,
  CEPC-SPPCStudyGroup:2015csa}.
 
In~\figref{Hinv-sinalpha-mk}, we see that when $g_D$ is large, the
sensitivity to $\sin \alpha$ is much stronger than $\epsilon$, because
the decay widths for $H_0 \to S S$ and $H_0 \to \tilde{K} \tilde{K}$
are much larger than $H_0 \to \tilde{K} \tilde{Z}$ due to light
$m_{K}$, as discussed previously. More importantly, for large $g_D$,
$\text{BR}(S\to \tilde{K}\tilde{K})$ and $\text{BR}(\tilde{K}\to
\bar{\chi}\chi)$ are close to $100\%$.  When $g_D < \sin\alpha$ or
$g_D < \epsilon$, $\text{BR}(S\to \tilde{K}\tilde{K})$ and
$\text{BR}(\tilde{K}\to \bar{\chi}\chi)$ will both be subdominant and
result in the decrease of $\text{BR}^{\text{eff}}_{\text{inv}}$ as in
the middle panel of~\figref{Hinv-sinalpha-mk}.  As shown in the right
panel of~\figref{Hinv-sinalpha-mk}, the constraint on $\sin \alpha$
from invisible Higgs decays can be relaxed by making $g_D$ smaller.

These exotic decays can also give fully visible and semi-visible
signatures~\cite{Martin:2011pd, Curtin:2013fra, Huang:2013ima,
  Huang:2014cla, Curtin:2014cca, Liu:2016zki} when the $\tilde{K}$ or
$S$ particle decays to SM final states, which provide additional
handles for Higgs collider phenomenology.  Those references
concentrate on the scenario where the decay to SM final states are
dominant, {\it e.g.}  $m_{\tilde{K}} < 2 m_\chi$.  Therefore, such
constraints should be modified by the relevant branching ratios in the
DDP model because the DDP model includes a DM decay.  If visible
decays of $\tilde{K}$ dominate, though, the search for $h \to 2a \to
4\mu$~\cite{CMS:2013lea} constrains the process $H_0 \to \tilde{K}
\tilde{K}$, and bounds $\lambda'_{HP} \lesssim 0.01$ for
$m_{\tilde{K}} \lesssim 10$~GeV~\cite{Curtin:2014cca}, while the bound
strengthens to $\lambda'_{HP} \lesssim 0.001$~\cite{Curtin:2014cca}
for $m_{\tilde{K}} \gtrsim 10$~GeV from recasting the differential
distributions in 8~TeV $h \to Z Z^* \to 4\ell$ data~\cite{CMS:xwa}.
The coupling $\lambda'_{HP} = \lambda_{HP} m_{H_0}^2/ | m^2_{H_0} -
m_{S}^2|$, is roughly the same as $\lambda_{HP}$ if $m_S$ is not close
to or much larger than $m_{H_0}$.  The high luminosity LHC (HL-LHC)
with $3$ ab$^{-1}$ of 14~TeV luminosity is expected to be sensitive to
$\lambda'_{HP} \lesssim (\text{few}) \times 10^{-5}$ in this same
channel, depending on the $m_{\tilde{K}}$ mass.  The four lepton final
state has also been used to constrain the exotic decay $H_0 \to
\tilde{Z} \tilde{K}$, which gives sensitivity to $\epsilon$
from~\eqref{eqn:SVVinteractions}.  The current bound using 8~TeV data
is weak, with the strongest sensitivity for $m_{\tilde{K}} \approx
30$~GeV giving $\epsilon \lesssim 0.05$, while the improvement at the
HL-LHC is expected to reach $\epsilon \lesssim
0.01$~\cite{Curtin:2014cca}.  These gains are mainly limited by the
statistics afforded by Higgs production rates.  We remark that for
very small $\epsilon$ and $m_{\chi} > m_{\tilde{K}} /2$, as discussed
in~\secref{DDIDpheno}, the hidden photon will have a displaced decay
to SM states, which provides a new set of challenges to trigger and
detect at colliders.  Current exclusions and future prospects for
displaced decays can be found, {\it e.g.}, in
Refs.~\cite{Essig:2013lka, Curtin:2014cca, Alexander:2016aln}.

\subsubsection{Modifications to Drell-Yan processes}
\label{subsec:DrellYan}

The $\tilde{K}$ decay to SM final states can be dominant, if the decay
to DM pairs is kinematically forbidden or $g_D \ll \epsilon$.  In this
case, at the LHC, the Drell-Yan process $pp \to \tilde{Z}, \tilde{K}
\to \ell^+ \ell^-$ can be used to constrain the kinetic mixing
parameter $\epsilon$, since this process has been studied with
exquisite precision by the ATLAS and CMS experiments.  Both ATLAS and
CMS have searched for dilepton resonances at high mass, $m_{\tilde{K}}
\gtrsim 200$~GeV, using $20$~fb$^{-1}$ of 8~TeV
data~\cite{Aad:2014cka, Khachatryan:2014fba}, which restricts
$\epsilon \lesssim 0.01$ at $m_{\tilde{K}} = 200$~GeV and weakens to
$\epsilon \lesssim 0.05$ at $m_{\tilde{K}} =
1000$~GeV~\cite{Cline:2014dwa, Hoenig:2014dsa}.  For $m_{\tilde{K}}$
between $10$ and $80$~GeV, the Drell-Yan search using 7~TeV data by
CMS~\cite{Chatrchyan:2013tia} and the corresponding sensitivity using
8~TeV data gives $\epsilon \lesssim 0.005$, stronger than the current
electroweak precision constraints~\cite{Hoenig:2014dsa, Curtin:2014cca}.

The HL-LHC is expected to constrain $\epsilon \lesssim 0.001$ for
$m_{\tilde{K}}$ between $10$ and $80$~GeV using Drell-Yan data, while
high mass $\tilde{K}$ can be probed at the $\epsilon \sim 0.002$ level
for $m_{\tilde{K}} = 200$~GeV and the $\epsilon \sim 0.01$ level for
$m_{\tilde{K}} = 1000$~GeV~\cite{Cline:2014dwa,
  Hoenig:2014dsa, Curtin:2014cca}.  The recent 13~TeV, $3.2$~fb$^{-1}$
search for high mass dilepton resonances by
ATLAS~\cite{Aaboud:2016cth} also constrains the kinetic mixing
parameter $\epsilon \lesssim 0.04$ for $m_{\tilde{K}} >
100$~GeV~\cite{Liu:2017jzn}, but this result is hampered by the small
statistics.  We see that as long as $\tilde{K}$ has an appreciable
branching fraction to SM final states, in particular leptons, the
Drell-Yan process at the LHC and HL-LHC will provide stronger
sensitivity to $\epsilon$ compared to electroweak precision
observables.  For $g_D / \epsilon \gg 1$, the decay $\tilde{K} \to
\bar{\chi} \chi$ is dominant and the situation reverses, and then
Drell-Yan constraints will not compete with the electroweak precision
observables.  After rescaling by the appropriate visible branching
ratio, we plot these Drell-Yan constraints as the ``LHC-DY'' contours
in~\figref{combined-epsilon-constraint} and ~\figref{combined-epsilon-constraint2}.

\subsubsection{Radiative return processes and dark matter production at the LHC}
\label{subsec:radreturn}

The radiative return process, $e^+ e^- \to \gamma X$, enables on-shell
production of new particles at fixed $\sqrt{s}$ colliders by using an
extra radiated photon to conserve four-momentum.  At hadron colliders,
since the colliding objects are composite, dark matter production via
radiative return is more commonly known as monojet or monophoton
processes, recognizing the fact that the partonic center of mass
energy is not constant on an event-by-event basis.

As a result, the visible decays $\tilde{K} \to \ell^+ \ell^-$
discussed in the Higgs to four leptons and the Drell-Yan contexts are
complemented by the LHC searches for dark matter production in monojet
and monophoton processes.  We remark that in our DDP model, we will
assume that $\tilde{K} \to \bar{\chi} \chi$ is the dominant decay
channel, leading to an overall $\epsilon^2 / g_D^2$ suppression in the
above visible decay rates.  Both ATLAS and CMS have searches for dark
matter production using 8~TeV data~\cite{Aad:2015zva,
  Khachatryan:2014rra}, sensitive to mediator masses as low as
$10$~GeV~\cite{Aad:2015zva}.  The corresponding 13~TeV
searches~\cite{Aaboud:2016tnv, Sirunyan:2017hci} have yet to achieve
the same sensivity at low masses.  In the DDP model, the $\tilde{K}$
mediator is produced on-shell and decays dominantly to $\bar{\chi}
\chi$, and calculating the results for on-shell mediator production at
the LHC, we obtain $\epsilon \lesssim 0.07$, similar to previous
studies~\cite{Jacques:2015zha, Chala:2015ama, Brennan:2016xjh}.  It is
also possible to search for the dark bremsstrahlung of $\tilde{K}$
from the DM pair~\cite{Gupta:2015lfa, Buschmann:2015awa}, as a probe
of $\epsilon$, although these rates are negligible in our model.

\section{Prospects for future colliders}
\label{sec:futureprospects}

We have established that significant room remains to be explored in
both the $\epsilon$ and $\lambda_{HP}$ portal couplings.  We will now
demonstrate that a future $e^+ e^-$ collider, currently envisioned as
a Higgs factory, will have leading sensitivity to probing both
couplings simultaneously through the production of new, light, hidden
states $\tilde{K}$ and $S$.  The primary motivation for the $\sqrt{s}
\sim 240-250$~GeV center-of-mass energy of such a collider is to
optimize the expected $\sigma(e^+ e^- \to Zh)$ SM Higgsstrahlung cross
section, taking into account the possible polarization of the incoming
electron-positron beams.  Such high energies, however, also enable
production of the new states $\tilde{K}$ and $S$ from radiative return
processes, exotic Higgs decays, and exotic Higgsstrahlung diagrams.

A few different variations exist for next-generation $e^+ e^-$
machines, namely the International Linear Collider
(ILC)~\cite{Djouadi:2007ik}, an $e^+ e^-$ Future Circular Collider
(FCC-ee), which shares strong overlap with
TLEP~\cite{Gomez-Ceballos:2013zzn}, or a Circular Electron-Positron
Collider (CEPC)~\cite{CEPC-SPPCStudyGroup:2015csa}.  Since the physics
we discuss will only depend very mildly on the particular $\sqrt{s}$
of the future machine and possible polarization of the incoming
electron and positron beams, we will adopt a $\sqrt{s} = 250$~GeV
machine colliding unpolarized $e^+$ and $e^-$ beams as our reference
machine with a total integrated luminosity of $L = 5$ ab$^{-1}$ in our
collider studies.  For comparison, we also show future expectations
for a possible $\sqrt{s} = 500$~GeV machine with $L = 5$ ab$^{-1}$
total integrated luminosity.  Our work will complement and extend
previous $\epsilon$ and $\sin \alpha$ sensitivity estimates made for
various specific collider environments, which we review first.

\subsection{Electroweak precision tests, Higgsstrahlung, and invisible Higgs decays at future $e^+ e^-$ colliders}
\label{subsec:EWPTfuture}

Because the SM $Z$ and the dark vector $K$ mix, the $\tilde{Z}$ mass
eigenstate develops a new invisible decay channel, $\tilde{Z} \to
\bar{\chi} \chi$, if kinematically allowed.  This invisible decay
width can be accurately measured at a future $e^+ e^-$ machine.  At
FCC-ee, for example, running on the $W^+ W^-$ threshold using the
radiative return process $e^+ e^- \to Z \gamma$ ($L_{\text{int}} =
15.2$~ab$^{-1}$), complemented by additional runs at $\sqrt{s} =
240$~GeV ($L_{\text{int}} = 10.44$~ab$^{-1}$) and $\sqrt{s} = 350$~GeV
($L_{\text{int}} = 0.42$~ab$^{-1}$), can constrain the number of
active neutrino species, assuming statistical uncertainties, down to
$\Delta N_{\nu} \approx 0.001$~\cite{Dam:2016ebi, dEnterria:2016sca}.
This leads to the possible constraint $\epsilon < 0.01$ for
$m_{\tilde{K}} < m_{\tilde{Z}} / 2$.  Measurements directly on the
$Z$-pole are not expected to compete with this constraint because of
theory uncertainties on the small-angle Bhabha-scattering cross
section remain too large~\cite{Dam:2016ebi}.  This constraint also
applies to the $\tilde{Z} \to \tilde{K} S \to 6 \chi$ exotic decay, if
$m_{\tilde{K}} + m_S < m_{\tilde{Z}}$.  For a lepton collider running
on the $Z$-pole, though, other electroweak precision observables will
have greatly enhanced precision.  The combination of improved
electroweak precision observables can constrain $\epsilon \lesssim
0.004$ for $m_{\tilde{K}} < m_{\tilde{Z}}$, although the $\epsilon$
constraint is much weaker for $m_{\tilde{K}} >
m_{\tilde{Z}}$~\cite{Curtin:2014cca}.

The $e^+ e^-$ Higgs factory is expected to have a precision
measurement of the Higgsstrahlung process $e^+ e^- \to Zh$, with
accuracies ranging from $\mathcal{O}(0.3\%-0.7\%)$ expected, using
$5-10$~ab$^{-1}$ of luminosity~\cite{Gomez-Ceballos:2013zzn,
  CEPC-SPPCStudyGroup:2015csa, Ruan:2014xxa}.  These rates imply that
the scalar mixing angle is probed to $\sin \alpha \lesssim
0.055-0.084$, simply from the observation of the Higgsstrahlung
process.

Aside from precision Higgs measurements, a future $e^+ e^-$ machine
will have leading sensitivity to an invisible decay of the 125~GeV
Higgs.  As reviewed in~\subsecref{Higgschanges}, the current
constraint on the Higgs invisible decay branching ratio is
BR$_{\text{inv}} < 0.23$~\cite{Aad:2015pla, Khachatryan:2016whc},
while the limit at FCC-ee is expected to be BR$_{\text{inv}} <
0.005$~\cite{Gomez-Ceballos:2013zzn}.  We have discussed the two main
invisible decays, $H_0 \to S S \to 8\chi$ and $H_0 \to \tilde{K}
\tilde{K} \to 4\chi$, as well as the irreducible signal from $H_0 \to
\tilde{Z} \tilde{K} \to \bar{\nu} \nu \bar{\chi} \chi$
in~\subsecref{Higgschanges}.  We also show the corresponding sensivity
in the $\sin \alpha$ vs.~$\epsilon$ and $\sin \alpha$
vs.~$m_{\tilde{K}}$ planes in~\figref{Hinv-sinalpha-mk}.  We again
emphasize that these limits can be stronger or weaker because of their
dependence on $g_D$, as seen in~\figref{Hinv-sinalpha-mk}.

\subsection{Production of new, light states at future $e^+ e^-$ colliders}
\label{subsec:NPproduction}

As outlined in~\secref{ColliderLEPLHC} and shown in~\figref{diagrams},
many new possibilities open up for production of new, light, hidden
particles in the Double Dark Portal model.  We can classify the new
physics processes into vector + scalar, radiative return, and massive
diboson production topologies.  For small $\epsilon$ and scalar mixing
angle $\alpha$, the leading processes are:
\begin{description}
\item[$e^+ e^- \to \tilde{Z} H_0$] The usual Higgsstrahlung diagram
  is suppressed by $\cos^2 \alpha$, with an additional contribution
  from intermediate $\tilde{K}^*$ that can interfere with
  $\tilde{Z}^*$ exchange.
\item[$e^+ e^- \to \tilde{Z} S$] This new process can be probed by the
  usual recoil mass method for well-reconstructed $\tilde{Z}$ decays,
  studying the entire recoil mass differential distribution.
\item[$e^+ e^- \to \tilde{K} S$] This exotic production process
  involves two non-standard objects, and is dominantly produced via
  $\tilde{K}^*$, with a rate proportional to $\epsilon^2 \cos^2
  \alpha$.  Since $\tilde{K}$ and $S$ dominantly decay to dark matter,
  though, we would require an additional photon or a visible decay of
  $\tilde{K}$ or $S$ in order to tag the event.
\item[$e^+ e^- \to \tilde{A} \tilde{K}$] The radiative return process
  produces $\tilde{K}$ in association with a hard photon $\tilde{A}
  \sim \gamma$, giving direct sensitivity to $\epsilon$.  We remark
  that the $\tilde{K} \to \ell^+ \ell^-$ decay has been studied for
  $m_{\tilde{K}}$ between $10$~GeV to $240$~GeV at an $\sqrt{s} =
  250$~GeV machine with $10$~ab$^{-1}$~\cite{Karliner:2015tga}, giving
  $\epsilon \lesssim 5 \times 10^{-4}$, if the decay $\tilde{K} \to
  \ell^+ \ell^-$ is assumed dominant.
\item[$e^+ e^- \to \tilde{Z} \tilde{K}$] The massive diboson pair
  production process also provides direct sensitivity to $\epsilon$,
  but measuring the rate precisely will pay leptonic branching
  fractions of the $\tilde{Z}$.
\item[$e^+ e^- \to H_0 \tilde{K}$] This very interesting scenario can
  be probed by using the 125~GeV SM-like Higgs as a recoil candidate
  for the $\tilde{K}$ heavy vector.  The total rate gives sensitivity
  to both $\epsilon$ and $\alpha$ and highlights the power of
  considering the SM-like Higgs as a signal probe for new physics.
\end{description}

Having identified the main production modes for the $\tilde{K}$ and
$S$ states, we can match them to decay topologies illustrated in the
bottom row of~\figref{diagrams}.  We also include the underexplored
decay $H_0 \to \tilde{K} \tilde{Z}$, which gives an exotic decay of
the SM-like Higgs into the SM-like $\tilde{Z}$ boson and the hidden
photon $\tilde{K}$ sensitive to $\epsilon$.  As mentioned
in~\secref{DDIDpheno}, we focus on $S \to \tilde{K} \tilde{K} \to
4\chi$ and $\tilde{K} \to \bar{\chi} \chi$, and thus the dark portal
couplings must be tested by recoil mass techniques or mono-energetic
photon spectra searches.  We also demonstrate the importance of these
missing energy searches by explicitly considering leptonic decays of
$\tilde{K}$ in the $\tilde{Z} \tilde{K}$ and $\gamma \tilde{K}$
processes as well as the fully inclusive recoil mass distribution
targeting $\gamma \tilde{K}$ production.

Since the workhorse SM Higgsstrahlung process has been studied
extensively~\cite{Djouadi:2007ik, Gomez-Ceballos:2013zzn,
  CEPC-SPPCStudyGroup:2015csa}, we use these previous results to
recast the sensitivity for $\epsilon$ and $\alpha$.  We also ignore
the $\tilde{K} S$ production mode, since the dominant signature has
nothing visible to tag the event.  The $H_0 \tilde{K}$ process is
interesting to consider for future work, but it requires optimizing
the $H_0$ decay channel to gain maximum sensitivity to the recoil mass
of the rest of the event.

This leaves the $\tilde{Z} H_0$, $\tilde{Z} S$, $\gamma \tilde{K}$ and
$\tilde{Z} \tilde{K}$ processes as new opportunities to revisit or
study.  We simulate each process using MadGraph5
v2.4.3~\cite{Alwall:2014hca}, Pythia v6.4~\cite{Sjostrand:2006za} for
showering and hadronization, and Delphes
v3.2~\cite{deFavereau:2013fsa} for detector simulation.  Detector
performance parameters were taken from the preliminary validated CEPC
Delphes card~\cite{DelphesCEPC}.  Backgrounds for each process are
generated including up to one additional photon to account for initial
state and final state radiation effects.  Events are required to pass
preselection cuts of $|\eta| < 2.3$ for all visible particles, while
photons and charged leptons must have $E > 5$~GeV, jets must have $E >
10$~GeV, and missing transverse energy must satisfy $\slashed{E} >
10$~GeV.  Our analysis is insensitive to the dark matter mass as long
as $\tilde{K}$ and $S$ give missing energy signatures.

\subsection{Testing $\epsilon$ and $\sin \alpha$ with new particle production}
\subsubsection{$\tilde{Z} \tilde{K}$ production}
\label{subsec:ZKsearch}

\begin{figure*}[tb!]
\includegraphics[width=0.48\textwidth]{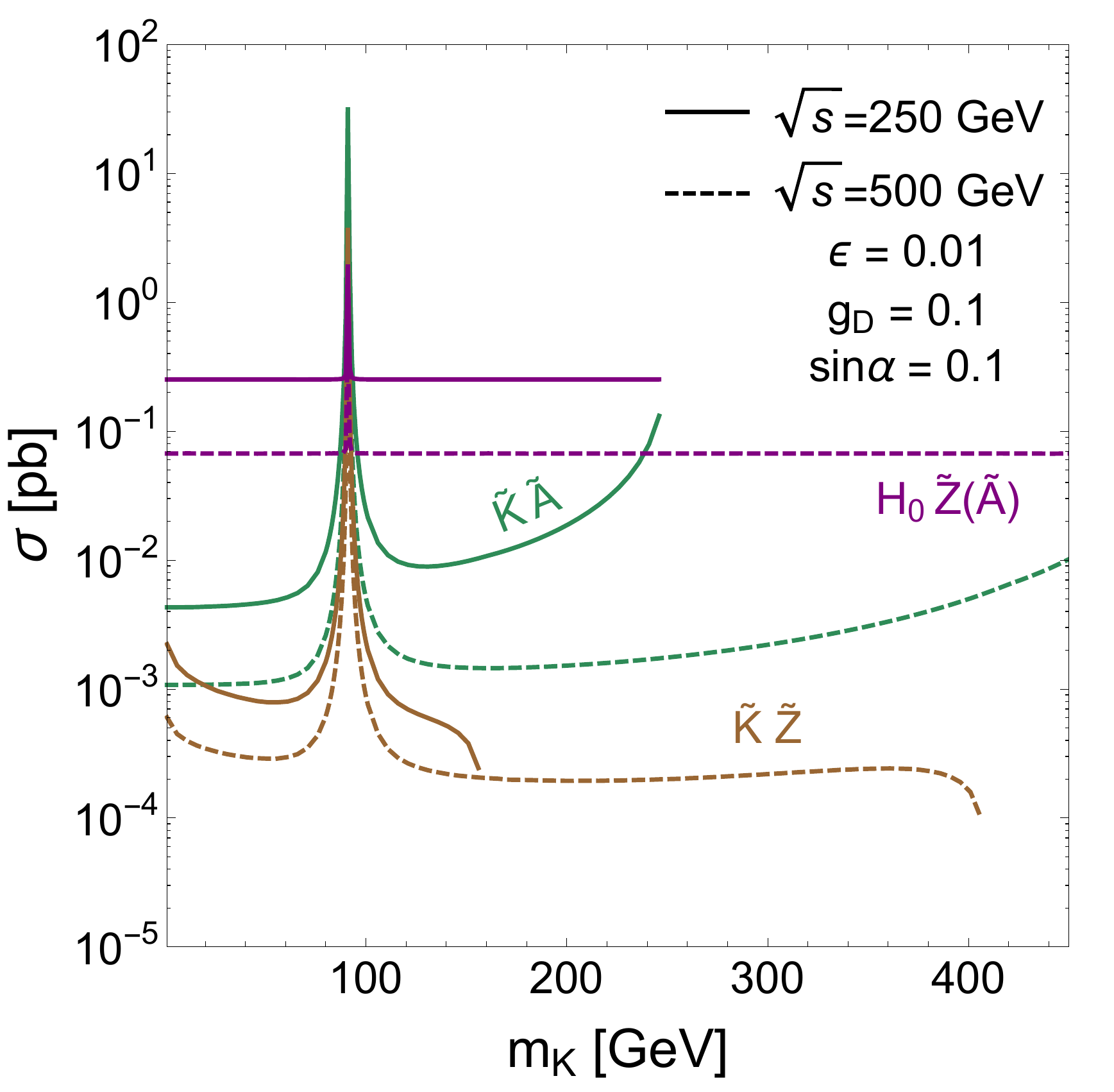}
\includegraphics[width=0.48\textwidth]{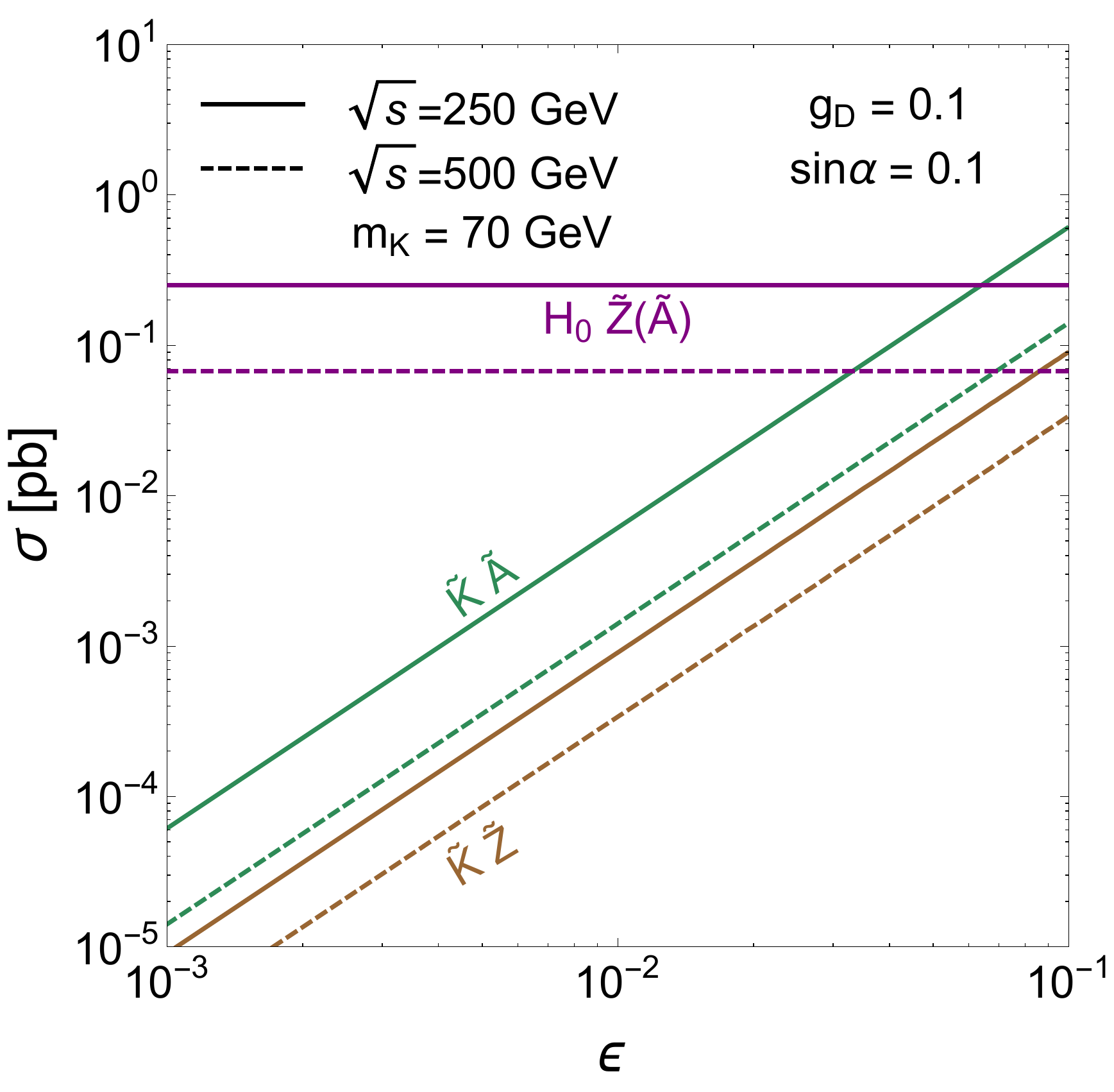}
\caption{Cross sections for $\tilde{Z} H_0$, $\tilde{A} \tilde{K}$,
  and $\tilde{Z} \tilde{K}$ processes as a function of (left panel)
  $m_K$ or (right panel) $\epsilon$.  Solid lines correspond to $e^+
  e^-$ machines operating at $\sqrt{s} = 250$~GeV with unpolarized
  beams, while dashed lines correspond to $\sqrt{s} = 500$~GeV.  The
  mass of $\tilde{K}$ is derived from $m_K$ using~\eqnref{KZmasses}. }
\label{fig:combined-XS}
\end{figure*}

The cross section for $\tilde{Z} \tilde{K}$ production is shown
in~\figref{combined-XS} for various choices of $m_K$ and $\epsilon$.
We see that $\tilde{Z} \tilde{K}$ production grows with $\epsilon^2$,
as expected.  We consider both $\tilde{K} \to \ell^+ \ell^-$ and
$\tilde{K} \to \bar{\chi} \chi$ decays, for $\ell = e$ or $\mu$, where
the missing energy branching ratio dominates by $g_D^2 / \epsilon^2$.
We also study $\tilde{Z} \to \ell^+ \ell^-$ with the SM branching
fraction of $6.8\%$~\cite{Olive:2016xmw}.  We show the background
cross sections for the corresponding $2\ell 2\nu$ and $4\ell$ final
states after the preselection cuts described
in~\subsecref{NPproduction} in~\tableref{cut-table}.  The $2\ell 2\nu$
background includes a combination of $Z\nu \nu$, $ZZ$ and $W^+ W^-$
processes, while the $4\ell$ background is mainly attributed to $ZZ /
Z\gamma^*$ production.

\begin{table}[tb!]
\centering
\scriptsize

\begin{tabular}{|M{0.09\textwidth}|M{0.05 \textwidth}|M{0.18\textwidth}|M{0.06\textwidth}|M{0.18\textwidth}|M{0.25\textwidth}|} \hline 
  Parameter  &   \multicolumn{2}{c|}{Signal process}  &  \multicolumn{2}{c|}{Background (pb)}   &  Signal region   \\ \hline
  
\multirow{16}{*}{$\epsilon$} &  
\multirow{4}{*}{$\tilde{Z} \tilde{K}$} 
& \multirow{2}{*}{$\tilde{Z} \to \bar{\ell} \ell$, $\tilde{K} \to \bar{\chi} \chi$} 
& \multirow{2}{*}{$\bar{\ell} \ell \bar{\nu} \nu$} 
        & 0.929 (250 GeV)  & $N_\ell \geq 2$, $|m_{\ell \ell} - m_Z| < 10$ GeV,  \\  \cline{5-5}
&  & &  & 0.545 (500 GeV)  & and $|m_{\text{recoil}} - m_{\tilde{K}}| < 2.5$ GeV \\  \cline{3-6}
& & \multirow{2}{*}{$\tilde{Z}\to \bar{\ell} \ell$, $\tilde{K}\to \bar{\ell} \ell$} 
& \multirow{2}{*}{$\bar{\ell} \ell \bar{\ell} \ell$} & 0.055 (250 GeV) & $N_\ell \geq 4$, $|m_{\ell \ell}-m_Z| < 10$ GeV, \\  \cline{5-5}
        &  &   & & 0.023 (500 GeV)   & and $|m_{\ell \ell} - m_{\tilde{K}}|< 2.5$ GeV \\  \cline{2-6}
& \multirow{6}{*}{ $\tilde{A} \tilde{K}$ }
& \multirow{2}{*}{$\tilde{K}$ inclusive decay}  
&  \multirow{2}{*}{$\gamma \bar{f} f$} & 23.14 (250 GeV) &  $N_\gamma \geq 1$, and   \\  \cline{5-5}
& &  &  &  8.88 (250 GeV) &   $|E_\gamma-(\frac{\sqrt{s}}{2} - \frac{m_{\tilde{K}}^2}{2\sqrt{s}})| < 2.5$ GeV  \\  \cline{3-6}
&  & \multirow{2}{*}{$\tilde{K} \to \bar{\ell} \ell$} &  \multirow{2}{*}{$\gamma \bar{\ell} \ell$}  &  12.67 (250 GeV) & $N_\gamma \geq 1$, $N_{\ell} \geq 2$, $|E_\gamma-(\frac{\sqrt{s}}{2}-\frac{m_{\tilde{K}}^2}{2\sqrt{s}})| < 2.5$ GeV, \\  \cline{5-5}
& &  &  &  4.38 (500 GeV) & and $|m_{\ell \ell}-m_{\tilde{K}}|< 5$ GeV \\  \cline{3-6}
& & \multirow{2}{*}{$\tilde{K} \to \bar{\chi} \chi$} & \multirow{2}{*}{$\gamma \bar{\nu} \nu $}  &  3.45 (250 GeV) & $N_\gamma \geq 1$, $|E_\gamma-(\frac{\sqrt{s}}{2}-\frac{m_{\tilde{K}}^2}{2\sqrt{s}})| < 2.5$ GeV, \\  \cline{5-5}
& &  &  &  2.92 (500 GeV) & and $\slashed{E} > 50$~GeV \\  \cline{2-6}
&  \multirow{2}{*}{$\tilde{Z} H_0$}  & $H_0 \to \tilde{K} \tilde{Z}$ with &  \multirow{2}{*}{$\bar{\ell} \bar{\ell} \ell \ell \bar{\nu} \nu$} & $1.8 \times 10^{-5}$ (250 GeV) & $N_\ell \geq 4$, $|m_{\ell \ell} - m_Z| < 10$ GeV, \\   \cline{5-5}
&  &  $\tilde{K}\to \bar{\chi} \chi$, $\tilde{Z} \to \bar{\ell} \ell$   & & $3.5\times 10^{-4}$ (500 GeV) & and $|m_{\text{recoil}} - m_{\tilde{K}}| < 2.5$ GeV \\ \hline
\multirow{2}{*}{ $\sin\alpha$} 
& \multirow{2}{*}{$\tilde{Z}S$} & $\tilde{Z}\to \bar{\ell} \ell$ & \multirow{2}{*}{$\bar{\ell} \ell \bar{\nu} \nu$}  &  0.87 (250 GeV) & $N_\ell \geq 2$, $|m_{\ell \ell} - m_Z| < 10$ GeV,  \\  \cline{5-5}
&  &    $S \to \tilde{K} \tilde{K} \to 4\chi$ &   &  0.87 (250 GeV) & and $|m_{\text{recoil}} - m_S| < 2.5$ GeV \\  \cline{5-5}
   \hline
\end{tabular}
\caption{Summary of the different vector + scalar and vector + vector
  production modes studied, along with the most salient cuts to
  identify the individual signals.  All background processes include
  up to one additional photon to account for initial and final state
  radiation.  Background rates are given for $\sqrt{s} = 250$~GeV or
  $500$~GeV, and visible particles are required to satisfy
  preselection cuts given in the main text.}
\label{table:cut-table}
\end{table}

\begin{figure*}[tbh!]
  \includegraphics[width=0.8\textwidth]{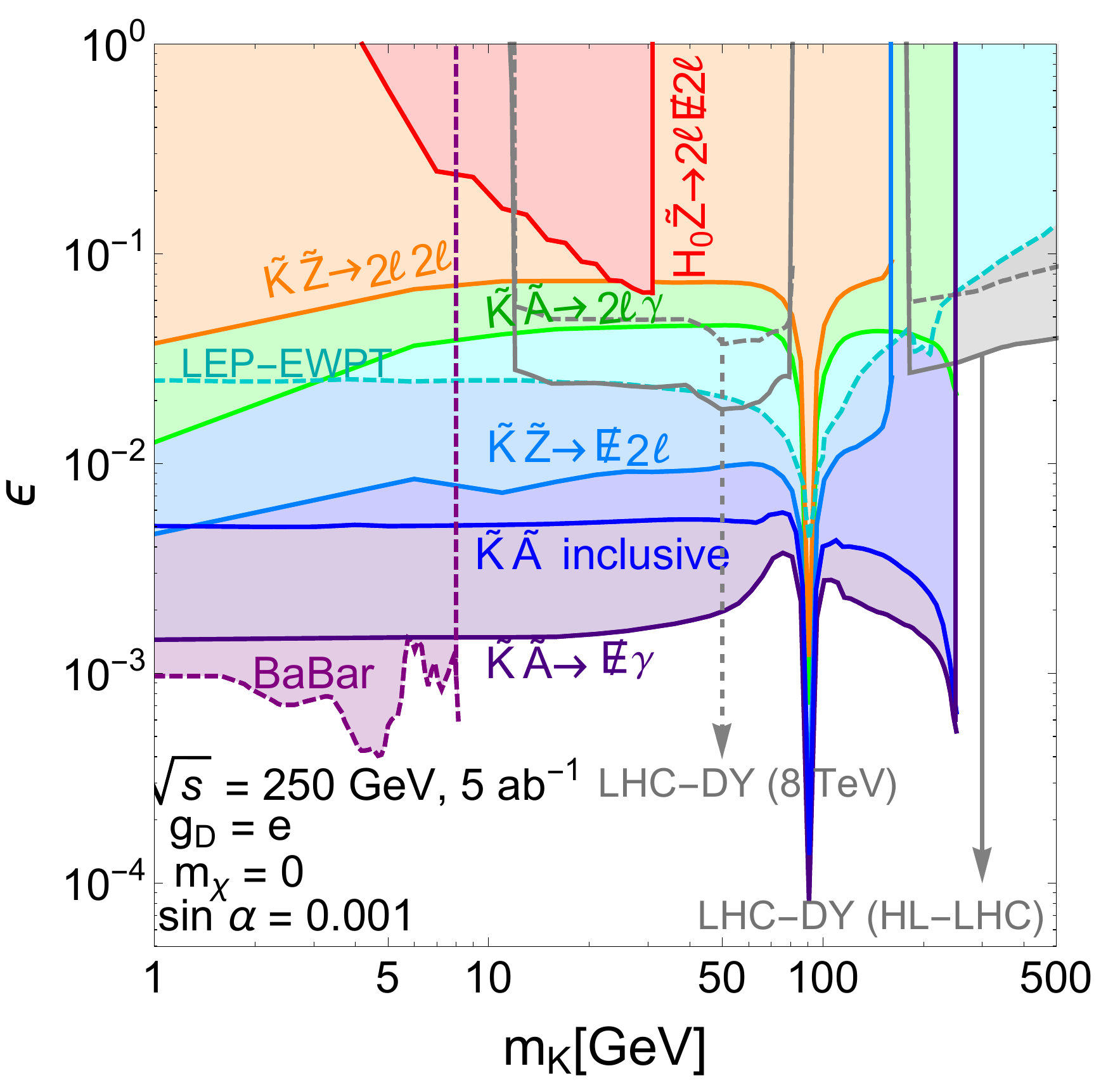}
\caption{Projected exclusion regions in the $\epsilon$ vs.~$m_K$ plane
  from multiple complementary searches of $\tilde{K}$ production.
  Solid lines enclose expected exclusion regions with $L =
  5$~ab$^{-1}$ of $\sqrt{s} = 250$~GeV $e^+ e^-$ machine data. Dashed
  lines indicate existing limits from the LEP $e^-e^+ \to
  \ell^-\ell^+$ contact operator search, LEP electroweak precision
  tests (LEP-EWPT), BaBar $\tilde{K}$ invisible decay search (BaBar) 
  and LHC Drell-Yan constraints (LHC-DY). The $3$~ab$^{-1}$
  HL-LHC projection for Drell-Yan constraints is also shown as a solid
  line.  Note $m_K$ is approximately the $m_{\tilde{K}}$ mass
  eigenvalue according to~\eqnref{KZmasses}.}
\label{fig:combined-epsilon-constraint}
\end{figure*}

\begin{figure*}[tbh!]
  \includegraphics[width=0.8\textwidth]{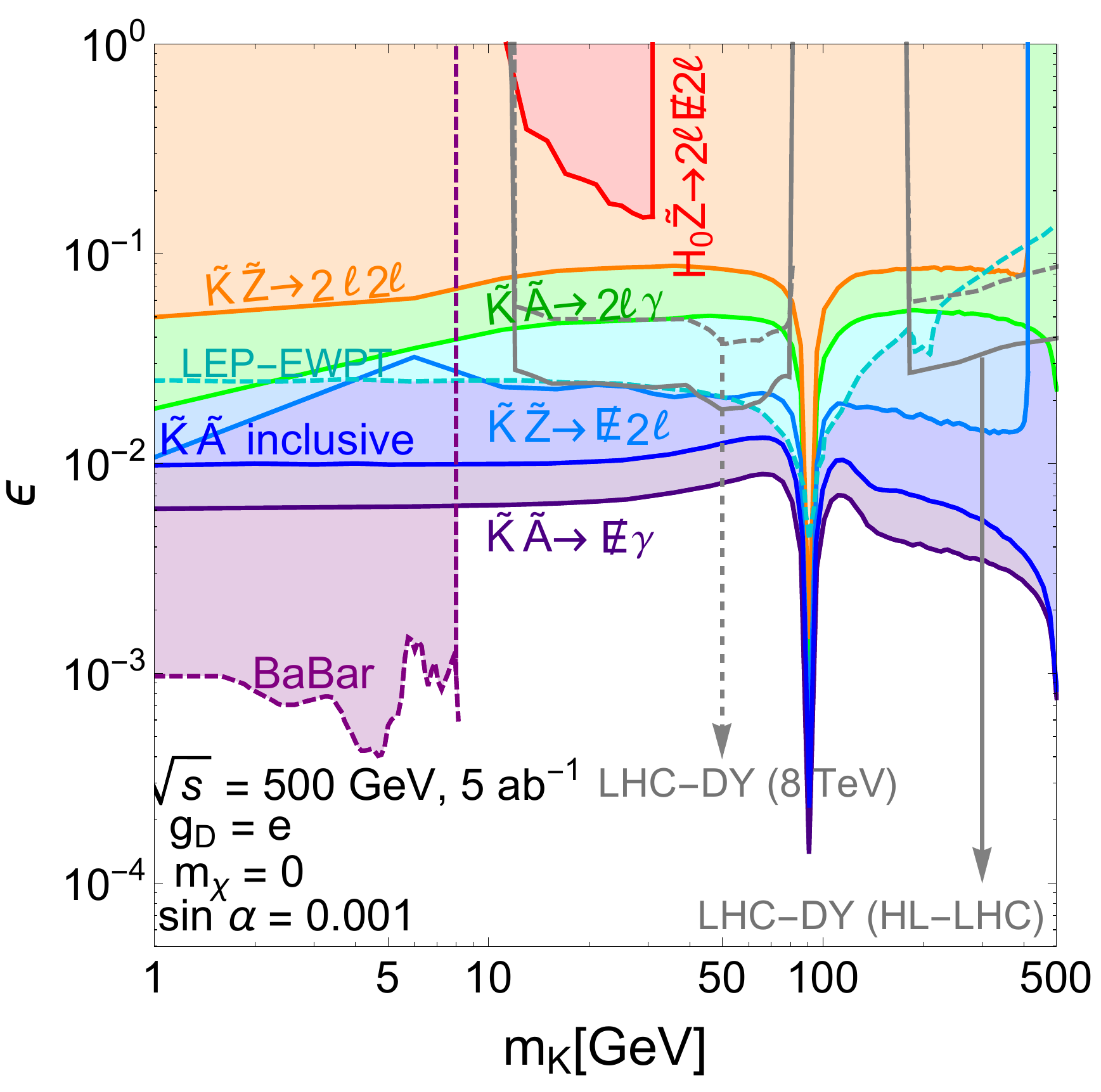}
  \caption{Same as~\figref{combined-epsilon-constraint}, except the
    $e^+ e^-$ projections are made for a $\sqrt{s} = 500$~GeV
    machine.}
\label{fig:combined-epsilon-constraint2}
\end{figure*}

For the $2\ell + \slashed{E}$ final state, we require a $Z$-candidate
with $|m_{\ell \ell} - m_Z| < 10$~GeV and then look for a peak in the
recoil mass distribution, see~\eqnref{recoilmass}.  For the $4\ell$
final state, we identify the $Z$-candidate from the opposite-sign,
same-flavor dilepton pair whose invariant mass is closest to the $Z$
mass and then study the invariant mass distribution of the remaining
dilepton pair.  The $\tilde{K}$ signal is tested for each signal mass
point in the corresponding mass distributions, and we draw 95\%
C.L.~exclusion regions for each channel in the $\epsilon$ vs.~$m_K$
plane for an integrated luminosity $L = 5$~ab$^{-1}$
in~\figref{combined-epsilon-constraint} and~\figref{combined-epsilon-constraint2}.  
The relative weight between
the $2\ell 2\chi$ and $4\ell$ final states is fixed by choosing $g_D =
e \approx 0.3$.  We see that the fully visible
$4\ell$ final state performs worse than the $2\ell 2\nu$ signal
selection, simply reflecting the dominant signal statistics in the
missing energy channel.

\subsubsection{$\tilde{A} \tilde{K}$ production}
\label{subsubsec:gammaK}

We study the radiative return process, $e^+ e^- \to \tilde{A}
\tilde{K}$, for $\tilde{K} \to \bar{\chi} \chi$, $\ell^+ \ell^-$, and
inclusive decays.  While each search will use the same observable,
namely a monochromatic peak in the photon energy as
in~\eqnref{visibleE}, the different contributions of SM backgrounds in
each event selection will result in the best sensitivity for the
$\tilde{K} \to \bar{\chi} \chi$ decay.  Background rates and signal
regions are shown in~\tableref{cut-table}.  For the inclusive decay of
$\tilde{K}$, the background $\gamma \bar{f} f$ is generated where $f$
is a SM fermion, including neutrinos.  As mentioned
in~\subsecref{NPproduction}, the visible energy distribution is
technically equivalent to the recoil mass distribution, and this
equivalence is sharpest when the visible SM state is a single photon.

From the results in~\figref{combined-epsilon-constraint} 
and~\figref{combined-epsilon-constraint2}, we see that
the most sensitive decay channel is $\tilde{K} \to \bar{\chi} \chi$,
again reflecting the dominant statistics in this final state and the
affordable reduction of SM backgrounds by the $\slashed{E}$ and
mono-chromatic photon requirements.  

The single photon in the background $\gamma \bar{\nu} \nu$ generally
comes from initial state radiation and hence tends to be soft except
when produced in the on-shell $\gamma \tilde{Z} \to \gamma
\bar{\nu}\nu$ process.  As long as $m_K \neq m_{\tilde{Z}}$, however,
the signal peak will not run overlap the background peak, and thus we
have a flat sensitivity to $\epsilon$ when $m_K < m_{\tilde{Z}}$.
There are two spikes in $\epsilon$ exclusion sensitivity.  The first
is for $m_K \sim m_{\tilde{Z}}$, when the production cross section is
greatly enhanced due to maximal $\tilde{K}-\tilde{Z}$ mixing, and the
second is for $m_K \sim \sqrt{s}$, when the production is enhanced by
soft, infrared divergent photon emission.  Note the exclusion can only
reach $m_K \sim \sqrt{s} -5$~GeV because of the preselection cut on
the photon energy.

\subsubsection{$\tilde{Z} H_0$, $H_0 \to \tilde{K} \tilde{Z}$ exotic decay}
\label{subsubsec:HtoKZexotic}

The next process we consider is the exotic Higgs decay, $H_0 \to
\tilde{K} \tilde{Z}$, with $\tilde{K} \to \bar{\chi} \chi$ and
$\tilde{Z} \to \ell^+ \ell^-$.  This Higgs exotic decay partial width,
from~\eqnref{HKZwidth}, is proportional to $\epsilon^2 \cos^2 \alpha$,
as long as $m_{\tilde{K}} \lesssim 34$~GeV and $\sin \alpha$ is
neglected.  The signal process thus has 2 $Z$ candidates balancing an
invisible $\tilde{K}$ particle, which we identify from the peak in the
recoil mass distribution.  Our event selection cuts, summarized
in~\tableref{cut-table}, require two pairs of opposite sign and same
flavor lepton with invariant masses in a window around
$m_{\tilde{Z}}$, and the recoil mass from the four visible charged
leptons should be in a window around the test variable
$m_{\tilde{K}}$.  The resulting sensitivity, as seen
in~\figref{combined-epsilon-constraint} and~\figref{combined-epsilon-constraint2},
 is not competitive with the
other $\tilde{K}$ production processes, given the limited Higgs
production statistics and the suppression of the small leptomic decay
branching ratio of $\tilde{Z}$.  We remark that this decay can also be
probed via $H_0 \to $ invisible searches using the SM rate for $Z \to
\bar{\nu} \nu$, which was discussed in~\subsecref{Higgschanges}.

\begin{figure*}[tb!]
\includegraphics[width=0.48\textwidth]{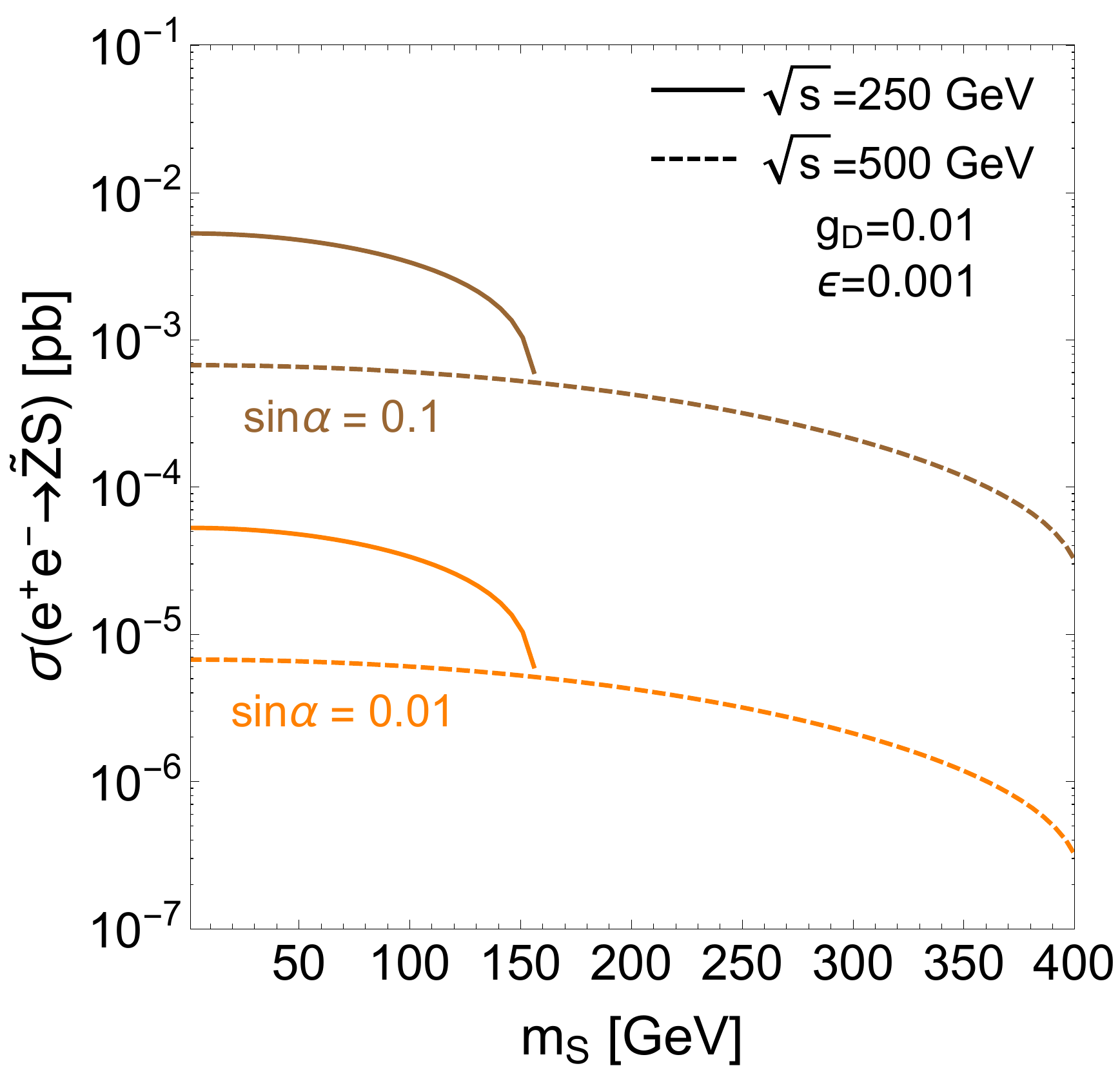}
\includegraphics[width=0.48\textwidth]{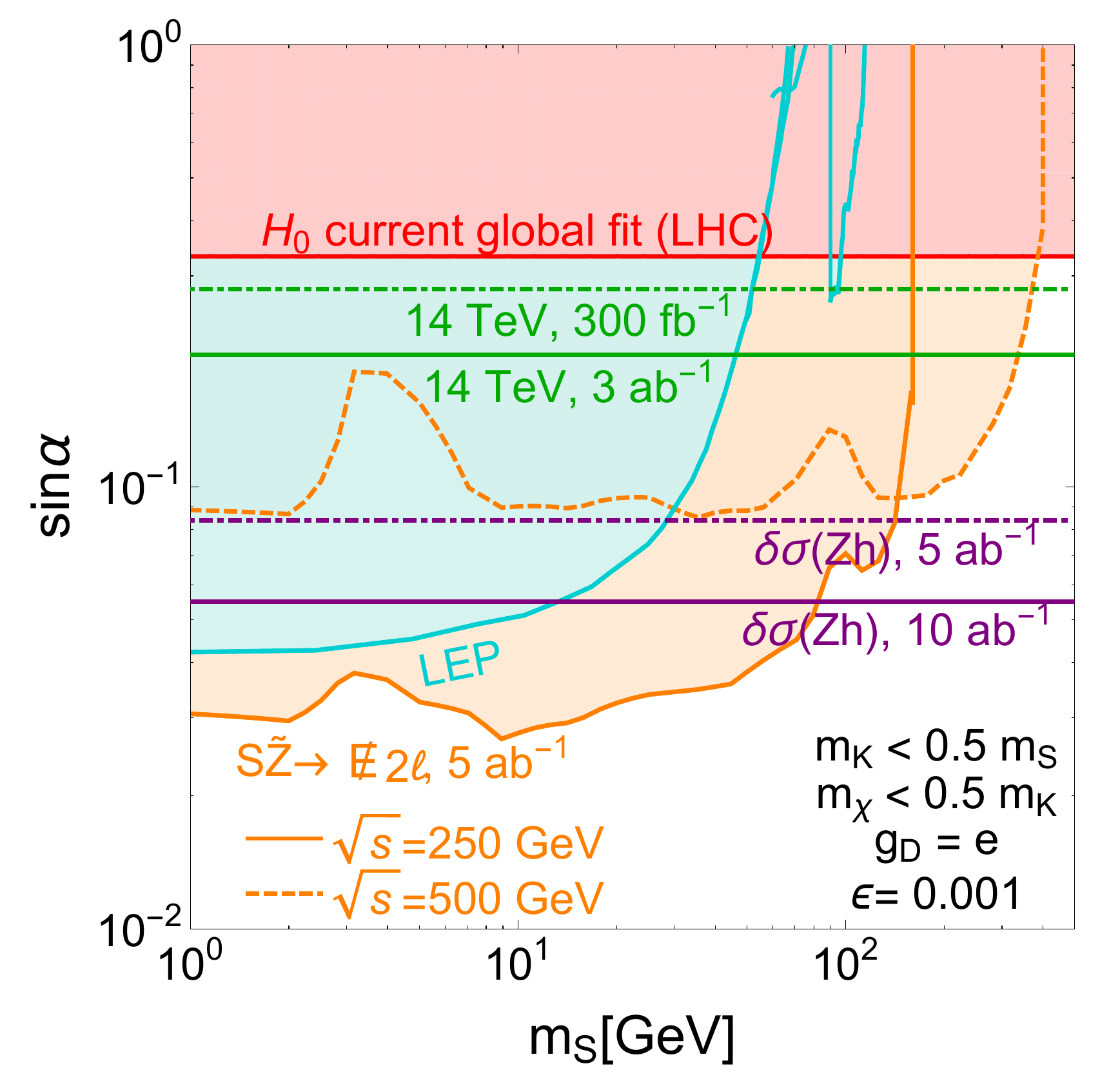}
\caption{ (Left panel) Cross section rates for the $e^+ e^- \to
  \tilde{Z} S$ process at $\sqrt{s} = 250$~GeV and $500$~GeV as a
  function of $m_S$, with $\sin \alpha = 0.1$ and $0.01$.  (Right
  panel) Exclusion reach from the $\tilde{Z} S$, $\tilde{Z} \to \ell^+
  \ell^-$ search in the recoil mass distribution for invisible $S$
  decays in the $\sin \alpha$ vs.~$m_S$ plane using $5$ ab$^{-1}$ of
  $e^+ e^-$ data at $\sqrt{s} = 250$~GeV or $500$~GeV.  We also show
  comparisons to the current fit, $\sin \alpha <
  0.33$~\cite{Khachatryan:2016vau}, future LHC projections of 0.28 (0.20)
  using 300 fb$^{-1}$ (3 ab$^{-1}$) luminosity~\cite{Dawson:2013bba},
  and precision $\delta\sigma(Zh)$ measurements constraining 0.084 (0.055)
  using 5 ab$^{-1}$ (10 ab$^{-1}$)~\cite{Gomez-Ceballos:2013zzn,
    CEPC-SPPCStudyGroup:2015csa, Ruan:2014xxa}.  We plot the excluded
  region from LEP searches for invisible low mass Higgs in $ZS$
  channel in cyan~\cite{Acciarri:1996um, Acciarri:1997tr,
    Abreu:1999vu, Searches:2001ab}.}
\label{fig:sig-XS-ZS}
\end{figure*}

\subsubsection{$\tilde{Z} S$ production}
\label{subsubsec:ZS}

Lastly, we can also probe the scalar mixing angle $\sin \alpha$ in
$\tilde{Z} S$ production.  This search is exactly analogous to the
previous search at LEP-II for a purely invisible decaying
Higgs~\cite{Acciarri:1996um}, where the visible $\tilde{Z} \to \ell^+
\ell^-$ decay is used to construct the recoil mass distribution.  The
$\tilde{Z} S$ cross section is proportional to $\sin^2 \alpha$ if we
neglect $\epsilon$, and $\sigma_{\tilde{Z}S}$ is shown
in~\figref{sig-XS-ZS} for $\sin \alpha = 0.1$ and $0.01$ at $\sqrt{s}
= 250$~GeV and $\sqrt{s} = 500$~GeV.  To maximize sensitivity to
$\alpha$, we study $\tilde{Z} \to \ell^+ \ell^-$ and $S$ decaying
invisibly.  The signal region is summarized in~\tableref{cut-table}
and focuses on selecting a dilepton $Z$ candidate and reconstructing
the recoil mass distribution to identify the $S$ peak.  From this
analysis, we find that $\sin \alpha = 0.03$ can be probed for light
$m_K$ using $L = 5$~ab$^{-1}$ luminosity for $\sqrt{s} = 250$~GeV, as
shown in~\figref{sig-XS-ZS}.  This result would significantly improve
on the current global fit to Higgs data by ATLAS, which constrains
$\sin \alpha < 0.33$~\cite{Khachatryan:2016vau}.  This sensitivity also
exceeds the projected LHC reach of $\sin \alpha < 0.28$ ($0.20$) using
300 fb$^{-1}$ (3 ab$^{-1}$) data and critical reductions in
theoretical uncertainties~\cite{Dawson:2013bba}.  We remark that
improved sensivity can be obtained by varying the $\sqrt{s}$ of the
collider to maximize the $\sigma(e^+ e^- \to \tilde{Z} S)$ rate for
the test $S$ mass (see also Ref.~\cite{Jia:2016pbe}).

\subsection{Summary}
\label{subsec:summaryplots}

\begin{figure*}[tb!]
\includegraphics[width=0.48\textwidth]{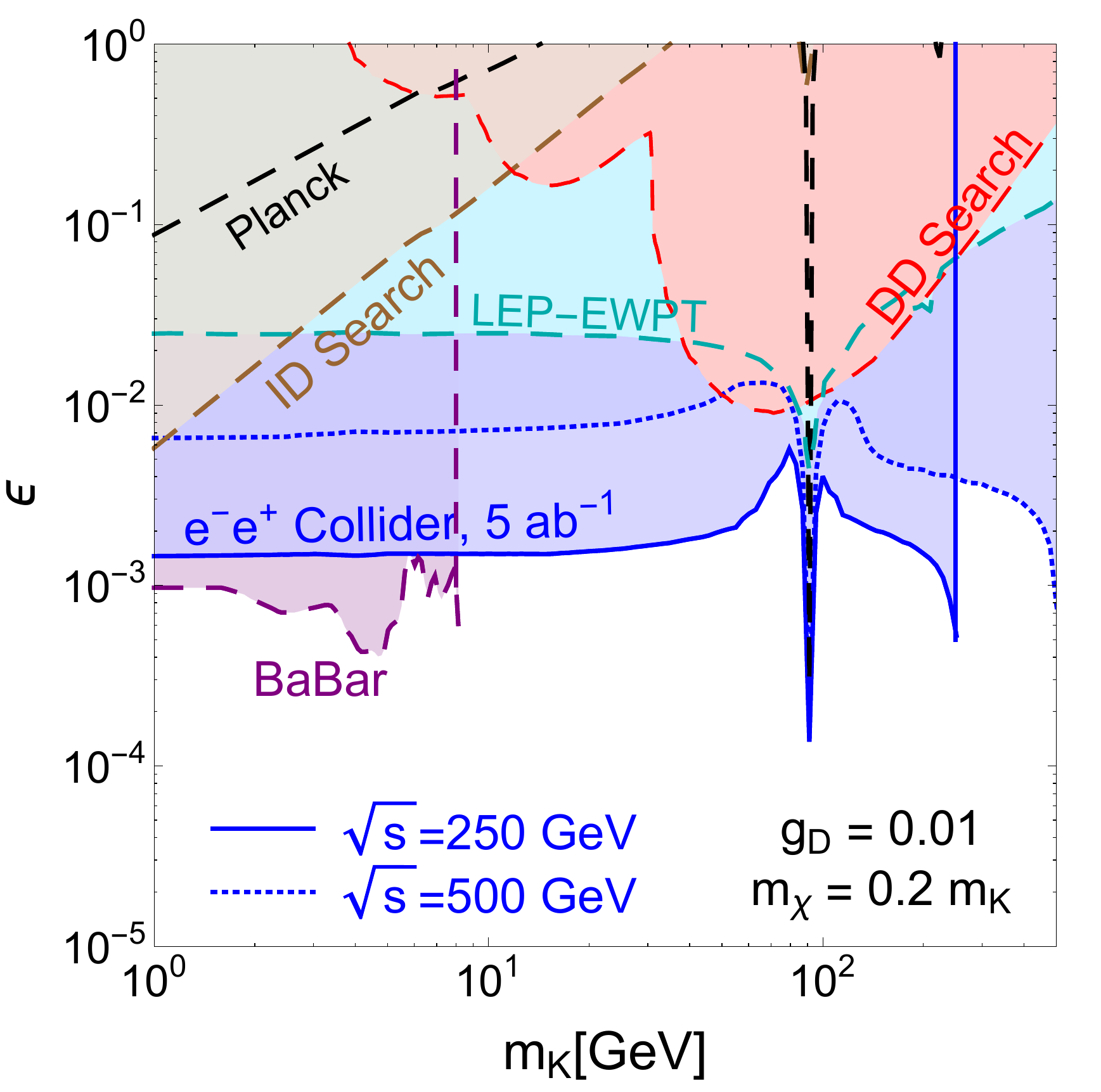}
\includegraphics[width=0.48\textwidth]{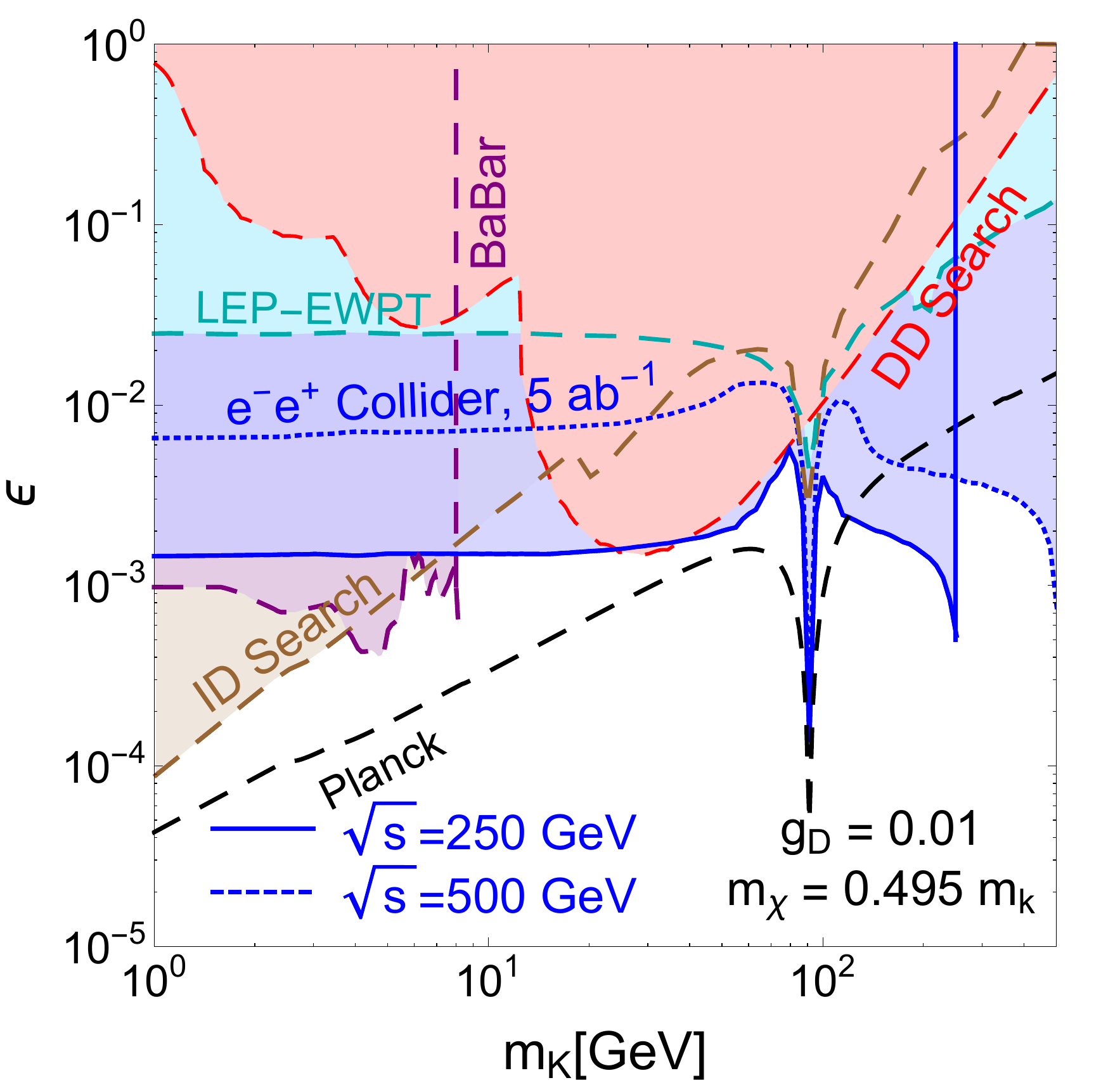}
\caption{Combined results of direct detection (red), indirect
  detection (brown), and $e^-e^+$ collider searches (blue) in
  the $\epsilon$ vs.~$m_K$ plane.  We choose $g_D = 0.01$, $m_\chi=0.2
  m_K$ (left panel) and $m_\chi=0.495 m_K$ (right panel).  We also
  show the contours when $\chi$ satisfies the relic density
  measurement by the Planck collaboration~\cite{Ade:2015xua} as black
  dashed lines.  The collider constraint is adapted
  from~\figref{combined-epsilon-constraint} 
  and~\figref{combined-epsilon-constraint2}, taking into account the
  changes in the $\tilde{K}$ branching fractions.  We also include
  existing constraints from LEP electroweak precision searches
  (LEP-EWPT) and the BaBar search for the $\tilde{K}$ invisible decay
  (BaBar).}
\label{fig:summary}
\end{figure*}

We summarize the sensitivity to $\epsilon$ in different channels at a
future $e^+e^-$ collider running at $\sqrt{s} = 250~(500)$ GeV with $L
= 5$~ab$^{-1}$ in~\figref{summary}, and we compare the collider
searches with constraints from direct detection and indirect detection
experiments.  In~\figref{summary}, the dark green shaded region is the
exclusion limit from the strongest of the $e^+e^-$ collider searches
presented in~\figref{combined-epsilon-constraint} 
and~\figref{combined-epsilon-constraint2}.  We also show the
strongest limit from direct detection and indirect detection
experiments from~\figref{DMdirectdetection}
and~\figref{DMindirectdetection}, as well as the contour satisfying
the correct dark matter relic abundance measured by
Planck~\cite{Ade:2015xua}.  While the constraints from dark matter
detection experiments depend sensitively on the dark matter mass, the
collider prospects are insensitive to the dark matter mass, as long as
the decay to $\chi$ is kinematically allowed and $g_D \gg \epsilon$.
We note that for $m_K$ around $m_{\tilde{Z}}$, the best limit comes
from the inclusive $\tilde{A} \tilde{K}$ search, which is insensitive
to $g_D$, while for $m_K$ larger or smaller than $m_{\tilde{Z}}$, the
best sensitivity comes from the monochromatic photon search with
$\slashed{E}$.

On the other hand, the indirect detection sensitivity and the relic
abundance contour both change significantly with dark matter mass.
When $m_{\chi} = 0.495 m_K$, the dark matter resonantly annihilates,
improving the reach for indirect searches and dramatically lowering
the required $\epsilon$ to satisfy the relic density measurement.
During thermal freeze-out, the finite temperature of the $\chi$
velocity distribution gives a strong boost to the annihilation cross
section, and thus only very small $\epsilon$ is needed.  For $m_{\chi}
= 0.2 m_K$, however, the limits from indirect detection exclude the
relic abundance contour, and the parameter space is instead
characterized by an overabundance of the dark matter relic density.
For this region to satisfy the Planck bound, additional mediators or
new dark matter dynamics controlling the freeze-out behavior are
needed.  Direct detection experiments also lose sensitivity to dark
matter signals for light $m_{\chi}$, since the nuclear recoil spectrum
is too soft to pass the fiducial energy threshold.  In addition, the
decreasing sensitivity for heavy $m_{\chi}$ comes from the fall off in
the scattering cross section scaling as $\mu_{\chi p}^2/
m_{\tilde{K}}^2$, see~\eqnref{DDxsec}.

We also emphasize that the collider constraint is not sensitive to
varying $g_D$ as long as $g_D \gg \epsilon$, which ensures the
invisible decay of $\tilde{K}$ dominates.  Hence, the collider
constraints from~\figref{combined-epsilon-constraint} and~\figref{combined-epsilon-constraint2}
and~\figref{summary} are essentially unchanged, since changing $g_D$
from $e$ to $0.01$ does not significantly change the invisible
branching fraction, except for the trade off between inclusive
$\tilde{K}$ decays and invisible $\tilde{K}$ decays around $m_K
\approx m_{\tilde{Z}}$.  On the other hand, the direct detection and
indirect detection rates scale with $g_D^2$, and thus collider
searches will have better sensitivity for small $g_D$.

From~\figref{summary}, we see that the prospective collider limits,
corresponding to the radiative return process $e^+ e^- \to \tilde{A}
\tilde{K}$, are expected to overtake the current bounds from direct
detection and indirect detection experiments.  In the case where dark
matter mass is light, $m_{\chi} = 0.2 m_K$, the collider limits are
typically at least one order of magnitude stronger than the current
limits, especially in the high mass region, and hence out of the reach
of next generation 1-ton scale direct detection experiments.  For dark
matter close to half the mediator mass, $m_{\chi} = 0.495 m_K$, the
thermal relic abundance measured by Planck~\cite{Ade:2015xua} offers
an attractive target parameter space for experimental probes.  The
projected $e^+ e^-$ sensitivity exceeds the current experimental
sensitivity around $m_K \sim 10$~GeV and $m_K > 100$~GeV, and while
improvements in the dark matter experiments will also challenge the
open parameter space for $m_K \sim 10$~GeV, the striking sensitivity
of $e^+ e^-$ radiative return processes for $m_K > 100$~GeV is
expected to be unmatched.  Thus, results from a future $e^+e^-$
collider will both complement and supersede the reach from dark matter
searches, stemming from its ability to produce directly the mediators
of dark matter interactions.

\section{Conclusion}
\label{sec:conclusions}

We have presented a comprehensive discussion of the phenomenology of
the Double Dark Portal model, which addresses the simultaneous
possibility of a kinetic mixing $\epsilon$ parameter with a scalar
Higgs portal $\lambda$.  We emphasize that these Lagrangian parameters
are generic in any $U(1)$ extension of the SM when the additional
gauge symmetry is Higgsed.  An additional motivation for considering
such a $U(1)$ extension is the fact that such a symmetry readily
stabilizes the lightest dark sector fermion $\chi$, making this model
a natural framework to study possible dark matter interactions in
tandem with updated precision $Z$ and Higgs constraints anticipated at
future colliders.  This study also demonstrates the ability of a
future $e^+ e^-$ machine to produce new particles, which are not
probed with the current dark matter and LHC experiments.

We work out the interactions in the mass eigenstate basis of neutral
vector bosons and Higgses.  The direct detection limits for this model
have been studied, along with indirect detection constraints from CMB
measurements, gamma ray measurements, and $e^{\pm}$ measurements,
where we have explored both the non-resonant and resonant dark matter
parameter regions.  For collider constraints, we discussed the
existing bounds from by electroweak precision and $Z$-pole
observables, Higgs measurements, Drell-Yan measurements, and radiative
return processes.  Previous constraints have mostly focused on the
visible decay, $\tilde{K} \to \ell^+ \ell^-$, and leaving the
prospects and sensitivity estimates for the invisible decay,
$\tilde{K} \to \bar{\chi} \chi$ largely unexplored.

We studied both the Higgs bremsstrahlung and radiative return
processes for a future $e^+ e^-$ collider, emphasizing that a future
lepton collider not only has vital Higgs precision capabilities but
also new possibilities for producing light new particles, $\tilde{K}$
and $S$.  Since both $\tilde{K}$ and $S$ decays are dominantly
invisible, the recoil mass method afforded by an $e^+ e^-$ machine is
crucial.  We also highlight that the recoil mass method can be
simplified to a monochromatic photon study in the case that the new
particle is produced in the radiative return process, which simplifies
the search procedure and enhances the importance for upcoming
calorimeters to have a precise, high-resolution energy determination
for photons.  The various Higgsstrahlung and radiative return
processes we study are listed in~\tableref{cut-table}, and we obtain
the best sensitivity on $\epsilon$ and $\sin\alpha$ from the radiative
return process $\tilde{A} \tilde{K}$ and Higgsstrahlung process
$\tilde{Z} S$, respectively.

In comparing $\epsilon$ prospects, we analyzed the future collider
reach with direct detection, indirect detection and relic abundance
sensitivities.  The collider prospects are less affected by DM mass
$m_\chi$, and surpass the other experimental probes for small $g_D$.
Since $\tilde{K}$ decays invisibly, the most relevant current
constraints are from electroweak precision measurements and LHC
mono-jet searches, but they are not as strong as the radiative return
process $\tilde{A} \tilde{K}$ reach.  Therefore, a future $e^+e^-$
collider provides an important and complementary sensitivity test of
the DDP model.

For $\sin \alpha$, the best constraints come from studying the singlet
bremsstrahlung process $\tilde{Z} S$, the Higgs invisible decay rate,
and precision measurements of SM Higgs production rates.  We studied
the $\tilde{Z} S$ process with $S$ decaying invisibly for a future
$e^+e^-$ collider and estimated the sensitivity to be $\sin \alpha
\sim 0.03$.  This compares favorably with earlier LEP studies for
light $m_S$, and readily provides leading sensitivity for heavy $S$.
We also recasted bounds using the Higgs invisible decay channel, where
the current LHC constraint is BR$_{\text{inv}} <
0.23$~\cite{Aad:2015pla, Khachatryan:2016whc} and the future $e^+ e^-$
collider reach is BR$_{\text{inv}} <
0.005$~\cite{Gomez-Ceballos:2013zzn}.  In the DDP model, these bounds
simultaneously constrain the three exotic processes, $H_0 \to S S$,
$H_0 \to \tilde{K} \tilde{K}$, and $H_0 \to \tilde{Z} \tilde{K}$ when
$\tilde{Z}$ decays to neutrinos.  While the constraints on $\sin
\alpha$ can be strong, these limits also depend sensitively on $g_D$
and are insignificant for small $g_D$.  The future $\sigma(e^+ e^- \to
Zh)$ precision measurement readily constrains $\cos^2\alpha$, but this
projection is weaker than the direct $\tilde{Z} S$ search.

In summary, the Double Dark Portal model predicts new dark sector
particles, $\tilde{K}$, $S$, and $\chi$, whose vector and scalar
portal interactions with the Standard Model can be uniquely tested at
a future $e^+ e^-$ collider.  We explicitly propose and study
radiative return and Higgsstrahlung processes to find the invisible
decays of the $\tilde{K}$ and $S$ mediators.  An additional benefit of
the $e^+ e^-$ search strategies discussed in this work is that, in the
event of a discovery, the $\tilde{K}$ or $S$ mass is immediately
measured in the recoil mass distribution.  Hence, a future $e^+ e^-$
collider not only has exciting prospects for determining the precise
properties of the 125~GeV Higgs boson, but also has a unique and
promising new physics program founded on the production of new, light,
hidden particles.

\section*{Acknowledgments}
\label{sec:acknowledgments}

This research is supported by the Cluster of Excellence Precision
Physics, Fundamental Interactions and Structure of Matter (PRISMA-EXC
1098).  FY would like to thank the hospitality of the CERN theory
group while this work was being completed.  The work of JL and XPW is
also supported by the German Research Foundation (DFG) under Grants
No. \mbox{KO 4820/1--1}, and No. FOR 2239, and from the European
Research Council (ERC) under the European Union’s Horizon 2020
research and innovation program (Grant No. 637506,
``$\nu$Directions'').

\begin{appendix}

\section{Two limiting cases for $\tilde{K}$, $\tilde{Z}$, and $\tilde{A}$ mixing}
\label{sec:limits}
From~\subsecref{neutralvectors}, we decompose the gauge eigenstate
vectors into their mass eigenstate components according to
\begin{align}
& \left( \begin{array}{c} 
Z_{\mu, \text{ SM}} \\ 
A_{\mu, \text{ SM}} \\
K_\mu  
\end{array} \right) =  U_1 U_2 R_M
 \left( \begin{array}{c} 
\tilde{Z}_{\mu} \\ 
\tilde{A}_{\mu} \\
\tilde{K}_\mu  \end{array} \right) \ ,
\end{align}
where the expressions for $U_1$, $U_2$ and $R_M$ have been given
in~\eqnref{U1U2def}, and~\eqnref{RMdef}, respectively.  We will
consider the two limiting cases, $m_K \to 0$ and $m_K \to
m_{Z,\text{SM}}$, and study the corresponding changes for the kinetic
and mass mixing matrices.

For $m_K \to 0$, the gauge boson masses are
\begin{align}
m_{\tilde A} &= m_{\tilde K } = 0 \ ,  \\
m_{\tilde{Z}}^2 &= m_{Z,\text{ SM}}^2 \frac{1-\epsilon^2}{1 - \epsilon^2 c_W^{-2}}
\approx m_{Z,\text{ SM}}^2 \left(1 + \epsilon^2 t_W^2 \right) + \mathcal{O}(\epsilon^3) \ ,
\end{align}
and the field redefinition is
\begin{align}
U_1 U_2
&= \left(
\begin{array}{ccc}
\sqrt{\frac{1-\epsilon^2}{1 - \epsilon^2 c_W^{-2} }} & 0 & 0 \\
-\frac{\epsilon^2 t_W}{\sqrt{(1-\epsilon^2)(1 - \epsilon^2 c_W^{-2})}} & 1 & \frac{\epsilon}{\sqrt{1-\epsilon^2}}    \\
-\frac{\epsilon t_W}{\sqrt{(1-\epsilon^2)(1 - \epsilon^2 c_W^{-2})}} & 0 & \frac{1}{\sqrt{1-\epsilon^2}}   
\end{array}
\right) \approx \left(
\begin{array}{ccc}
1 + \frac{1}{2} \epsilon^2 t_W^2 & 0 & 0 \\
-\epsilon^2 t_W & 1 & \epsilon    \\
-\epsilon t_W & 0 & 1+ \frac{1}{2} \epsilon^2 
\end{array}
\right) + \mathcal{O}(\epsilon^3)  \ .
\label{eqn:mkto0}
\end{align}
The Jacobi rotation $R_M$, from~\eqnref{RMdef}, is now ill-defined in
the lower right two-by-two block, since $\tilde{A}$ and $\tilde{K}$
can be rotated into each other keeping both the kinetic terms and
masses unchanged.  This simply reflects the residual unbroken
$U(1)_{\text{em}} \times U(1)_D$ gauge symmetry.  For $R_M = I_3$, the
currents are
\begin{align}
\mathcal{L} &\supset
\tilde{Z}_\mu
\left( g J_Z^\mu - g_D t_W \epsilon J_D^\mu + \dfrac{1}{2} g t_W^2 \epsilon^2 J_Z^\mu - e t_W \epsilon^2 J_{\text{em}}^\mu \right) \nonumber \\
&+ \tilde{K}_\mu \left( g_D J_D^\mu 
+ e \epsilon J_{\text{em}}^\mu 
+ \dfrac{1}{2} g_D \epsilon^2 J_D^\mu \right) \nonumber \\
&+ \tilde{A}_\mu e J_{\text{em}}^\mu \ ,
\label{eqn:mkto0currents}
\end{align}
but under a unitary rotation $U_X$ where $(\tilde{A}',~ \tilde{K}')^T
= U_X (\tilde{A},~ \tilde{K})^T$, the dark matter $\chi$ and the SM
fermions will generally have nonzero charges mediated by both
$\tilde{A}'$ and $\tilde{K}'$, leading to photon and dark
photon-mediated electric and dark millicharges.

For $m_K \to m_{Z, \text{ SM}}$, the masses of the three vector bosons
are
\begin{align}
m_{\tilde{A}} &= 0, \quad 
m_{\tilde{K},~ \tilde{Z}}^2 = m_{Z, \text{ SM}}^2 \left(
1 \mp \epsilon t_W + \frac{1}{2} \epsilon^2 \left( 1 + 2 t_W^2 \right) \right) \ ,
\end{align}
and the field redefinition required, to $\mathcal{O}(\epsilon^2)$, is
\begin{align}
U_1 U_2 R_M &= \dfrac{1}{\sqrt{2}} \left(
\begin{array}{ccc}
1 \mp \frac{\epsilon}{4} (t_W^{-1} -2 t_W) & 0 & 
\pm 1 + \frac{\epsilon}{4} (t_W^{-1} -2 t_W) \\
\mp \epsilon & \sqrt{2} & \epsilon \\
\mp 1 - \frac{\epsilon}{4} (t_W^{-1} + 2 t_W) & 0 & 
1 \mp \frac{\epsilon}{4} (t_W^{-1} + 2 t_W) 
\end{array} \right) \ ,
\label{eqn:mktomZ}
\end{align}
where the top and bottom signs correspond to $m_K \to m_{Z, \text{
    SM}}^\mp$.  We see that the mixing between $Z_\mu$ and $K_\mu$ is
nearly maximal, $45^{\circ}$, while the discontinuous behavior for
$m_K$ below and above $m_{Z, \text{ SM}}$ reflects the level crossing
in the mass eigenvalues.  We remark that as long as $\epsilon \neq 0$,
this maximal mixing feature remains, dictated by the structure of the
symmetric mass matrix in~\eqnref{intermediateKZmasses}.  If $\epsilon
= 0$ and $m_K = m_Z$, then the rotation matrix in~\eqnref{RMdef}
becomes ill-defined and the maximal mixing feature is lost.

\section{Cancellation effect in multiple kinetic mixing terms}
\label{sec:cancellation}

We observe that the $\tilde{Z}$ and $\tilde{K}$ mediated couplings
in~\eqnref{currents} show a cancellation effect when mediating DM
interactions with SM fermions.  This feature can be generalized to the
situation with multiple $U(1)$ gauge groups with multiple kinetic
mixing terms between each other.  Explicitly, we analyze the
Lagrangian
\begin{align}
\mathcal{L} \supset \frac{1}{4} V^T \mathcal{K} V + \frac{1}{2} V^T
M^2 V \ ,
\end{align}
where $\mathcal{K}_{ab} = \delta_{ab} + \mathcal{O}(\epsilon)
(1-\delta_{ab})$ is the kinetic mixing matrix and $M^2$ is the
diagonal mass matrix, with $a$, $b$ as indices.  Then, we define the
field redefinition matrix $U$ such that $U^T \mathcal{K} U =
\mathbb{I}$, which also gives $\tilde{M}^2 = U^T M^2\, U$ as the mass
matrix corresponding to the mass eigenstates, $\tilde{V} = U^{-1} V$.
Moreover, the gauge currents now become
\begin{align}
\mathcal{L} \supset g_i V_{\mu}^i J_i^{\mu} = g_i U_{ik} \tilde{V}_{k,
  \mu} J_i^{\mu} ,
\end{align}
in the mass basis.  As a result, scattering rates between two currents
$J_a$ and $J_b$ (which represent the corresponding fermion bilinears)
are schematically
\begin{align}
\mathcal{M} &\propto (g_a J_a^{\mu}) \otimes (g_b J_b^{\nu}) \left[
U_{ak} U_{bk} \dfrac{ g_{\mu \nu} - q_\mu q_\nu / m_{\tilde{V}_k}^2}{Q^2 - m_{\tilde{V}_k}^2} \right] \nonumber \\
&\sim (g_a J_a^{\mu}) \otimes (g_b J_b^{\nu}) \left[
U_{ak} (U^T)_{kb} \left( \dfrac{-g_{\mu \nu}}{m_{\tilde{V}_k}^2} + g_{\mu \nu} \, \mathcal{O}(\frac{Q^2}{m_{\tilde{V}_k}^4}) \right) \right] \ .
\end{align}
The $-g_{\mu \nu} / m_{\tilde{V}_k}^2$ term in the parentheses,
however, vanishes, when including the sum over $U_{ak} (U^T)_{kb}$,
because these transformations are controlled by the diagonalization
requirement of the two mass matrices, specifically $U \tilde{M}^{-2}
\,U^T = M^{-2}$.  The leading contribution in the amplitude is then
proportional to $\epsilon Q^2 / m_{\tilde{V}}^2$.

\section{Annihilation cross sections}
\label{sec:annXS}
In this section, we present the annihilation cross sections for the
processes $\bar{\chi} \chi \to \bar{f} f$, $W^+ W^-$, where $f$ is a
SM fermion.  We focus on the case with $m_\chi < m_{\tilde{K}}$, since
otherwise the direct annihilation of dark matter to dark vectors
$\tilde{K} \tilde{K}$ opens up and does not depend on $\epsilon$.  In
this setup, the annihilation cross section is proportional to $g_D^2
\epsilon^2$.  The diagrams include $s$-channel $\tilde{K}$ and
$\tilde{Z}$ exchange.  The annihilation cross sections before thermal
averaging are
\begin{align}
\sigma v &\left( \bar \chi \chi \to \ell^+ \ell^- \right) = \frac{e^2 \epsilon ^2 g_D^2 \sqrt{s-4 m_{\ell}^2} \left(2 m_{\chi }^2+s\right) }{48 \pi s^{3/2} c_W^4 \left(m_{\tilde{Z}}^2-m_{\tilde{K}}^2\right)^2\left( m_{\tilde{Z}}^2 \Gamma_{\tilde{Z}}^2 + \left(s - m_{\tilde{Z}}^2\right)^2 \right) \left(m_{\tilde{K}}^2 \Gamma_{\tilde{K}}^2 + \left( s - m_{\tilde{K}}^2 \right)^2 \right)} \nonumber \\
&\times \left[ \left(5s+7m_{\ell}^2\right) \left(s^2(m_{\tilde{Z}}^2-m_{\tilde{K}}^2)^2+m_{\tilde{Z}}^2m_{\tilde{K}}^2(m_{\tilde{K}}\Gamma_{\tilde{Z}}-m_{\tilde{Z}}\Gamma_{\tilde{K}})^2\right) \right. \nonumber \\
&-12c_W^2 (s+2m_{\ell}^2)m_{\tilde{Z}}^2\left(s(m_{\tilde{Z}}^2-m_{\tilde{K}}^2)^2+m_{\tilde{Z}}m_{\tilde K}(-m_{\tilde K}\Gamma_{\tilde{Z}}+m_{\tilde{Z}}\Gamma_{\tilde K})(-m_{\tilde{Z}}\Gamma_{\tilde{Z}}+m_{\tilde K}\Gamma_{\tilde K})\right) \nonumber \\
&\left. +8c_W^2 (s+2m_{\ell}^2)m_{\tilde{Z}}^4\left(m_{\tilde K}^4+m_{\tilde{Z}}^2(m_{\tilde{Z}}^2+\Gamma_{\tilde{Z}}^2)-2m_{\tilde{Z}}m_{\tilde K}\Gamma_{\tilde{Z}}\Gamma_{\tilde K}+m_{\tilde K}^2(-2m_{\tilde{Z}}^2+\Gamma_{\tilde K}^2) \right) \right] \ ,
\end{align}
\begin{align}
\sigma v &\left( \bar \chi \chi \to \bar u u \right) = \frac{e^2 \epsilon ^2 g_D^2 \sqrt{s-4 m_u^2} \left(2 m_{\chi }^2+s\right) }{144 \pi s^{3/2} c_W^4 \left(m_{\tilde{Z}}^2-m_{\tilde K}^2\right)^2\left( m_{\tilde{Z}}^2 \Gamma_{\tilde{Z}}^2 + \left(s - m_{\tilde{Z}}^2\right)^2 \right) \left(m_{\tilde{K}}^2 \Gamma_{\tilde{K}}^2 + \left( s - m_{\tilde{K}}^2 \right)^2 \right)} \nonumber \\
&\times \left[ \left(17s+7m_u^2\right)\left(s^2(m_{\tilde{Z}}^2-m_{\tilde K}^2)^2+m_{\tilde{Z}}^2m_{\tilde K}^2(m_{\tilde K}\Gamma_{\tilde{Z}}-m_{\tilde{Z}}\Gamma_{\tilde K})^2\right)\right. \nonumber \\
&-40c_W^2 (s+2m_u^2)m_{\tilde{Z}}^2\left(s(m_{\tilde{Z}}^2-m_{\tilde K}^2)^2+m_{\tilde{Z}}m_{\tilde K}(-m_{\tilde K}\Gamma_{\tilde{Z}}+m_{\tilde{Z}}\Gamma_{\tilde K})(-m_{\tilde{Z}}\Gamma_{\tilde{Z}}+m_{\tilde K}\Gamma_{\tilde K})\right) \nonumber \\
&+\left. 32 c_W^4(s+2m_u^2)m_{\tilde{Z}}^4\left(m_{\tilde K}^4+m_{\tilde{Z}}^2(m_{\tilde{Z}}^2+\Gamma_{\tilde{Z}}^2)-2m_{\tilde{Z}}m_{\tilde K}\Gamma_{\tilde{Z}}\Gamma_{\tilde K}+m_{\tilde K}^2(-2m_{\tilde{Z}}^2+\Gamma_{\tilde K}^2)\right)\right] \ ,
\end{align}
\begin{align}
\sigma v &\left( \bar \chi \chi \to \bar d d \right) = \frac{e^2 \epsilon ^2 g_D^2 \sqrt{s-4 m_d^2} \left(2 m_{\chi }^2+s\right) }{144 \pi s^{3/2} c_W^4 \left(m_{\tilde{Z}}^2-m_{\tilde K}^2\right)^2\left( m_{\tilde{Z}}^2 \Gamma_{\tilde{Z}}^2 + \left(s - m_{\tilde{Z}}^2\right)^2 \right) \left(m_{\tilde{K}}^2 \Gamma_{\tilde{K}}^2 + \left( s - m_{\tilde{K}}^2 \right)^2 \right)} \nonumber \\
&\times \left[ \left(5s-17m_d^2\right)\left(s^2(m_{\tilde{Z}}^2-m_{\tilde K}^2)^2+m_{\tilde{Z}}^2m_{\tilde K}^2(m_{\tilde K}\Gamma_{\tilde{Z}}-m_{\tilde{Z}}\Gamma_{\tilde K})^2\right)\right. \nonumber \\
&-4c_W^2 (s+2m_d^2)m_{\tilde{Z}}^2\left(s(m_{\tilde{Z}}^2-m_{\tilde K}^2)^2+m_{\tilde{Z}}m_{\tilde K}(-m_{\tilde K}\Gamma_{\tilde{Z}}+m_{\tilde{Z}}\Gamma_{\tilde K})(-m_{\tilde{Z}}\Gamma_{\tilde{Z}}+m_{\tilde K}\Gamma_{\tilde K})\right) \nonumber \\
&+\left. 8 c_W^4(s+2m_d^2)m_{\tilde{Z}}^4\left(m_{\tilde K}^4+m_{\tilde{Z}}^2(m_{\tilde{Z}}^2+\Gamma_{\tilde{Z}}^2)-2m_{\tilde{Z}}m_{\tilde K}\Gamma_{\tilde{Z}}\Gamma_{\tilde K}+m_{\tilde K}^2(-2m_{\tilde{Z}}^2+\Gamma_{\tilde K}^2) \right) \right] \ ,
\end{align}
\begin{align}
\sigma v &\left( \bar \chi \chi \to \bar \nu_{\ell} \nu_{\ell}\right) = \frac{e^2 \epsilon ^2 g_D^2 \left(2 m_{\chi }^2+s\right)\left(s^2(m_{\tilde{Z}}^2-m_{\tilde K}^2)^2+m_{\tilde{Z}}^2m_{\tilde K}^2(m_{\tilde K}\Gamma_{\tilde{Z}}-m_{\tilde{Z}}\Gamma_{\tilde K})^2\right)}{48 \pi c_W^4 \left(m_{\tilde{Z}}^2-m_{\tilde K}^2\right)^2 
\left( m_{\tilde{Z}}^2 \Gamma_{\tilde{Z}}^2 + \left(s - m_{\tilde{Z}}^2\right)^2 \right) \left(m_{\tilde{K}}^2 \Gamma_{\tilde{K}}^2 + \left( s - m_{\tilde{K}}^2 \right)^2 \right)} \ ,
\end{align}
\begin{align}
\sigma v &\left( \bar \chi \chi \to W^+ W^- \right) = \frac{e^2 \epsilon ^2 g_D^2 \left(2 m_{\chi }^2+s\right) \left(s-4 m_W^2 \right){}^{3/2} \left(20 s m_W^2 +12 m_W^4+s^2\right)}{96 \pi s^{3/2} c_W^4 \left( m_{\tilde{Z}}^2-m_{\tilde K}^2 \right)^2
\left( m_{\tilde{Z}}^2 \Gamma_{\tilde{Z}}^2 + \left(s - m_{\tilde{Z}}^2 \right)^2 \right)
\left( m_{\tilde{K}}^2 \Gamma_{\tilde{K}}^2 + \left(s - m_{\tilde{K}}^2 \right)^2 \right)} \nonumber \\
&\times \left(m_{\tilde{Z}}^4+m_{\tilde{Z}}^2(-2m_{\tilde K}^2+\Gamma_{\tilde{Z}}^2)-2m_{\tilde{Z}} m_{\tilde K}\Gamma_{\tilde{Z}}\Gamma_{\tilde K}+m_{\tilde K}^2(m_{\tilde K}^2+\Gamma_{\tilde K}^2)\right) \ ,
\end{align}
where $s$ is the Mandelstam parameter for the center-of-mass energy
squared.
\end{appendix}

\bibliography{ref}

\begin{thebibliography}{99}%
\makeatletter
\providecommand \@ifxundefined [1]{%
 \@ifx{#1\undefined}
}%
\providecommand \@ifnum [1]{%
 \ifnum #1\expandafter \@firstoftwo
 \else \expandafter \@secondoftwo
 \fi
}%
\providecommand \@ifx [1]{%
 \ifx #1\expandafter \@firstoftwo
 \else \expandafter \@secondoftwo
 \fi
}%
\providecommand \natexlab [1]{#1}%
\providecommand \enquote  [1]{``#1''}%
\providecommand \bibnamefont  [1]{#1}%
\providecommand \bibfnamefont [1]{#1}%
\providecommand \citenamefont [1]{#1}%
\providecommand \href@noop [0]{\@secondoftwo}%
\providecommand \href [0]{\begingroup \@sanitize@url \@href}%
\providecommand \@href[1]{\@@startlink{#1}\@@href}%
\providecommand \@@href[1]{\endgroup#1\@@endlink}%
\providecommand \@sanitize@url [0]{\catcode `\\12\catcode `\$12\catcode
  `\&12\catcode `\#12\catcode `\^12\catcode `\_12\catcode `\%12\relax}%
\providecommand \@@startlink[1]{}%
\providecommand \@@endlink[0]{}%
\providecommand \url  [0]{\begingroup\@sanitize@url \@url }%
\providecommand \@url [1]{\endgroup\@href {#1}{\urlprefix }}%
\providecommand \urlprefix  [0]{URL }%
\providecommand \Eprint [0]{\href }%
\providecommand \doibase [0]{http://dx.doi.org/}%
\providecommand \selectlanguage [0]{\@gobble}%
\providecommand \bibinfo  [0]{\@secondoftwo}%
\providecommand \bibfield  [0]{\@secondoftwo}%
\providecommand \translation [1]{[#1]}%
\providecommand \BibitemOpen [0]{}%
\providecommand \bibitemStop [0]{}%
\providecommand \bibitemNoStop [0]{.\EOS\space}%
\providecommand \EOS [0]{\spacefactor3000\relax}%
\providecommand \BibitemShut  [1]{\csname bibitem#1\endcsname}%
\let\auto@bib@innerbib\@empty
\bibitem [{\citenamefont {Dawson}\ \emph {et~al.}(2013)\citenamefont {Dawson}
  \emph {et~al.}}]{Dawson:2013bba}%
  \BibitemOpen
  \bibfield  {author} {\bibinfo {author} {\bibfnamefont {Sally}\ \bibnamefont
  {Dawson}} \emph {et~al.},\ }\bibfield  {title} {\enquote {\bibinfo {title}
  {{Working Group Report: Higgs Boson}},}\ }in\ \href
  {http://inspirehep.net/record/1262795/files/arXiv:1310.8361.pdf} {\emph
  {\bibinfo {booktitle} {{Proceedings, 2013 Community Summer Study on the
  Future of U.S. Particle Physics: Snowmass on the Mississippi (CSS2013):
  Minneapolis, MN, USA, July 29-August 6, 2013}}}}\ (\bibinfo {year} {2013})\
  \Eprint {http://arxiv.org/abs/1310.8361} {arXiv:1310.8361 [hep-ex]}
  \BibitemShut {NoStop}%
\bibitem [{\citenamefont {Aarons}\ \emph {et~al.}(2007)\citenamefont {Aarons}
  \emph {et~al.}}]{Djouadi:2007ik}%
  \BibitemOpen
  \bibfield  {author} {\bibinfo {author} {\bibfnamefont {Gerald}\ \bibnamefont
  {Aarons}} \emph {et~al.} (\bibinfo {collaboration} {ILC}),\ }\bibfield
  {title} {\enquote {\bibinfo {title} {{International Linear Collider Reference
  Design Report Volume 2: Physics at the ILC}},}\ }\href@noop {} {\  (\bibinfo
  {year} {2007})},\ \Eprint {http://arxiv.org/abs/0709.1893} {arXiv:0709.1893
  [hep-ph]} \BibitemShut {NoStop}%
\bibitem [{\citenamefont {Bicer}\ \emph {et~al.}(2014)\citenamefont {Bicer}
  \emph {et~al.}}]{Gomez-Ceballos:2013zzn}%
  \BibitemOpen
  \bibfield  {author} {\bibinfo {author} {\bibfnamefont {M.}~\bibnamefont
  {Bicer}} \emph {et~al.} (\bibinfo {collaboration} {TLEP Design Study Working
  Group}),\ }\bibfield  {title} {\enquote {\bibinfo {title} {{First Look at the
  Physics Case of TLEP}},}\ }\bibfield  {booktitle} {\emph {\bibinfo
  {booktitle} {{Proceedings, Community Summer Study 2013: Snowmass on the
  Mississippi (CSS2013): Minneapolis, MN, USA, July 29-August 6, 2013}}},\
  }\href {\doibase 10.1007/JHEP01(2014)164} {\bibfield  {journal} {\bibinfo
  {journal} {JHEP}\ }\textbf {\bibinfo {volume} {01}},\ \bibinfo {pages} {164}
  (\bibinfo {year} {2014})},\ \Eprint {http://arxiv.org/abs/1308.6176}
  {arXiv:1308.6176 [hep-ex]} \BibitemShut {NoStop}%
\bibitem [{\citenamefont {Group}(2015)}]{CEPC-SPPCStudyGroup:2015csa}%
  \BibitemOpen
  \bibfield  {author} {\bibinfo {author} {\bibfnamefont {CEPC-SPPC~Study}\
  \bibnamefont {Group}},\ }\bibfield  {title} {\enquote {\bibinfo {title}
  {{CEPC-SPPC Preliminary Conceptual Design Report. 1. Physics and
  Detector}},}\ }\href@noop {} {\  (\bibinfo {year} {2015})}\BibitemShut
  {NoStop}%
\bibitem [{\citenamefont {Yin}\ \emph {et~al.}(2009)\citenamefont {Yin},
  \citenamefont {Liu},\ and\ \citenamefont {Zhu}}]{Yin:2009mc}%
  \BibitemOpen
  \bibfield  {author} {\bibinfo {author} {\bibfnamefont {Peng-fei}\
  \bibnamefont {Yin}}, \bibinfo {author} {\bibfnamefont {Jia}\ \bibnamefont
  {Liu}}, \ and\ \bibinfo {author} {\bibfnamefont {Shou-hua}\ \bibnamefont
  {Zhu}},\ }\bibfield  {title} {\enquote {\bibinfo {title} {{Detecting light
  leptophilic gauge boson at BESIII detector}},}\ }\href {\doibase
  10.1016/j.physletb.2009.07.014} {\bibfield  {journal} {\bibinfo  {journal}
  {Phys. Lett.}\ }\textbf {\bibinfo {volume} {B679}},\ \bibinfo {pages}
  {362--368} (\bibinfo {year} {2009})},\ \Eprint
  {http://arxiv.org/abs/0904.4644} {arXiv:0904.4644 [hep-ph]} \BibitemShut
  {NoStop}%
\bibitem [{\citenamefont {Li}\ and\ \citenamefont {Luo}(2010)}]{Li:2009wz}%
  \BibitemOpen
  \bibfield  {author} {\bibinfo {author} {\bibfnamefont {Hai-Bo}\ \bibnamefont
  {Li}}\ and\ \bibinfo {author} {\bibfnamefont {Tao}\ \bibnamefont {Luo}},\
  }\bibfield  {title} {\enquote {\bibinfo {title} {{Probing Dark force at
  BES-III/BEPCII}},}\ }\href {\doibase 10.1016/j.physletb.2010.02.059}
  {\bibfield  {journal} {\bibinfo  {journal} {Phys. Lett.}\ }\textbf {\bibinfo
  {volume} {B686}},\ \bibinfo {pages} {249--253} (\bibinfo {year} {2010})},\
  \Eprint {http://arxiv.org/abs/0911.2067} {arXiv:0911.2067 [hep-ph]}
  \BibitemShut {NoStop}%
\bibitem [{\citenamefont {Lees}\ \emph {et~al.}(2014)\citenamefont {Lees} \emph
  {et~al.}}]{Lees:2014xha}%
  \BibitemOpen
  \bibfield  {author} {\bibinfo {author} {\bibfnamefont {J.~P.}\ \bibnamefont
  {Lees}} \emph {et~al.} (\bibinfo {collaboration} {BaBar}),\ }\bibfield
  {title} {\enquote {\bibinfo {title} {{Search for a Dark Photon in $e^+e^-$
  Collisions at BaBar}},}\ }\href {\doibase 10.1103/PhysRevLett.113.201801}
  {\bibfield  {journal} {\bibinfo  {journal} {Phys. Rev. Lett.}\ }\textbf
  {\bibinfo {volume} {113}},\ \bibinfo {pages} {201801} (\bibinfo {year}
  {2014})},\ \Eprint {http://arxiv.org/abs/1406.2980} {arXiv:1406.2980
  [hep-ex]} \BibitemShut {NoStop}%
\bibitem [{\citenamefont {Curtin}\ \emph {et~al.}(2015)\citenamefont {Curtin},
  \citenamefont {Essig}, \citenamefont {Gori},\ and\ \citenamefont
  {Shelton}}]{Curtin:2014cca}%
  \BibitemOpen
  \bibfield  {author} {\bibinfo {author} {\bibfnamefont {David}\ \bibnamefont
  {Curtin}}, \bibinfo {author} {\bibfnamefont {Rouven}\ \bibnamefont {Essig}},
  \bibinfo {author} {\bibfnamefont {Stefania}\ \bibnamefont {Gori}}, \ and\
  \bibinfo {author} {\bibfnamefont {Jessie}\ \bibnamefont {Shelton}},\
  }\bibfield  {title} {\enquote {\bibinfo {title} {{Illuminating Dark Photons
  with High-Energy Colliders}},}\ }\href {\doibase 10.1007/JHEP02(2015)157}
  {\bibfield  {journal} {\bibinfo  {journal} {JHEP}\ }\textbf {\bibinfo
  {volume} {02}},\ \bibinfo {pages} {157} (\bibinfo {year} {2015})},\ \Eprint
  {http://arxiv.org/abs/1412.0018} {arXiv:1412.0018 [hep-ph]} \BibitemShut
  {NoStop}%
\bibitem [{\citenamefont {Anastasi}\ \emph {et~al.}(2015)\citenamefont
  {Anastasi} \emph {et~al.}}]{Babusci:2015zda}%
  \BibitemOpen
  \bibfield  {author} {\bibinfo {author} {\bibfnamefont {A.}~\bibnamefont
  {Anastasi}} \emph {et~al.} (\bibinfo {collaboration} {KLOE-2}),\ }\bibfield
  {title} {\enquote {\bibinfo {title} {{Search for dark Higgsstrahlung in
  $e^+e^- \to \mu^+ \mu^-$ and missing energy events with the KLOE
  experiment}},}\ }\href {\doibase 10.1016/j.physletb.2015.06.015} {\bibfield
  {journal} {\bibinfo  {journal} {Phys. Lett.}\ }\textbf {\bibinfo {volume}
  {B747}},\ \bibinfo {pages} {365--372} (\bibinfo {year} {2015})},\ \Eprint
  {http://arxiv.org/abs/1501.06795} {arXiv:1501.06795 [hep-ex]} \BibitemShut
  {NoStop}%
\bibitem [{\citenamefont {Karliner}\ \emph {et~al.}(2015)\citenamefont
  {Karliner}, \citenamefont {Low}, \citenamefont {Rosner},\ and\ \citenamefont
  {Wang}}]{Karliner:2015tga}%
  \BibitemOpen
  \bibfield  {author} {\bibinfo {author} {\bibfnamefont {Marek}\ \bibnamefont
  {Karliner}}, \bibinfo {author} {\bibfnamefont {Matthew}\ \bibnamefont {Low}},
  \bibinfo {author} {\bibfnamefont {Jonathan~L.}\ \bibnamefont {Rosner}}, \
  and\ \bibinfo {author} {\bibfnamefont {Lian-Tao}\ \bibnamefont {Wang}},\
  }\bibfield  {title} {\enquote {\bibinfo {title} {{Radiative return
  capabilities of a high-energy, high-luminosity $e^+e^-$ collider}},}\ }\href
  {\doibase 10.1103/PhysRevD.92.035010} {\bibfield  {journal} {\bibinfo
  {journal} {Phys. Rev.}\ }\textbf {\bibinfo {volume} {D92}},\ \bibinfo {pages}
  {035010} (\bibinfo {year} {2015})},\ \Eprint
  {http://arxiv.org/abs/1503.07209} {arXiv:1503.07209 [hep-ph]} \BibitemShut
  {NoStop}%
\bibitem [{\citenamefont {Prasad}\ \emph {et~al.}(2015)\citenamefont {Prasad},
  \citenamefont {Li},\ and\ \citenamefont {Lou}}]{Prasad:2015mxa}%
  \BibitemOpen
  \bibfield  {author} {\bibinfo {author} {\bibfnamefont {Vindhyawasini}\
  \bibnamefont {Prasad}}, \bibinfo {author} {\bibfnamefont {Haibo}\
  \bibnamefont {Li}}, \ and\ \bibinfo {author} {\bibfnamefont {Xinchou}\
  \bibnamefont {Lou}},\ }\bibfield  {title} {\enquote {\bibinfo {title}
  {{Search for low-mass Higgs and dark photons at BESIII}},}\ }in\ \href
  {https://inspirehep.net/record/1391142/files/arXiv:1508.07659.pdf} {\emph
  {\bibinfo {booktitle} {{7th International Workshop on Charm Physics (Charm
  2015) Detroit, MI, USA, May 18-22, 2015}}}}\ (\bibinfo {year} {2015})\
  \Eprint {http://arxiv.org/abs/1508.07659} {arXiv:1508.07659 [hep-ex]}
  \BibitemShut {NoStop}%
\bibitem [{\citenamefont {Hochberg}\ \emph {et~al.}(2016)\citenamefont
  {Hochberg}, \citenamefont {Kuflik},\ and\ \citenamefont
  {Murayama}}]{Hochberg:2015vrg}%
  \BibitemOpen
  \bibfield  {author} {\bibinfo {author} {\bibfnamefont {Yonit}\ \bibnamefont
  {Hochberg}}, \bibinfo {author} {\bibfnamefont {Eric}\ \bibnamefont {Kuflik}},
  \ and\ \bibinfo {author} {\bibfnamefont {Hitoshi}\ \bibnamefont {Murayama}},\
  }\bibfield  {title} {\enquote {\bibinfo {title} {{SIMP Spectroscopy}},}\
  }\href {\doibase 10.1007/JHEP05(2016)090} {\bibfield  {journal} {\bibinfo
  {journal} {JHEP}\ }\textbf {\bibinfo {volume} {05}},\ \bibinfo {pages} {090}
  (\bibinfo {year} {2016})},\ \Eprint {http://arxiv.org/abs/1512.07917}
  {arXiv:1512.07917 [hep-ph]} \BibitemShut {NoStop}%
\bibitem [{\citenamefont {Won}\ \emph {et~al.}(2016)\citenamefont {Won} \emph
  {et~al.}}]{Won:2016pjz}%
  \BibitemOpen
  \bibfield  {author} {\bibinfo {author} {\bibfnamefont {E.}~\bibnamefont
  {Won}} \emph {et~al.} (\bibinfo {collaboration} {Belle}),\ }\bibfield
  {title} {\enquote {\bibinfo {title} {{Search for a dark vector gauge boson
  decaying to $\pi^+ \pi^-$ using $\eta \rightarrow \pi^+\pi^- \gamma$
  decays}},}\ }\href {\doibase 10.1103/PhysRevD.94.092006} {\bibfield
  {journal} {\bibinfo  {journal} {Phys. Rev.}\ }\textbf {\bibinfo {volume}
  {D94}},\ \bibinfo {pages} {092006} (\bibinfo {year} {2016})},\ \Eprint
  {http://arxiv.org/abs/1609.05599} {arXiv:1609.05599 [hep-ex]} \BibitemShut
  {NoStop}%
\bibitem [{\citenamefont {Kaneta}\ and\ \citenamefont
  {Shimomura}(2016)}]{Kaneta:2016uyt}%
  \BibitemOpen
  \bibfield  {author} {\bibinfo {author} {\bibfnamefont {Yuya}\ \bibnamefont
  {Kaneta}}\ and\ \bibinfo {author} {\bibfnamefont {Takashi}\ \bibnamefont
  {Shimomura}},\ }\bibfield  {title} {\enquote {\bibinfo {title} {{On the
  possibility of search for $L_\mu - L_\tau$ gauge boson at Belle-II and
  neutrino beam experiments}},}\ }\href@noop {} {\  (\bibinfo {year} {2016})},\
  \Eprint {http://arxiv.org/abs/1701.00156} {arXiv:1701.00156 [hep-ph]}
  \BibitemShut {NoStop}%
\bibitem [{\citenamefont {He}\ \emph {et~al.}(2017)\citenamefont {He},
  \citenamefont {He},\ and\ \citenamefont {Huang}}]{He:2017ord}%
  \BibitemOpen
  \bibfield  {author} {\bibinfo {author} {\bibfnamefont {Min}\ \bibnamefont
  {He}}, \bibinfo {author} {\bibfnamefont {Xiao-Gang}\ \bibnamefont {He}}, \
  and\ \bibinfo {author} {\bibfnamefont {Cheng-Kai}\ \bibnamefont {Huang}},\
  }\bibfield  {title} {\enquote {\bibinfo {title} {{Dark Photon Search at A
  Circular $e^+e^-$ Collider}},}\ }\href@noop {} {\  (\bibinfo {year}
  {2017})},\ \Eprint {http://arxiv.org/abs/1701.08614} {arXiv:1701.08614
  [hep-ph]} \BibitemShut {NoStop}%
\bibitem [{\citenamefont {Araki}\ \emph {et~al.}(2017)\citenamefont {Araki},
  \citenamefont {Hoshino}, \citenamefont {Ota}, \citenamefont {Sato},\ and\
  \citenamefont {Shimomura}}]{Araki:2017wyg}%
  \BibitemOpen
  \bibfield  {author} {\bibinfo {author} {\bibfnamefont {Takeshi}\ \bibnamefont
  {Araki}}, \bibinfo {author} {\bibfnamefont {Shihori}\ \bibnamefont
  {Hoshino}}, \bibinfo {author} {\bibfnamefont {Toshihiko}\ \bibnamefont
  {Ota}}, \bibinfo {author} {\bibfnamefont {Joe}\ \bibnamefont {Sato}}, \ and\
  \bibinfo {author} {\bibfnamefont {Takashi}\ \bibnamefont {Shimomura}},\
  }\bibfield  {title} {\enquote {\bibinfo {title} {{Cornering the
  $L_{\mu}-L_{\tau}$ gauge boson at Belle-II}},}\ }\href@noop {} {\  (\bibinfo
  {year} {2017})},\ \Eprint {http://arxiv.org/abs/1702.01497} {arXiv:1702.01497
  [hep-ph]} \BibitemShut {NoStop}%
\bibitem [{\citenamefont {Biswas}\ \emph {et~al.}(2017)\citenamefont {Biswas},
  \citenamefont {Gabrielli}, \citenamefont {Heikinheimo},\ and\ \citenamefont
  {Mele}}]{Biswas:2017lyg}%
  \BibitemOpen
  \bibfield  {author} {\bibinfo {author} {\bibfnamefont {Sanjoy}\ \bibnamefont
  {Biswas}}, \bibinfo {author} {\bibfnamefont {Emidio}\ \bibnamefont
  {Gabrielli}}, \bibinfo {author} {\bibfnamefont {Matti}\ \bibnamefont
  {Heikinheimo}}, \ and\ \bibinfo {author} {\bibfnamefont {Barbara}\
  \bibnamefont {Mele}},\ }\bibfield  {title} {\enquote {\bibinfo {title}
  {{Dark-photon searches via $ZH$ production at $e^+e^-$ colliders}},}\
  }\href@noop {} {\  (\bibinfo {year} {2017})},\ \Eprint
  {http://arxiv.org/abs/1703.00402} {arXiv:1703.00402 [hep-ph]} \BibitemShut
  {NoStop}%
\bibitem [{\citenamefont {Acciarri}\ \emph {et~al.}(1996)\citenamefont
  {Acciarri} \emph {et~al.}}]{Acciarri:1996um}%
  \BibitemOpen
  \bibfield  {author} {\bibinfo {author} {\bibfnamefont {M.}~\bibnamefont
  {Acciarri}} \emph {et~al.} (\bibinfo {collaboration} {L3}),\ }\bibfield
  {title} {\enquote {\bibinfo {title} {{Search for neutral Higgs boson
  production through the process e+ e- $\to$ Z* H0}},}\ }\href {\doibase
  10.1016/0370-2693(96)00987-2} {\bibfield  {journal} {\bibinfo  {journal}
  {Phys. Lett.}\ }\textbf {\bibinfo {volume} {B385}},\ \bibinfo {pages}
  {454--470} (\bibinfo {year} {1996})}\BibitemShut {NoStop}%
\bibitem [{\citenamefont {Acciarri}\ \emph {et~al.}(1998)\citenamefont
  {Acciarri} \emph {et~al.}}]{Acciarri:1997tr}%
  \BibitemOpen
  \bibfield  {author} {\bibinfo {author} {\bibfnamefont {M.}~\bibnamefont
  {Acciarri}} \emph {et~al.} (\bibinfo {collaboration} {L3}),\ }\bibfield
  {title} {\enquote {\bibinfo {title} {{Missing mass spectra in hadronic events
  from e+ e- collisions at s**(1/2) = 161-GeV - 172-GeV and limits on invisible
  Higgs decay}},}\ }\href {\doibase 10.1016/S0370-2693(97)01394-4} {\bibfield
  {journal} {\bibinfo  {journal} {Phys. Lett.}\ }\textbf {\bibinfo {volume}
  {B418}},\ \bibinfo {pages} {389--398} (\bibinfo {year} {1998})}\BibitemShut
  {NoStop}%
\bibitem [{\citenamefont {Abreu}\ \emph {et~al.}(1999)\citenamefont {Abreu}
  \emph {et~al.}}]{Abreu:1999vu}%
  \BibitemOpen
  \bibfield  {author} {\bibinfo {author} {\bibfnamefont {P.}~\bibnamefont
  {Abreu}} \emph {et~al.} (\bibinfo {collaboration} {DELPHI}),\ }\bibfield
  {title} {\enquote {\bibinfo {title} {{A Search for invisible Higgs bosons
  produced in e+ e- interactions at LEP-2 energies}},}\ }\href {\doibase
  10.1016/S0370-2693(99)00597-3} {\bibfield  {journal} {\bibinfo  {journal}
  {Phys. Lett.}\ }\textbf {\bibinfo {volume} {B459}},\ \bibinfo {pages}
  {367--381} (\bibinfo {year} {1999})}\BibitemShut {NoStop}%
\bibitem [{Sea(2001)}]{Searches:2001ab}%
  \BibitemOpen
  \bibfield  {title} {\enquote {\bibinfo {title} {{Searches for invisible Higgs
  bosons: Preliminary combined results using LEP data collected at energies up
  to 209-GeV}},}\ }in\ \href
  {https://inspirehep.net/record/559835/files/arXiv:hep-ex_0107032.pdf} {\emph
  {\bibinfo {booktitle} {{Lepton and photon interactions at high energies.
  Proceedings, 20th International Symposium, LP 2001, Rome, Italy, July 23-28,
  2001}}}}\ (\bibinfo {year} {2001})\ \Eprint
  {http://arxiv.org/abs/hep-ex/0107032} {arXiv:hep-ex/0107032 [hep-ex]}
  \BibitemShut {NoStop}%
\bibitem [{\citenamefont {Schael}\ \emph {et~al.}(2006)\citenamefont {Schael}
  \emph {et~al.}}]{Schael:2006cr}%
  \BibitemOpen
  \bibfield  {author} {\bibinfo {author} {\bibfnamefont {S.}~\bibnamefont
  {Schael}} \emph {et~al.} (\bibinfo {collaboration} {DELPHI, OPAL, ALEPH, LEP
  Working Group for Higgs Boson Searches, L3}),\ }\bibfield  {title} {\enquote
  {\bibinfo {title} {{Search for neutral MSSM Higgs bosons at LEP}},}\ }\href
  {\doibase 10.1140/epjc/s2006-02569-7} {\bibfield  {journal} {\bibinfo
  {journal} {Eur. Phys. J.}\ }\textbf {\bibinfo {volume} {C47}},\ \bibinfo
  {pages} {547--587} (\bibinfo {year} {2006})},\ \Eprint
  {http://arxiv.org/abs/hep-ex/0602042} {arXiv:hep-ex/0602042 [hep-ex]}
  \BibitemShut {NoStop}%
\bibitem [{\citenamefont {Flacke}\ \emph {et~al.}(2016)\citenamefont {Flacke},
  \citenamefont {Frugiuele}, \citenamefont {Fuchs}, \citenamefont {Gupta},\
  and\ \citenamefont {Perez}}]{Flacke:2016szy}%
  \BibitemOpen
  \bibfield  {author} {\bibinfo {author} {\bibfnamefont {Thomas}\ \bibnamefont
  {Flacke}}, \bibinfo {author} {\bibfnamefont {Claudia}\ \bibnamefont
  {Frugiuele}}, \bibinfo {author} {\bibfnamefont {Elina}\ \bibnamefont
  {Fuchs}}, \bibinfo {author} {\bibfnamefont {Rick~S.}\ \bibnamefont {Gupta}},
  \ and\ \bibinfo {author} {\bibfnamefont {Gilad}\ \bibnamefont {Perez}},\
  }\bibfield  {title} {\enquote {\bibinfo {title} {{Phenomenology of
  relaxion-Higgs mixing}},}\ }\href@noop {} {\  (\bibinfo {year} {2016})},\
  \Eprint {http://arxiv.org/abs/1610.02025} {arXiv:1610.02025 [hep-ph]}
  \BibitemShut {NoStop}%
\bibitem [{\citenamefont {Jaegle}(2015)}]{TheBelle:2015mwa}%
  \BibitemOpen
  \bibfield  {author} {\bibinfo {author} {\bibfnamefont {I.}~\bibnamefont
  {Jaegle}} (\bibinfo {collaboration} {Belle}),\ }\bibfield  {title} {\enquote
  {\bibinfo {title} {{Search for the dark photon and the dark Higgs boson at
  Belle}},}\ }\href {\doibase 10.1103/PhysRevLett.114.211801} {\bibfield
  {journal} {\bibinfo  {journal} {Phys. Rev. Lett.}\ }\textbf {\bibinfo
  {volume} {114}},\ \bibinfo {pages} {211801} (\bibinfo {year} {2015})},\
  \Eprint {http://arxiv.org/abs/1502.00084} {arXiv:1502.00084 [hep-ex]}
  \BibitemShut {NoStop}%
\bibitem [{\citenamefont {Biswas}\ \emph {et~al.}(2015)\citenamefont {Biswas},
  \citenamefont {Gabrielli}, \citenamefont {Heikinheimo},\ and\ \citenamefont
  {Mele}}]{Biswas:2015sha}%
  \BibitemOpen
  \bibfield  {author} {\bibinfo {author} {\bibfnamefont {Sanjoy}\ \bibnamefont
  {Biswas}}, \bibinfo {author} {\bibfnamefont {Emidio}\ \bibnamefont
  {Gabrielli}}, \bibinfo {author} {\bibfnamefont {Matti}\ \bibnamefont
  {Heikinheimo}}, \ and\ \bibinfo {author} {\bibfnamefont {Barbara}\
  \bibnamefont {Mele}},\ }\bibfield  {title} {\enquote {\bibinfo {title}
  {{Higgs-boson production in association with a dark photon in e+e-
  collisions}},}\ }\href {\doibase 10.1007/JHEP06(2015)102} {\bibfield
  {journal} {\bibinfo  {journal} {JHEP}\ }\textbf {\bibinfo {volume} {06}},\
  \bibinfo {pages} {102} (\bibinfo {year} {2015})},\ \Eprint
  {http://arxiv.org/abs/1503.05836} {arXiv:1503.05836 [hep-ph]} \BibitemShut
  {NoStop}%
\bibitem [{\citenamefont {Angelescu}\ \emph {et~al.}(2017)\citenamefont
  {Angelescu}, \citenamefont {Moreau},\ and\ \citenamefont
  {Richard}}]{Angelescu:2017jyj}%
  \BibitemOpen
  \bibfield  {author} {\bibinfo {author} {\bibfnamefont {Andrei}\ \bibnamefont
  {Angelescu}}, \bibinfo {author} {\bibfnamefont {Grégory}\ \bibnamefont
  {Moreau}}, \ and\ \bibinfo {author} {\bibfnamefont {François}\ \bibnamefont
  {Richard}},\ }\bibfield  {title} {\enquote {\bibinfo {title} {{Scalar
  Production in Association with a Z Boson at LHC and ILC: the Mixed
  Higgs-Radion Case of Warped Models}},}\ }\href@noop {} {\  (\bibinfo {year}
  {2017})},\ \Eprint {http://arxiv.org/abs/1702.03984} {arXiv:1702.03984
  [hep-ph]} \BibitemShut {NoStop}%
\bibitem [{\citenamefont {Dreiner}\ \emph {et~al.}(2013)\citenamefont
  {Dreiner}, \citenamefont {Huck}, \citenamefont {Krämer}, \citenamefont
  {Schmeier},\ and\ \citenamefont {Tattersall}}]{Dreiner:2012xm}%
  \BibitemOpen
  \bibfield  {author} {\bibinfo {author} {\bibfnamefont {Herbert}\ \bibnamefont
  {Dreiner}}, \bibinfo {author} {\bibfnamefont {Moritz}\ \bibnamefont {Huck}},
  \bibinfo {author} {\bibfnamefont {Michael}\ \bibnamefont {Krämer}}, \bibinfo
  {author} {\bibfnamefont {Daniel}\ \bibnamefont {Schmeier}}, \ and\ \bibinfo
  {author} {\bibfnamefont {Jamie}\ \bibnamefont {Tattersall}},\ }\bibfield
  {title} {\enquote {\bibinfo {title} {{Illuminating Dark Matter at the
  ILC}},}\ }\href {\doibase 10.1103/PhysRevD.87.075015} {\bibfield  {journal}
  {\bibinfo  {journal} {Phys. Rev.}\ }\textbf {\bibinfo {volume} {D87}},\
  \bibinfo {pages} {075015} (\bibinfo {year} {2013})},\ \Eprint
  {http://arxiv.org/abs/1211.2254} {arXiv:1211.2254 [hep-ph]} \BibitemShut
  {NoStop}%
\bibitem [{\citenamefont {Alexander}\ \emph {et~al.}(2016)\citenamefont
  {Alexander} \emph {et~al.}}]{Alexander:2016aln}%
  \BibitemOpen
  \bibfield  {author} {\bibinfo {author} {\bibfnamefont {Jim}\ \bibnamefont
  {Alexander}} \emph {et~al.},\ }\bibfield  {title} {\enquote {\bibinfo {title}
  {{Dark Sectors 2016 Workshop: Community Report}},}\ \ }(\bibinfo {year}
  {2016})\ \Eprint {http://arxiv.org/abs/1608.08632} {arXiv:1608.08632
  [hep-ph]} \BibitemShut {NoStop}%
\bibitem [{\citenamefont {Lees}\ \emph {et~al.}(2017)\citenamefont {Lees} \emph
  {et~al.}}]{Lees:2017lec}%
  \BibitemOpen
  \bibfield  {author} {\bibinfo {author} {\bibfnamefont {J.~P.}\ \bibnamefont
  {Lees}} \emph {et~al.} (\bibinfo {collaboration} {BaBar}),\ }\bibfield
  {title} {\enquote {\bibinfo {title} {{Search for invisible decays of a dark
  photon produced in e+e- collisions at BaBar}},}\ }\href@noop {} {\  (\bibinfo
  {year} {2017})},\ \Eprint {http://arxiv.org/abs/1702.03327} {arXiv:1702.03327
  [hep-ex]} \BibitemShut {NoStop}%
\bibitem [{\citenamefont {Agnese}\ \emph {et~al.}(2016)\citenamefont {Agnese}
  \emph {et~al.}}]{Agnese:2015nto}%
  \BibitemOpen
  \bibfield  {author} {\bibinfo {author} {\bibfnamefont {R.}~\bibnamefont
  {Agnese}} \emph {et~al.} (\bibinfo {collaboration} {SuperCDMS}),\ }\bibfield
  {title} {\enquote {\bibinfo {title} {{New Results from the Search for
  Low-Mass Weakly Interacting Massive Particles with the CDMS Low Ionization
  Threshold Experiment}},}\ }\href {\doibase 10.1103/PhysRevLett.116.071301}
  {\bibfield  {journal} {\bibinfo  {journal} {Phys. Rev. Lett.}\ }\textbf
  {\bibinfo {volume} {116}},\ \bibinfo {pages} {071301} (\bibinfo {year}
  {2016})},\ \Eprint {http://arxiv.org/abs/1509.02448} {arXiv:1509.02448
  [astro-ph.CO]} \BibitemShut {NoStop}%
\bibitem [{\citenamefont {Angloher}\ \emph {et~al.}(2016)\citenamefont
  {Angloher} \emph {et~al.}}]{Angloher:2015ewa}%
  \BibitemOpen
  \bibfield  {author} {\bibinfo {author} {\bibfnamefont {G.}~\bibnamefont
  {Angloher}} \emph {et~al.} (\bibinfo {collaboration} {CRESST}),\ }\bibfield
  {title} {\enquote {\bibinfo {title} {{Results on light dark matter particles
  with a low-threshold CRESST-II detector}},}\ }\href {\doibase
  10.1140/epjc/s10052-016-3877-3} {\bibfield  {journal} {\bibinfo  {journal}
  {Eur. Phys. J.}\ }\textbf {\bibinfo {volume} {C76}},\ \bibinfo {pages} {25}
  (\bibinfo {year} {2016})},\ \Eprint {http://arxiv.org/abs/1509.01515}
  {arXiv:1509.01515 [astro-ph.CO]} \BibitemShut {NoStop}%
\bibitem [{\citenamefont {Tan}\ \emph {et~al.}(2016)\citenamefont {Tan} \emph
  {et~al.}}]{Tan:2016zwf}%
  \BibitemOpen
  \bibfield  {author} {\bibinfo {author} {\bibfnamefont {Andi}\ \bibnamefont
  {Tan}} \emph {et~al.} (\bibinfo {collaboration} {PandaX-II}),\ }\bibfield
  {title} {\enquote {\bibinfo {title} {{Dark Matter Results from First 98.7
  Days of Data from the PandaX-II Experiment}},}\ }\href {\doibase
  10.1103/PhysRevLett.117.121303} {\bibfield  {journal} {\bibinfo  {journal}
  {Phys. Rev. Lett.}\ }\textbf {\bibinfo {volume} {117}},\ \bibinfo {pages}
  {121303} (\bibinfo {year} {2016})},\ \Eprint
  {http://arxiv.org/abs/1607.07400} {arXiv:1607.07400 [hep-ex]} \BibitemShut
  {NoStop}%
\bibitem [{\citenamefont {Akerib}\ \emph {et~al.}(2017)\citenamefont {Akerib}
  \emph {et~al.}}]{Akerib:2016vxi}%
  \BibitemOpen
  \bibfield  {author} {\bibinfo {author} {\bibfnamefont {D.~S.}\ \bibnamefont
  {Akerib}} \emph {et~al.} (\bibinfo {collaboration} {LUX}),\ }\bibfield
  {title} {\enquote {\bibinfo {title} {{Results from a search for dark matter
  in the complete LUX exposure}},}\ }\href {\doibase
  10.1103/PhysRevLett.118.021303} {\bibfield  {journal} {\bibinfo  {journal}
  {Phys. Rev. Lett.}\ }\textbf {\bibinfo {volume} {118}},\ \bibinfo {pages}
  {021303} (\bibinfo {year} {2017})},\ \Eprint
  {http://arxiv.org/abs/1608.07648} {arXiv:1608.07648 [astro-ph.CO]}
  \BibitemShut {NoStop}%
\bibitem [{\citenamefont {Ade}\ \emph {et~al.}(2015)\citenamefont {Ade} \emph
  {et~al.}}]{Ade:2015xua}%
  \BibitemOpen
  \bibfield  {author} {\bibinfo {author} {\bibfnamefont {P.~A.~R.}\
  \bibnamefont {Ade}} \emph {et~al.} (\bibinfo {collaboration} {Planck}),\
  }\bibfield  {title} {\enquote {\bibinfo {title} {{Planck 2015 results. XIII.
  Cosmological parameters}},}\ }\href@noop {} {\  (\bibinfo {year} {2015})},\
  \Eprint {http://arxiv.org/abs/1502.01589} {arXiv:1502.01589 [astro-ph.CO]}
  \BibitemShut {NoStop}%
\bibitem [{\citenamefont {Ackermann}\ \emph {et~al.}(2015)\citenamefont
  {Ackermann} \emph {et~al.}}]{Ackermann:2015zua}%
  \BibitemOpen
  \bibfield  {author} {\bibinfo {author} {\bibfnamefont {M.}~\bibnamefont
  {Ackermann}} \emph {et~al.} (\bibinfo {collaboration} {Fermi-LAT}),\
  }\bibfield  {title} {\enquote {\bibinfo {title} {{Searching for Dark Matter
  Annihilation from MilkyÂ Way Dwarf Spheroidal Galaxies with Six Years of
  Fermi Large Area Telescope Data}},}\ }\href {\doibase
  10.1103/PhysRevLett.115.231301} {\bibfield  {journal} {\bibinfo  {journal}
  {Phys. Rev. Lett.}\ }\textbf {\bibinfo {volume} {115}},\ \bibinfo {pages}
  {231301} (\bibinfo {year} {2015})},\ \Eprint
  {http://arxiv.org/abs/1503.02641} {arXiv:1503.02641 [astro-ph.HE]}
  \BibitemShut {NoStop}%
\bibitem [{\citenamefont {Ahnen}\ \emph {et~al.}(2016)\citenamefont {Ahnen}
  \emph {et~al.}}]{Ahnen:2016qkx}%
  \BibitemOpen
  \bibfield  {author} {\bibinfo {author} {\bibfnamefont {M.~L.}\ \bibnamefont
  {Ahnen}} \emph {et~al.} (\bibinfo {collaboration} {Fermi-LAT, MAGIC}),\
  }\bibfield  {title} {\enquote {\bibinfo {title} {{Limits to dark matter
  annihilation cross-section from a combined analysis of MAGIC and Fermi-LAT
  observations of dwarf satellite galaxies}},}\ }\href {\doibase
  10.1088/1475-7516/2016/02/039} {\bibfield  {journal} {\bibinfo  {journal}
  {JCAP}\ }\textbf {\bibinfo {volume} {1602}},\ \bibinfo {pages} {039}
  (\bibinfo {year} {2016})},\ \Eprint {http://arxiv.org/abs/1601.06590}
  {arXiv:1601.06590 [astro-ph.HE]} \BibitemShut {NoStop}%
\bibitem [{\citenamefont {Massari}\ \emph {et~al.}(2015)\citenamefont
  {Massari}, \citenamefont {Izaguirre}, \citenamefont {Essig}, \citenamefont
  {Albert}, \citenamefont {Bloom},\ and\ \citenamefont
  {Gomez-Vargas}}]{Massari:2015xea}%
  \BibitemOpen
  \bibfield  {author} {\bibinfo {author} {\bibfnamefont {Andrea}\ \bibnamefont
  {Massari}}, \bibinfo {author} {\bibfnamefont {Eder}\ \bibnamefont
  {Izaguirre}}, \bibinfo {author} {\bibfnamefont {Rouven}\ \bibnamefont
  {Essig}}, \bibinfo {author} {\bibfnamefont {Andrea}\ \bibnamefont {Albert}},
  \bibinfo {author} {\bibfnamefont {Elliott}\ \bibnamefont {Bloom}}, \ and\
  \bibinfo {author} {\bibfnamefont {German~Arturo}\ \bibnamefont
  {Gomez-Vargas}},\ }\bibfield  {title} {\enquote {\bibinfo {title} {{Strong
  Optimized Conservative $Fermi$-LAT Constraints on Dark Matter Models from the
  Inclusive Photon Spectrum}},}\ }\href {\doibase 10.1103/PhysRevD.91.083539}
  {\bibfield  {journal} {\bibinfo  {journal} {Phys. Rev.}\ }\textbf {\bibinfo
  {volume} {D91}},\ \bibinfo {pages} {083539} (\bibinfo {year} {2015})},\
  \Eprint {http://arxiv.org/abs/1503.07169} {arXiv:1503.07169 [hep-ph]}
  \BibitemShut {NoStop}%
\bibitem [{\citenamefont {Aguilar}\ \emph {et~al.}(2014)\citenamefont {Aguilar}
  \emph {et~al.}}]{Aguilar:2014mma}%
  \BibitemOpen
  \bibfield  {author} {\bibinfo {author} {\bibfnamefont {M.}~\bibnamefont
  {Aguilar}} \emph {et~al.} (\bibinfo {collaboration} {AMS}),\ }\bibfield
  {title} {\enquote {\bibinfo {title} {{Electron and Positron Fluxes in Primary
  Cosmic Rays Measured with the Alpha Magnetic Spectrometer on the
  International Space Station}},}\ }\href {\doibase
  10.1103/PhysRevLett.113.121102} {\bibfield  {journal} {\bibinfo  {journal}
  {Phys. Rev. Lett.}\ }\textbf {\bibinfo {volume} {113}},\ \bibinfo {pages}
  {121102} (\bibinfo {year} {2014})}\BibitemShut {NoStop}%
\bibitem [{\citenamefont {Aaboud}\ \emph
  {et~al.}(2016{\natexlab{a}})\citenamefont {Aaboud} \emph
  {et~al.}}]{Aaboud:2016tnv}%
  \BibitemOpen
  \bibfield  {author} {\bibinfo {author} {\bibfnamefont {Morad}\ \bibnamefont
  {Aaboud}} \emph {et~al.} (\bibinfo {collaboration} {ATLAS}),\ }\bibfield
  {title} {\enquote {\bibinfo {title} {{Search for new phenomena in final
  states with an energetic jet and large missing transverse momentum in $pp$
  collisions at $\sqrt{s}=13$~TeV using the ATLAS detector}},}\ }\href
  {\doibase 10.1103/PhysRevD.94.032005} {\bibfield  {journal} {\bibinfo
  {journal} {Phys. Rev.}\ }\textbf {\bibinfo {volume} {D94}},\ \bibinfo {pages}
  {032005} (\bibinfo {year} {2016}{\natexlab{a}})},\ \Eprint
  {http://arxiv.org/abs/1604.07773} {arXiv:1604.07773 [hep-ex]} \BibitemShut
  {NoStop}%
\bibitem [{\citenamefont {Sirunyan}\ \emph {et~al.}(2017)\citenamefont
  {Sirunyan} \emph {et~al.}}]{Sirunyan:2017hci}%
  \BibitemOpen
  \bibfield  {author} {\bibinfo {author} {\bibfnamefont {Albert~M}\
  \bibnamefont {Sirunyan}} \emph {et~al.} (\bibinfo {collaboration} {CMS}),\
  }\bibfield  {title} {\enquote {\bibinfo {title} {{Search for dark matter
  produced with an energetic jet or a hadronically decaying W or Z boson at
  $\sqrt{s}$ = 13 TeV}},}\ }\href@noop {} {\  (\bibinfo {year} {2017})},\
  \Eprint {http://arxiv.org/abs/1703.01651} {arXiv:1703.01651 [hep-ex]}
  \BibitemShut {NoStop}%
\bibitem [{\citenamefont {Liu}\ \emph {et~al.}(2015)\citenamefont {Liu},
  \citenamefont {Weiner},\ and\ \citenamefont {Xue}}]{Liu:2014cma}%
  \BibitemOpen
  \bibfield  {author} {\bibinfo {author} {\bibfnamefont {Jia}\ \bibnamefont
  {Liu}}, \bibinfo {author} {\bibfnamefont {Neal}\ \bibnamefont {Weiner}}, \
  and\ \bibinfo {author} {\bibfnamefont {Wei}\ \bibnamefont {Xue}},\ }\bibfield
   {title} {\enquote {\bibinfo {title} {{Signals of a Light Dark Force in the
  Galactic Center}},}\ }\href {\doibase 10.1007/JHEP08(2015)050} {\bibfield
  {journal} {\bibinfo  {journal} {JHEP}\ }\textbf {\bibinfo {volume} {08}},\
  \bibinfo {pages} {050} (\bibinfo {year} {2015})},\ \Eprint
  {http://arxiv.org/abs/1412.1485} {arXiv:1412.1485 [hep-ph]} \BibitemShut
  {NoStop}%
\bibitem [{\citenamefont {Griest}\ and\ \citenamefont
  {Seckel}(1991)}]{Griest:1990kh}%
  \BibitemOpen
  \bibfield  {author} {\bibinfo {author} {\bibfnamefont {Kim}\ \bibnamefont
  {Griest}}\ and\ \bibinfo {author} {\bibfnamefont {David}\ \bibnamefont
  {Seckel}},\ }\bibfield  {title} {\enquote {\bibinfo {title} {{Three
  exceptions in the calculation of relic abundances}},}\ }\href {\doibase
  10.1103/PhysRevD.43.3191} {\bibfield  {journal} {\bibinfo  {journal} {Phys.
  Rev.}\ }\textbf {\bibinfo {volume} {D43}},\ \bibinfo {pages} {3191--3203}
  (\bibinfo {year} {1991})}\BibitemShut {NoStop}%
\bibitem [{\citenamefont {Baker}\ \emph {et~al.}(2015)\citenamefont {Baker}
  \emph {et~al.}}]{Baker:2015qna}%
  \BibitemOpen
  \bibfield  {author} {\bibinfo {author} {\bibfnamefont {Michael~J.}\
  \bibnamefont {Baker}} \emph {et~al.},\ }\bibfield  {title} {\enquote
  {\bibinfo {title} {{The Coannihilation Codex}},}\ }\href {\doibase
  10.1007/JHEP12(2015)120} {\bibfield  {journal} {\bibinfo  {journal} {JHEP}\
  }\textbf {\bibinfo {volume} {12}},\ \bibinfo {pages} {120} (\bibinfo {year}
  {2015})},\ \Eprint {http://arxiv.org/abs/1510.03434} {arXiv:1510.03434
  [hep-ph]} \BibitemShut {NoStop}%
\bibitem [{\citenamefont {Buschmann}\ \emph {et~al.}(2016)\citenamefont
  {Buschmann}, \citenamefont {El~Hedri}, \citenamefont {Kaminska},
  \citenamefont {Liu}, \citenamefont {de~Vries}, \citenamefont {Wang},
  \citenamefont {Yu},\ and\ \citenamefont {Zurita}}]{Buschmann:2016hkc}%
  \BibitemOpen
  \bibfield  {author} {\bibinfo {author} {\bibfnamefont {Malte}\ \bibnamefont
  {Buschmann}}, \bibinfo {author} {\bibfnamefont {Sonia}\ \bibnamefont
  {El~Hedri}}, \bibinfo {author} {\bibfnamefont {Anna}\ \bibnamefont
  {Kaminska}}, \bibinfo {author} {\bibfnamefont {Jia}\ \bibnamefont {Liu}},
  \bibinfo {author} {\bibfnamefont {Maikel}\ \bibnamefont {de~Vries}}, \bibinfo
  {author} {\bibfnamefont {Xiao-Ping}\ \bibnamefont {Wang}}, \bibinfo {author}
  {\bibfnamefont {Felix}\ \bibnamefont {Yu}}, \ and\ \bibinfo {author}
  {\bibfnamefont {Jose}\ \bibnamefont {Zurita}},\ }\bibfield  {title} {\enquote
  {\bibinfo {title} {{Hunting for dark matter coannihilation by mixing dijet
  resonances and missing transverse energy}},}\ }\href {\doibase
  10.1007/JHEP09(2016)033} {\bibfield  {journal} {\bibinfo  {journal} {JHEP}\
  }\textbf {\bibinfo {volume} {09}},\ \bibinfo {pages} {033} (\bibinfo {year}
  {2016})},\ \Eprint {http://arxiv.org/abs/1605.08056} {arXiv:1605.08056
  [hep-ph]} \BibitemShut {NoStop}%
\bibitem [{\citenamefont {Alves}\ \emph {et~al.}(2016)\citenamefont {Alves},
  \citenamefont {Arcadi}, \citenamefont {Mambrini}, \citenamefont {Profumo},\
  and\ \citenamefont {Queiroz}}]{Alves:2016cqf}%
  \BibitemOpen
  \bibfield  {author} {\bibinfo {author} {\bibfnamefont {Alexandre}\
  \bibnamefont {Alves}}, \bibinfo {author} {\bibfnamefont {Giorgio}\
  \bibnamefont {Arcadi}}, \bibinfo {author} {\bibfnamefont {Yann}\ \bibnamefont
  {Mambrini}}, \bibinfo {author} {\bibfnamefont {Stefano}\ \bibnamefont
  {Profumo}}, \ and\ \bibinfo {author} {\bibfnamefont {Farinaldo~S.}\
  \bibnamefont {Queiroz}},\ }\bibfield  {title} {\enquote {\bibinfo {title}
  {{Augury of Darkness: The Low-Mass Dark Z' Portal}},}\ }\href@noop {} {\
  (\bibinfo {year} {2016})},\ \Eprint {http://arxiv.org/abs/1612.07282}
  {arXiv:1612.07282 [hep-ph]} \BibitemShut {NoStop}%
\bibitem [{\citenamefont {Essig}\ \emph {et~al.}(2013)\citenamefont {Essig}
  \emph {et~al.}}]{Essig:2013lka}%
  \BibitemOpen
  \bibfield  {author} {\bibinfo {author} {\bibfnamefont {Rouven}\ \bibnamefont
  {Essig}} \emph {et~al.},\ }\bibfield  {title} {\enquote {\bibinfo {title}
  {{Working Group Report: New Light Weakly Coupled Particles}},}\ }in\ \href
  {http://inspirehep.net/record/1263039/files/arXiv:1311.0029.pdf} {\emph
  {\bibinfo {booktitle} {{Proceedings, 2013 Community Summer Study on the
  Future of U.S. Particle Physics: Snowmass on the Mississippi (CSS2013):
  Minneapolis, MN, USA, July 29-August 6, 2013}}}}\ (\bibinfo {year} {2013})\
  \Eprint {http://arxiv.org/abs/1311.0029} {arXiv:1311.0029 [hep-ph]}
  \BibitemShut {NoStop}%
\bibitem [{\citenamefont {Ruegg}\ and\ \citenamefont
  {Ruiz-Altaba}(2004)}]{Ruegg:2003ps}%
  \BibitemOpen
  \bibfield  {author} {\bibinfo {author} {\bibfnamefont {Henri}\ \bibnamefont
  {Ruegg}}\ and\ \bibinfo {author} {\bibfnamefont {Marti}\ \bibnamefont
  {Ruiz-Altaba}},\ }\bibfield  {title} {\enquote {\bibinfo {title} {{The
  Stueckelberg field}},}\ }\href {\doibase 10.1142/S0217751X04019755}
  {\bibfield  {journal} {\bibinfo  {journal} {Int. J. Mod. Phys.}\ }\textbf
  {\bibinfo {volume} {A19}},\ \bibinfo {pages} {3265--3348} (\bibinfo {year}
  {2004})},\ \Eprint {http://arxiv.org/abs/hep-th/0304245}
  {arXiv:hep-th/0304245 [hep-th]} \BibitemShut {NoStop}%
\bibitem [{\citenamefont {Kors}\ and\ \citenamefont
  {Nath}(2005)}]{Kors:2005uz}%
  \BibitemOpen
  \bibfield  {author} {\bibinfo {author} {\bibfnamefont {Boris}\ \bibnamefont
  {Kors}}\ and\ \bibinfo {author} {\bibfnamefont {Pran}\ \bibnamefont {Nath}},\
  }\bibfield  {title} {\enquote {\bibinfo {title} {{Aspects of the Stueckelberg
  extension}},}\ }\href {\doibase 10.1088/1126-6708/2005/07/069} {\bibfield
  {journal} {\bibinfo  {journal} {JHEP}\ }\textbf {\bibinfo {volume} {07}},\
  \bibinfo {pages} {069} (\bibinfo {year} {2005})},\ \Eprint
  {http://arxiv.org/abs/hep-ph/0503208} {arXiv:hep-ph/0503208 [hep-ph]}
  \BibitemShut {NoStop}%
\bibitem [{\citenamefont {Kumar}\ \emph {et~al.}(2012)\citenamefont {Kumar},
  \citenamefont {Vega-Morales},\ and\ \citenamefont {Yu}}]{Kumar:2012ww}%
  \BibitemOpen
  \bibfield  {author} {\bibinfo {author} {\bibfnamefont {Kunal}\ \bibnamefont
  {Kumar}}, \bibinfo {author} {\bibfnamefont {Roberto}\ \bibnamefont
  {Vega-Morales}}, \ and\ \bibinfo {author} {\bibfnamefont {Felix}\
  \bibnamefont {Yu}},\ }\bibfield  {title} {\enquote {\bibinfo {title}
  {{Effects from New Colored States and the Higgs Portal on Gluon Fusion and
  Higgs Decays}},}\ }\href {\doibase 10.1103/PhysRevD.87.119903,
  10.1103/PhysRevD.86.113002} {\bibfield  {journal} {\bibinfo  {journal} {Phys.
  Rev.}\ }\textbf {\bibinfo {volume} {D86}},\ \bibinfo {pages} {113002}
  (\bibinfo {year} {2012})},\ \bibinfo {note} {[Erratum: Phys.
  Rev.D87,no.11,119903(2013)]},\ \Eprint {http://arxiv.org/abs/1205.4244}
  {arXiv:1205.4244 [hep-ph]} \BibitemShut {NoStop}%
\bibitem [{\citenamefont {Tucker-Smith}\ and\ \citenamefont
  {Weiner}(2001)}]{TuckerSmith:2001hy}%
  \BibitemOpen
  \bibfield  {author} {\bibinfo {author} {\bibfnamefont {David}\ \bibnamefont
  {Tucker-Smith}}\ and\ \bibinfo {author} {\bibfnamefont {Neal}\ \bibnamefont
  {Weiner}},\ }\bibfield  {title} {\enquote {\bibinfo {title} {{Inelastic dark
  matter}},}\ }\href {\doibase 10.1103/PhysRevD.64.043502} {\bibfield
  {journal} {\bibinfo  {journal} {Phys. Rev.}\ }\textbf {\bibinfo {volume}
  {D64}},\ \bibinfo {pages} {043502} (\bibinfo {year} {2001})},\ \Eprint
  {http://arxiv.org/abs/hep-ph/0101138} {arXiv:hep-ph/0101138 [hep-ph]}
  \BibitemShut {NoStop}%
\bibitem [{\citenamefont {Gondolo}\ and\ \citenamefont
  {Gelmini}(1991)}]{Gondolo:1990dk}%
  \BibitemOpen
  \bibfield  {author} {\bibinfo {author} {\bibfnamefont {Paolo}\ \bibnamefont
  {Gondolo}}\ and\ \bibinfo {author} {\bibfnamefont {Graciela}\ \bibnamefont
  {Gelmini}},\ }\bibfield  {title} {\enquote {\bibinfo {title} {{Cosmic
  abundances of stable particles: Improved analysis}},}\ }\href {\doibase
  10.1016/0550-3213(91)90438-4} {\bibfield  {journal} {\bibinfo  {journal}
  {Nucl. Phys.}\ }\textbf {\bibinfo {volume} {B360}},\ \bibinfo {pages}
  {145--179} (\bibinfo {year} {1991})}\BibitemShut {NoStop}%
\bibitem [{\citenamefont {Adams}\ \emph {et~al.}(1998)\citenamefont {Adams},
  \citenamefont {Sarkar},\ and\ \citenamefont {Sciama}}]{Adams:1998nr}%
  \BibitemOpen
  \bibfield  {author} {\bibinfo {author} {\bibfnamefont {Jennifer~A.}\
  \bibnamefont {Adams}}, \bibinfo {author} {\bibfnamefont {Subir}\ \bibnamefont
  {Sarkar}}, \ and\ \bibinfo {author} {\bibfnamefont {D.W.}\ \bibnamefont
  {Sciama}},\ }\bibfield  {title} {\enquote {\bibinfo {title} {{CMB anisotropy
  in the decaying neutrino cosmology}},}\ }\href {\doibase
  10.1046/j.1365-8711.1998.02017.x} {\bibfield  {journal} {\bibinfo  {journal}
  {Mon.Not.Roy.Astron.Soc.}\ }\textbf {\bibinfo {volume} {301}},\ \bibinfo
  {pages} {210--214} (\bibinfo {year} {1998})},\ \Eprint
  {http://arxiv.org/abs/astro-ph/9805108} {arXiv:astro-ph/9805108 [astro-ph]}
  \BibitemShut {NoStop}%
\bibitem [{\citenamefont {Padmanabhan}\ and\ \citenamefont
  {Finkbeiner}(2005)}]{Padmanabhan:2005es}%
  \BibitemOpen
  \bibfield  {author} {\bibinfo {author} {\bibfnamefont {Nikhil}\ \bibnamefont
  {Padmanabhan}}\ and\ \bibinfo {author} {\bibfnamefont {Douglas~P.}\
  \bibnamefont {Finkbeiner}},\ }\bibfield  {title} {\enquote {\bibinfo {title}
  {{Detecting Dark Matter Annihilation with CMB Polarization : Signatures and
  Experimental Prospects}},}\ }\href@noop {} {\bibfield  {journal} {\bibinfo
  {journal} {Phys. Rev.}\ }\textbf {\bibinfo {volume} {D72}},\ \bibinfo {pages}
  {023508} (\bibinfo {year} {2005})},\ \Eprint
  {http://arxiv.org/abs/astro-ph/0503486} {astro-ph/0503486} \BibitemShut
  {NoStop}%
\bibitem [{\citenamefont {Galli}\ \emph {et~al.}(2009)\citenamefont {Galli},
  \citenamefont {Iocco}, \citenamefont {Bertone},\ and\ \citenamefont
  {Melchiorri}}]{Galli:2009zc}%
  \BibitemOpen
  \bibfield  {author} {\bibinfo {author} {\bibfnamefont {Silvia}\ \bibnamefont
  {Galli}}, \bibinfo {author} {\bibfnamefont {Fabio}\ \bibnamefont {Iocco}},
  \bibinfo {author} {\bibfnamefont {Gianfranco}\ \bibnamefont {Bertone}}, \
  and\ \bibinfo {author} {\bibfnamefont {Alessandro}\ \bibnamefont
  {Melchiorri}},\ }\bibfield  {title} {\enquote {\bibinfo {title} {{CMB
  constraints on Dark Matter models with large annihilation cross-section}},}\
  }\href {\doibase 10.1103/PhysRevD.80.023505} {\bibfield  {journal} {\bibinfo
  {journal} {Phys. Rev.}\ }\textbf {\bibinfo {volume} {D80}},\ \bibinfo {pages}
  {023505} (\bibinfo {year} {2009})},\ \Eprint {http://arxiv.org/abs/0905.0003}
  {arXiv:0905.0003 [astro-ph.CO]} \BibitemShut {NoStop}%
\bibitem [{\citenamefont {Slatyer}\ \emph {et~al.}(2009)\citenamefont
  {Slatyer}, \citenamefont {Padmanabhan},\ and\ \citenamefont
  {Finkbeiner}}]{Slatyer:2009yq}%
  \BibitemOpen
  \bibfield  {author} {\bibinfo {author} {\bibfnamefont {Tracy~R.}\
  \bibnamefont {Slatyer}}, \bibinfo {author} {\bibfnamefont {Nikhil}\
  \bibnamefont {Padmanabhan}}, \ and\ \bibinfo {author} {\bibfnamefont
  {Douglas~P.}\ \bibnamefont {Finkbeiner}},\ }\bibfield  {title} {\enquote
  {\bibinfo {title} {{CMB Constraints on WIMP Annihilation: Energy Absorption
  During the Recombination Epoch}},}\ }\href {\doibase
  10.1103/PhysRevD.80.043526} {\bibfield  {journal} {\bibinfo  {journal} {Phys.
  Rev.}\ }\textbf {\bibinfo {volume} {D80}},\ \bibinfo {pages} {043526}
  (\bibinfo {year} {2009})},\ \Eprint {http://arxiv.org/abs/0906.1197}
  {arXiv:0906.1197 [astro-ph.CO]} \BibitemShut {NoStop}%
\bibitem [{\citenamefont {Finkbeiner}\ \emph {et~al.}(2012)\citenamefont
  {Finkbeiner}, \citenamefont {Galli}, \citenamefont {Lin},\ and\ \citenamefont
  {Slatyer}}]{Finkbeiner:2011dx}%
  \BibitemOpen
  \bibfield  {author} {\bibinfo {author} {\bibfnamefont {Douglas~P.}\
  \bibnamefont {Finkbeiner}}, \bibinfo {author} {\bibfnamefont {Silvia}\
  \bibnamefont {Galli}}, \bibinfo {author} {\bibfnamefont {Tongyan}\
  \bibnamefont {Lin}}, \ and\ \bibinfo {author} {\bibfnamefont {Tracy~R.}\
  \bibnamefont {Slatyer}},\ }\bibfield  {title} {\enquote {\bibinfo {title}
  {{Searching for Dark Matter in the CMB: A Compact Parameterization of Energy
  Injection from New Physics}},}\ }\href {\doibase 10.1103/PhysRevD.85.043522}
  {\bibfield  {journal} {\bibinfo  {journal} {Phys. Rev.}\ }\textbf {\bibinfo
  {volume} {D85}},\ \bibinfo {pages} {043522} (\bibinfo {year} {2012})},\
  \Eprint {http://arxiv.org/abs/1109.6322} {arXiv:1109.6322 [astro-ph.CO]}
  \BibitemShut {NoStop}%
\bibitem [{\citenamefont {Madhavacheril}\ \emph {et~al.}(2014)\citenamefont
  {Madhavacheril}, \citenamefont {Sehgal},\ and\ \citenamefont
  {Slatyer}}]{Madhavacheril:2013cna}%
  \BibitemOpen
  \bibfield  {author} {\bibinfo {author} {\bibfnamefont {Mathew~S.}\
  \bibnamefont {Madhavacheril}}, \bibinfo {author} {\bibfnamefont {Neelima}\
  \bibnamefont {Sehgal}}, \ and\ \bibinfo {author} {\bibfnamefont {Tracy~R.}\
  \bibnamefont {Slatyer}},\ }\bibfield  {title} {\enquote {\bibinfo {title}
  {{Current Dark Matter Annihilation Constraints from Cmb and Low-Redshift
  Data}},}\ }\href {\doibase 10.1103/PhysRevD.89.103508} {\bibfield  {journal}
  {\bibinfo  {journal} {Phys. Rev.}\ }\textbf {\bibinfo {volume} {D89}},\
  \bibinfo {pages} {103508} (\bibinfo {year} {2014})},\ \Eprint
  {http://arxiv.org/abs/1310.3815} {arXiv:1310.3815 [astro-ph.CO]} \BibitemShut
  {NoStop}%
\bibitem [{\citenamefont {Navarro}\ \emph {et~al.}(1996)\citenamefont
  {Navarro}, \citenamefont {Frenk},\ and\ \citenamefont
  {White}}]{Navarro:1995iw}%
  \BibitemOpen
  \bibfield  {author} {\bibinfo {author} {\bibfnamefont {Julio~F.}\
  \bibnamefont {Navarro}}, \bibinfo {author} {\bibfnamefont {Carlos~S.}\
  \bibnamefont {Frenk}}, \ and\ \bibinfo {author} {\bibfnamefont {Simon D.~M.}\
  \bibnamefont {White}},\ }\bibfield  {title} {\enquote {\bibinfo {title} {{The
  Structure of cold dark matter halos}},}\ }\href {\doibase 10.1086/177173}
  {\bibfield  {journal} {\bibinfo  {journal} {Astrophys. J.}\ }\textbf
  {\bibinfo {volume} {462}},\ \bibinfo {pages} {563--575} (\bibinfo {year}
  {1996})},\ \Eprint {http://arxiv.org/abs/astro-ph/9508025}
  {arXiv:astro-ph/9508025 [astro-ph]} \BibitemShut {NoStop}%
\bibitem [{\citenamefont {Elor}\ \emph {et~al.}(2016)\citenamefont {Elor},
  \citenamefont {Rodd}, \citenamefont {Slatyer},\ and\ \citenamefont
  {Xue}}]{Elor:2015bho}%
  \BibitemOpen
  \bibfield  {author} {\bibinfo {author} {\bibfnamefont {Gilly}\ \bibnamefont
  {Elor}}, \bibinfo {author} {\bibfnamefont {Nicholas~L.}\ \bibnamefont
  {Rodd}}, \bibinfo {author} {\bibfnamefont {Tracy~R.}\ \bibnamefont
  {Slatyer}}, \ and\ \bibinfo {author} {\bibfnamefont {Wei}\ \bibnamefont
  {Xue}},\ }\bibfield  {title} {\enquote {\bibinfo {title} {{Model-Independent
  Indirect Detection Constraints on Hidden Sector Dark Matter}},}\ }\href
  {\doibase 10.1088/1475-7516/2016/06/024} {\bibfield  {journal} {\bibinfo
  {journal} {JCAP}\ }\textbf {\bibinfo {volume} {1606}},\ \bibinfo {pages}
  {024} (\bibinfo {year} {2016})},\ \Eprint {http://arxiv.org/abs/1511.08787}
  {arXiv:1511.08787 [hep-ph]} \BibitemShut {NoStop}%
\bibitem [{\citenamefont {Jia}(2016)}]{Jia:2016pbe}%
  \BibitemOpen
  \bibfield  {author} {\bibinfo {author} {\bibfnamefont {Lian-Bao}\
  \bibnamefont {Jia}},\ }\bibfield  {title} {\enquote {\bibinfo {title} {{Study
  of WIMP annihilations into a pair of on-shell scalar mediators}},}\ }\href
  {\doibase 10.1103/PhysRevD.94.095028} {\bibfield  {journal} {\bibinfo
  {journal} {Phys. Rev.}\ }\textbf {\bibinfo {volume} {D94}},\ \bibinfo {pages}
  {095028} (\bibinfo {year} {2016})},\ \Eprint
  {http://arxiv.org/abs/1607.00737} {arXiv:1607.00737 [hep-ph]} \BibitemShut
  {NoStop}%
\bibitem [{\citenamefont {Chakrabarty}\ \emph {et~al.}(2015)\citenamefont
  {Chakrabarty}, \citenamefont {Han}, \citenamefont {Liu},\ and\ \citenamefont
  {Mukhopadhyaya}}]{Chakrabarty:2014pja}%
  \BibitemOpen
  \bibfield  {author} {\bibinfo {author} {\bibfnamefont {Nabarun}\ \bibnamefont
  {Chakrabarty}}, \bibinfo {author} {\bibfnamefont {Tao}\ \bibnamefont {Han}},
  \bibinfo {author} {\bibfnamefont {Zhen}\ \bibnamefont {Liu}}, \ and\ \bibinfo
  {author} {\bibfnamefont {Biswarup}\ \bibnamefont {Mukhopadhyaya}},\
  }\bibfield  {title} {\enquote {\bibinfo {title} {{Radiative Return for Heavy
  Higgs Boson at a Muon Collider}},}\ }\href {\doibase
  10.1103/PhysRevD.91.015008} {\bibfield  {journal} {\bibinfo  {journal} {Phys.
  Rev.}\ }\textbf {\bibinfo {volume} {D91}},\ \bibinfo {pages} {015008}
  (\bibinfo {year} {2015})},\ \Eprint {http://arxiv.org/abs/1408.5912}
  {arXiv:1408.5912 [hep-ph]} \BibitemShut {NoStop}%
\bibitem [{\citenamefont {Greco}\ \emph {et~al.}(2016)\citenamefont {Greco},
  \citenamefont {Han},\ and\ \citenamefont {Liu}}]{Greco:2016izi}%
  \BibitemOpen
  \bibfield  {author} {\bibinfo {author} {\bibfnamefont {Mario}\ \bibnamefont
  {Greco}}, \bibinfo {author} {\bibfnamefont {Tao}\ \bibnamefont {Han}}, \ and\
  \bibinfo {author} {\bibfnamefont {Zhen}\ \bibnamefont {Liu}},\ }\bibfield
  {title} {\enquote {\bibinfo {title} {{ISR effects for resonant Higgs
  production at future lepton colliders}},}\ }\href {\doibase
  10.1016/j.physletb.2016.10.078} {\bibfield  {journal} {\bibinfo  {journal}
  {Phys. Lett.}\ }\textbf {\bibinfo {volume} {B763}},\ \bibinfo {pages}
  {409--415} (\bibinfo {year} {2016})},\ \Eprint
  {http://arxiv.org/abs/1607.03210} {arXiv:1607.03210 [hep-ph]} \BibitemShut
  {NoStop}%
\bibitem [{\citenamefont {Alcaraz}\ \emph {et~al.}(2006)\citenamefont {Alcaraz}
  \emph {et~al.}}]{Alcaraz:2006mx}%
  \BibitemOpen
  \bibfield  {author} {\bibinfo {author} {\bibfnamefont {J.}~\bibnamefont
  {Alcaraz}} \emph {et~al.} (\bibinfo {collaboration} {DELPHI, OPAL, ALEPH, LEP
  Electroweak Working Group, L3}),\ }\bibfield  {title} {\enquote {\bibinfo
  {title} {{A Combination of preliminary electroweak measurements and
  constraints on the standard model}},}\ }\href@noop {} {\  (\bibinfo {year}
  {2006})},\ \Eprint {http://arxiv.org/abs/hep-ex/0612034}
  {arXiv:hep-ex/0612034 [hep-ex]} \BibitemShut {NoStop}%
\bibitem [{\citenamefont {Hook}\ \emph {et~al.}(2011)\citenamefont {Hook},
  \citenamefont {Izaguirre},\ and\ \citenamefont {Wacker}}]{Hook:2010tw}%
  \BibitemOpen
  \bibfield  {author} {\bibinfo {author} {\bibfnamefont {Anson}\ \bibnamefont
  {Hook}}, \bibinfo {author} {\bibfnamefont {Eder}\ \bibnamefont {Izaguirre}},
  \ and\ \bibinfo {author} {\bibfnamefont {Jay~G.}\ \bibnamefont {Wacker}},\
  }\bibfield  {title} {\enquote {\bibinfo {title} {{Model Independent Bounds on
  Kinetic Mixing}},}\ }\href {\doibase 10.1155/2011/859762} {\bibfield
  {journal} {\bibinfo  {journal} {Adv. High Energy Phys.}\ }\textbf {\bibinfo
  {volume} {2011}},\ \bibinfo {pages} {859762} (\bibinfo {year} {2011})},\
  \Eprint {http://arxiv.org/abs/1006.0973} {arXiv:1006.0973 [hep-ph]}
  \BibitemShut {NoStop}%
\bibitem [{\citenamefont {Aad}\ \emph {et~al.}(2012)\citenamefont {Aad} \emph
  {et~al.}}]{Aad:2012tfa}%
  \BibitemOpen
  \bibfield  {author} {\bibinfo {author} {\bibfnamefont {Georges}\ \bibnamefont
  {Aad}} \emph {et~al.} (\bibinfo {collaboration} {ATLAS}),\ }\bibfield
  {title} {\enquote {\bibinfo {title} {{Observation of a new particle in the
  search for the Standard Model Higgs boson with the ATLAS detector at the
  LHC}},}\ }\href {\doibase 10.1016/j.physletb.2012.08.020} {\bibfield
  {journal} {\bibinfo  {journal} {Phys. Lett.}\ }\textbf {\bibinfo {volume}
  {B716}},\ \bibinfo {pages} {1--29} (\bibinfo {year} {2012})},\ \Eprint
  {http://arxiv.org/abs/1207.7214} {arXiv:1207.7214 [hep-ex]} \BibitemShut
  {NoStop}%
\bibitem [{\citenamefont {Chatrchyan}\ \emph {et~al.}(2012)\citenamefont
  {Chatrchyan} \emph {et~al.}}]{Chatrchyan:2012xdj}%
  \BibitemOpen
  \bibfield  {author} {\bibinfo {author} {\bibfnamefont {Serguei}\ \bibnamefont
  {Chatrchyan}} \emph {et~al.} (\bibinfo {collaboration} {CMS}),\ }\bibfield
  {title} {\enquote {\bibinfo {title} {{Observation of a new boson at a mass of
  125 GeV with the CMS experiment at the LHC}},}\ }\href {\doibase
  10.1016/j.physletb.2012.08.021} {\bibfield  {journal} {\bibinfo  {journal}
  {Phys. Lett.}\ }\textbf {\bibinfo {volume} {B716}},\ \bibinfo {pages}
  {30--61} (\bibinfo {year} {2012})},\ \Eprint {http://arxiv.org/abs/1207.7235}
  {arXiv:1207.7235 [hep-ex]} \BibitemShut {NoStop}%
\bibitem [{\citenamefont {Aad}\ \emph {et~al.}(2015{\natexlab{a}})\citenamefont
  {Aad} \emph {et~al.}}]{Aad:2015pla}%
  \BibitemOpen
  \bibfield  {author} {\bibinfo {author} {\bibfnamefont {Georges}\ \bibnamefont
  {Aad}} \emph {et~al.} (\bibinfo {collaboration} {ATLAS}),\ }\bibfield
  {title} {\enquote {\bibinfo {title} {{Constraints on new phenomena via Higgs
  boson couplings and invisible decays with the ATLAS detector}},}\ }\href
  {\doibase 10.1007/JHEP11(2015)206} {\bibfield  {journal} {\bibinfo  {journal}
  {JHEP}\ }\textbf {\bibinfo {volume} {11}},\ \bibinfo {pages} {206} (\bibinfo
  {year} {2015}{\natexlab{a}})},\ \Eprint {http://arxiv.org/abs/1509.00672}
  {arXiv:1509.00672 [hep-ex]} \BibitemShut {NoStop}%
\bibitem [{\citenamefont {Khachatryan}\ \emph {et~al.}(2017)\citenamefont
  {Khachatryan} \emph {et~al.}}]{Khachatryan:2016whc}%
  \BibitemOpen
  \bibfield  {author} {\bibinfo {author} {\bibfnamefont {Vardan}\ \bibnamefont
  {Khachatryan}} \emph {et~al.} (\bibinfo {collaboration} {CMS}),\ }\bibfield
  {title} {\enquote {\bibinfo {title} {{Searches for invisible decays of the
  Higgs boson in pp collisions at sqrt(s) = 7, 8, and 13 TeV}},}\ }\href
  {\doibase 10.1007/JHEP02(2017)135} {\bibfield  {journal} {\bibinfo  {journal}
  {JHEP}\ }\textbf {\bibinfo {volume} {02}},\ \bibinfo {pages} {135} (\bibinfo
  {year} {2017})},\ \Eprint {http://arxiv.org/abs/1610.09218} {arXiv:1610.09218
  [hep-ex]} \BibitemShut {NoStop}%
\bibitem [{\citenamefont {Kopp}\ \emph {et~al.}(2016)\citenamefont {Kopp},
  \citenamefont {Liu}, \citenamefont {Slatyer}, \citenamefont {Wang},\ and\
  \citenamefont {Xue}}]{Kopp:2016yji}%
  \BibitemOpen
  \bibfield  {author} {\bibinfo {author} {\bibfnamefont {Joachim}\ \bibnamefont
  {Kopp}}, \bibinfo {author} {\bibfnamefont {Jia}\ \bibnamefont {Liu}},
  \bibinfo {author} {\bibfnamefont {Tracy~R.}\ \bibnamefont {Slatyer}},
  \bibinfo {author} {\bibfnamefont {Xiao-Ping}\ \bibnamefont {Wang}}, \ and\
  \bibinfo {author} {\bibfnamefont {Wei}\ \bibnamefont {Xue}},\ }\bibfield
  {title} {\enquote {\bibinfo {title} {{Impeded Dark Matter}},}\ }\href
  {\doibase 10.1007/JHEP12(2016)033} {\bibfield  {journal} {\bibinfo  {journal}
  {JHEP}\ }\textbf {\bibinfo {volume} {12}},\ \bibinfo {pages} {033} (\bibinfo
  {year} {2016})},\ \Eprint {http://arxiv.org/abs/1609.02147} {arXiv:1609.02147
  [hep-ph]} \BibitemShut {NoStop}%
\bibitem [{\citenamefont {Martin}\ and\ \citenamefont
  {Roy}(2011)}]{Martin:2011pd}%
  \BibitemOpen
  \bibfield  {author} {\bibinfo {author} {\bibfnamefont {Adam}\ \bibnamefont
  {Martin}}\ and\ \bibinfo {author} {\bibfnamefont {Tuhin~S.}\ \bibnamefont
  {Roy}},\ }\bibfield  {title} {\enquote {\bibinfo {title} {{The Gold-Plated
  Channel for Supersymmetric Higgs via Higgsphilic Z'}},}\ }\href@noop {} {\
  (\bibinfo {year} {2011})},\ \Eprint {http://arxiv.org/abs/1103.3504}
  {arXiv:1103.3504 [hep-ph]} \BibitemShut {NoStop}%
\bibitem [{\citenamefont {Curtin}\ \emph {et~al.}(2014)\citenamefont {Curtin}
  \emph {et~al.}}]{Curtin:2013fra}%
  \BibitemOpen
  \bibfield  {author} {\bibinfo {author} {\bibfnamefont {David}\ \bibnamefont
  {Curtin}} \emph {et~al.},\ }\bibfield  {title} {\enquote {\bibinfo {title}
  {{Exotic decays of the 125 GeV Higgs boson}},}\ }\href {\doibase
  10.1103/PhysRevD.90.075004} {\bibfield  {journal} {\bibinfo  {journal} {Phys.
  Rev.}\ }\textbf {\bibinfo {volume} {D90}},\ \bibinfo {pages} {075004}
  (\bibinfo {year} {2014})},\ \Eprint {http://arxiv.org/abs/1312.4992}
  {arXiv:1312.4992 [hep-ph]} \BibitemShut {NoStop}%
\bibitem [{\citenamefont {Huang}\ \emph
  {et~al.}(2014{\natexlab{a}})\citenamefont {Huang}, \citenamefont {Liu},
  \citenamefont {Wang},\ and\ \citenamefont {Yu}}]{Huang:2013ima}%
  \BibitemOpen
  \bibfield  {author} {\bibinfo {author} {\bibfnamefont {Jinrui}\ \bibnamefont
  {Huang}}, \bibinfo {author} {\bibfnamefont {Tao}\ \bibnamefont {Liu}},
  \bibinfo {author} {\bibfnamefont {Lian-Tao}\ \bibnamefont {Wang}}, \ and\
  \bibinfo {author} {\bibfnamefont {Felix}\ \bibnamefont {Yu}},\ }\bibfield
  {title} {\enquote {\bibinfo {title} {{Supersymmetric Exotic Decays of the 125
  GeV Higgs Boson}},}\ }\href {\doibase 10.1103/PhysRevLett.112.221803}
  {\bibfield  {journal} {\bibinfo  {journal} {Phys. Rev. Lett.}\ }\textbf
  {\bibinfo {volume} {112}},\ \bibinfo {pages} {221803} (\bibinfo {year}
  {2014}{\natexlab{a}})},\ \Eprint {http://arxiv.org/abs/1309.6633}
  {arXiv:1309.6633 [hep-ph]} \BibitemShut {NoStop}%
\bibitem [{\citenamefont {Huang}\ \emph
  {et~al.}(2014{\natexlab{b}})\citenamefont {Huang}, \citenamefont {Liu},
  \citenamefont {Wang},\ and\ \citenamefont {Yu}}]{Huang:2014cla}%
  \BibitemOpen
  \bibfield  {author} {\bibinfo {author} {\bibfnamefont {Jinrui}\ \bibnamefont
  {Huang}}, \bibinfo {author} {\bibfnamefont {Tao}\ \bibnamefont {Liu}},
  \bibinfo {author} {\bibfnamefont {Lian-Tao}\ \bibnamefont {Wang}}, \ and\
  \bibinfo {author} {\bibfnamefont {Felix}\ \bibnamefont {Yu}},\ }\bibfield
  {title} {\enquote {\bibinfo {title} {{Supersymmetric subelectroweak scale
  dark matter, the Galactic Center gamma-ray excess, and exotic decays of the
  125 GeV Higgs boson}},}\ }\href {\doibase 10.1103/PhysRevD.90.115006}
  {\bibfield  {journal} {\bibinfo  {journal} {Phys. Rev.}\ }\textbf {\bibinfo
  {volume} {D90}},\ \bibinfo {pages} {115006} (\bibinfo {year}
  {2014}{\natexlab{b}})},\ \Eprint {http://arxiv.org/abs/1407.0038}
  {arXiv:1407.0038 [hep-ph]} \BibitemShut {NoStop}%
\bibitem [{\citenamefont {Liu}\ \emph {et~al.}(2016)\citenamefont {Liu},
  \citenamefont {Wang},\ and\ \citenamefont {Zhang}}]{Liu:2016zki}%
  \BibitemOpen
  \bibfield  {author} {\bibinfo {author} {\bibfnamefont {Zhen}\ \bibnamefont
  {Liu}}, \bibinfo {author} {\bibfnamefont {Lian-Tao}\ \bibnamefont {Wang}}, \
  and\ \bibinfo {author} {\bibfnamefont {Hao}\ \bibnamefont {Zhang}},\
  }\bibfield  {title} {\enquote {\bibinfo {title} {{Exotic decays of the 125
  GeV Higgs boson at future $e^+e^-$ lepton colliders}},}\ }\href {\doibase
  10.1088/1674-1137/41/6/063102} {\  (\bibinfo {year} {2016}),\
  10.1088/1674-1137/41/6/063102},\ \Eprint {http://arxiv.org/abs/1612.09284}
  {arXiv:1612.09284 [hep-ph]} \BibitemShut {NoStop}%
\bibitem [{\citenamefont {Collaboration}(2013)}]{CMS:2013lea}%
  \BibitemOpen
  \bibfield  {author} {\bibinfo {author} {\bibfnamefont {CMS}\ \bibnamefont
  {Collaboration}} (\bibinfo {collaboration} {CMS}),\ }\bibfield  {title}
  {\enquote {\bibinfo {title} {{Search for a non-standard-model Higgs boson
  decaying to a pair of new light bosons in four-muon final states}},}\
  }\href@noop {} {\  (\bibinfo {year} {2013})}\BibitemShut {NoStop}%
\bibitem [{CMS(2013)}]{CMS:xwa}%
  \BibitemOpen
  \bibfield  {title} {\enquote {\bibinfo {title} {{Properties of the Higgs-like
  boson in the decay H to ZZ to 4l in pp collisions at sqrt s =7 and 8 TeV}},}\
  }\href@noop {} {\  (\bibinfo {year} {2013})}\BibitemShut {NoStop}%
\bibitem [{\citenamefont {Aad}\ \emph {et~al.}(2014)\citenamefont {Aad} \emph
  {et~al.}}]{Aad:2014cka}%
  \BibitemOpen
  \bibfield  {author} {\bibinfo {author} {\bibfnamefont {Georges}\ \bibnamefont
  {Aad}} \emph {et~al.} (\bibinfo {collaboration} {ATLAS}),\ }\bibfield
  {title} {\enquote {\bibinfo {title} {{Search for high-mass dilepton
  resonances in pp collisions at $\sqrt{s}=8$ TeV with the ATLAS detector}},}\
  }\href {\doibase 10.1103/PhysRevD.90.052005} {\bibfield  {journal} {\bibinfo
  {journal} {Phys. Rev.}\ }\textbf {\bibinfo {volume} {D90}},\ \bibinfo {pages}
  {052005} (\bibinfo {year} {2014})},\ \Eprint {http://arxiv.org/abs/1405.4123}
  {arXiv:1405.4123 [hep-ex]} \BibitemShut {NoStop}%
\bibitem [{\citenamefont {Khachatryan}\ \emph
  {et~al.}(2015{\natexlab{a}})\citenamefont {Khachatryan} \emph
  {et~al.}}]{Khachatryan:2014fba}%
  \BibitemOpen
  \bibfield  {author} {\bibinfo {author} {\bibfnamefont {Vardan}\ \bibnamefont
  {Khachatryan}} \emph {et~al.} (\bibinfo {collaboration} {CMS}),\ }\bibfield
  {title} {\enquote {\bibinfo {title} {{Search for physics beyond the standard
  model in dilepton mass spectra in proton-proton collisions at $ \sqrt{s}=8 $
  TeV}},}\ }\href {\doibase 10.1007/JHEP04(2015)025} {\bibfield  {journal}
  {\bibinfo  {journal} {JHEP}\ }\textbf {\bibinfo {volume} {04}},\ \bibinfo
  {pages} {025} (\bibinfo {year} {2015}{\natexlab{a}})},\ \Eprint
  {http://arxiv.org/abs/1412.6302} {arXiv:1412.6302 [hep-ex]} \BibitemShut
  {NoStop}%
\bibitem [{\citenamefont {Cline}\ \emph {et~al.}(2014)\citenamefont {Cline},
  \citenamefont {Dupuis}, \citenamefont {Liu},\ and\ \citenamefont
  {Xue}}]{Cline:2014dwa}%
  \BibitemOpen
  \bibfield  {author} {\bibinfo {author} {\bibfnamefont {James~M.}\
  \bibnamefont {Cline}}, \bibinfo {author} {\bibfnamefont {Grace}\ \bibnamefont
  {Dupuis}}, \bibinfo {author} {\bibfnamefont {Zuowei}\ \bibnamefont {Liu}}, \
  and\ \bibinfo {author} {\bibfnamefont {Wei}\ \bibnamefont {Xue}},\ }\bibfield
   {title} {\enquote {\bibinfo {title} {{The windows for kinetically mixed
  Z'-mediated dark matter and the galactic center gamma ray excess}},}\ }\href
  {\doibase 10.1007/JHEP08(2014)131} {\bibfield  {journal} {\bibinfo  {journal}
  {JHEP}\ }\textbf {\bibinfo {volume} {08}},\ \bibinfo {pages} {131} (\bibinfo
  {year} {2014})},\ \Eprint {http://arxiv.org/abs/1405.7691} {arXiv:1405.7691
  [hep-ph]} \BibitemShut {NoStop}%
\bibitem [{\citenamefont {Hoenig}\ \emph {et~al.}(2014)\citenamefont {Hoenig},
  \citenamefont {Samach},\ and\ \citenamefont {Tucker-Smith}}]{Hoenig:2014dsa}%
  \BibitemOpen
  \bibfield  {author} {\bibinfo {author} {\bibfnamefont {Isaac}\ \bibnamefont
  {Hoenig}}, \bibinfo {author} {\bibfnamefont {Gabriel}\ \bibnamefont
  {Samach}}, \ and\ \bibinfo {author} {\bibfnamefont {David}\ \bibnamefont
  {Tucker-Smith}},\ }\bibfield  {title} {\enquote {\bibinfo {title} {{Searching
  for dilepton resonances below the Z mass at the LHC}},}\ }\href {\doibase
  10.1103/PhysRevD.90.075016} {\bibfield  {journal} {\bibinfo  {journal} {Phys.
  Rev.}\ }\textbf {\bibinfo {volume} {D90}},\ \bibinfo {pages} {075016}
  (\bibinfo {year} {2014})},\ \Eprint {http://arxiv.org/abs/1408.1075}
  {arXiv:1408.1075 [hep-ph]} \BibitemShut {NoStop}%
\bibitem [{\citenamefont {Chatrchyan}\ \emph {et~al.}(2013)\citenamefont
  {Chatrchyan} \emph {et~al.}}]{Chatrchyan:2013tia}%
  \BibitemOpen
  \bibfield  {author} {\bibinfo {author} {\bibfnamefont {Serguei}\ \bibnamefont
  {Chatrchyan}} \emph {et~al.} (\bibinfo {collaboration} {CMS}),\ }\bibfield
  {title} {\enquote {\bibinfo {title} {{Measurement of the differential and
  double-differential Drell-Yan cross sections in proton-proton collisions at
  $\sqrt{s} =$ 7 TeV}},}\ }\href {\doibase 10.1007/JHEP12(2013)030} {\bibfield
  {journal} {\bibinfo  {journal} {JHEP}\ }\textbf {\bibinfo {volume} {12}},\
  \bibinfo {pages} {030} (\bibinfo {year} {2013})},\ \Eprint
  {http://arxiv.org/abs/1310.7291} {arXiv:1310.7291 [hep-ex]} \BibitemShut
  {NoStop}%
\bibitem [{\citenamefont {Aaboud}\ \emph
  {et~al.}(2016{\natexlab{b}})\citenamefont {Aaboud} \emph
  {et~al.}}]{Aaboud:2016cth}%
  \BibitemOpen
  \bibfield  {author} {\bibinfo {author} {\bibfnamefont {Morad}\ \bibnamefont
  {Aaboud}} \emph {et~al.} (\bibinfo {collaboration} {ATLAS}),\ }\bibfield
  {title} {\enquote {\bibinfo {title} {{Search for high-mass new phenomena in
  the dilepton final state using proton-proton collisions at $\sqrt{s}=13$ TeV
  with the ATLAS detector}},}\ }\href {\doibase 10.1016/j.physletb.2016.08.055}
  {\bibfield  {journal} {\bibinfo  {journal} {Phys. Lett.}\ }\textbf {\bibinfo
  {volume} {B761}},\ \bibinfo {pages} {372--392} (\bibinfo {year}
  {2016}{\natexlab{b}})},\ \Eprint {http://arxiv.org/abs/1607.03669}
  {arXiv:1607.03669 [hep-ex]} \BibitemShut {NoStop}%
\bibitem [{\citenamefont {Liu}\ \emph {et~al.}(2017)\citenamefont {Liu},
  \citenamefont {Yue},\ and\ \citenamefont {Guo}}]{Liu:2017jzn}%
  \BibitemOpen
  \bibfield  {author} {\bibinfo {author} {\bibfnamefont {Zhi-Cheng}\
  \bibnamefont {Liu}}, \bibinfo {author} {\bibfnamefont {Chong-Xing}\
  \bibnamefont {Yue}}, \ and\ \bibinfo {author} {\bibfnamefont {Yu-Chen}\
  \bibnamefont {Guo}},\ }\bibfield  {title} {\enquote {\bibinfo {title}
  {{Single production of dark photon at the LHC}},}\ }\href@noop {} {\
  (\bibinfo {year} {2017})},\ \Eprint {http://arxiv.org/abs/1703.00153}
  {arXiv:1703.00153 [hep-ph]} \BibitemShut {NoStop}%
\bibitem [{\citenamefont {Aad}\ \emph {et~al.}(2015{\natexlab{b}})\citenamefont
  {Aad} \emph {et~al.}}]{Aad:2015zva}%
  \BibitemOpen
  \bibfield  {author} {\bibinfo {author} {\bibfnamefont {Georges}\ \bibnamefont
  {Aad}} \emph {et~al.} (\bibinfo {collaboration} {ATLAS}),\ }\bibfield
  {title} {\enquote {\bibinfo {title} {{Search for new phenomena in final
  states with an energetic jet and large missing transverse momentum in pp
  collisions at $\sqrt{s}=$8 TeV with the ATLAS detector}},}\ }\href {\doibase
  10.1140/epjc/s10052-015-3517-3, 10.1140/epjc/s10052-015-3639-7} {\bibfield
  {journal} {\bibinfo  {journal} {Eur. Phys. J.}\ }\textbf {\bibinfo {volume}
  {C75}},\ \bibinfo {pages} {299} (\bibinfo {year} {2015}{\natexlab{b}})},\
  \bibinfo {note} {[Erratum: Eur. Phys. J.C75,no.9,408(2015)]},\ \Eprint
  {http://arxiv.org/abs/1502.01518} {arXiv:1502.01518 [hep-ex]} \BibitemShut
  {NoStop}%
\bibitem [{\citenamefont {Khachatryan}\ \emph
  {et~al.}(2015{\natexlab{b}})\citenamefont {Khachatryan} \emph
  {et~al.}}]{Khachatryan:2014rra}%
  \BibitemOpen
  \bibfield  {author} {\bibinfo {author} {\bibfnamefont {Vardan}\ \bibnamefont
  {Khachatryan}} \emph {et~al.} (\bibinfo {collaboration} {CMS}),\ }\bibfield
  {title} {\enquote {\bibinfo {title} {{Search for dark matter, extra
  dimensions, and unparticles in monojet events in proton–proton collisions
  at $\sqrt{s} = 8$ TeV}},}\ }\href {\doibase 10.1140/epjc/s10052-015-3451-4}
  {\bibfield  {journal} {\bibinfo  {journal} {Eur. Phys. J.}\ }\textbf
  {\bibinfo {volume} {C75}},\ \bibinfo {pages} {235} (\bibinfo {year}
  {2015}{\natexlab{b}})},\ \Eprint {http://arxiv.org/abs/1408.3583}
  {arXiv:1408.3583 [hep-ex]} \BibitemShut {NoStop}%
\bibitem [{\citenamefont {Jacques}\ and\ \citenamefont
  {Nordstrom}(2015)}]{Jacques:2015zha}%
  \BibitemOpen
  \bibfield  {author} {\bibinfo {author} {\bibfnamefont {Thomas}\ \bibnamefont
  {Jacques}}\ and\ \bibinfo {author} {\bibfnamefont {Karl}\ \bibnamefont
  {Nordstrom}},\ }\bibfield  {title} {\enquote {\bibinfo {title} {{Mapping
  monojet constraints onto Simplified Dark Matter Models}},}\ }\href {\doibase
  10.1007/JHEP06(2015)142} {\bibfield  {journal} {\bibinfo  {journal} {JHEP}\
  }\textbf {\bibinfo {volume} {06}},\ \bibinfo {pages} {142} (\bibinfo {year}
  {2015})},\ \Eprint {http://arxiv.org/abs/1502.05721} {arXiv:1502.05721
  [hep-ph]} \BibitemShut {NoStop}%
\bibitem [{\citenamefont {Chala}\ \emph {et~al.}(2015)\citenamefont {Chala},
  \citenamefont {Kahlhoefer}, \citenamefont {McCullough}, \citenamefont
  {Nardini},\ and\ \citenamefont {Schmidt-Hoberg}}]{Chala:2015ama}%
  \BibitemOpen
  \bibfield  {author} {\bibinfo {author} {\bibfnamefont {Mikael}\ \bibnamefont
  {Chala}}, \bibinfo {author} {\bibfnamefont {Felix}\ \bibnamefont
  {Kahlhoefer}}, \bibinfo {author} {\bibfnamefont {Matthew}\ \bibnamefont
  {McCullough}}, \bibinfo {author} {\bibfnamefont {Germano}\ \bibnamefont
  {Nardini}}, \ and\ \bibinfo {author} {\bibfnamefont {Kai}\ \bibnamefont
  {Schmidt-Hoberg}},\ }\bibfield  {title} {\enquote {\bibinfo {title}
  {{Constraining Dark Sectors with Monojets and Dijets}},}\ }\href {\doibase
  10.1007/JHEP07(2015)089} {\bibfield  {journal} {\bibinfo  {journal} {JHEP}\
  }\textbf {\bibinfo {volume} {07}},\ \bibinfo {pages} {089} (\bibinfo {year}
  {2015})},\ \Eprint {http://arxiv.org/abs/1503.05916} {arXiv:1503.05916
  [hep-ph]} \BibitemShut {NoStop}%
\bibitem [{\citenamefont {Brennan}\ \emph {et~al.}(2016)\citenamefont
  {Brennan}, \citenamefont {McDonald}, \citenamefont {Gramling},\ and\
  \citenamefont {Jacques}}]{Brennan:2016xjh}%
  \BibitemOpen
  \bibfield  {author} {\bibinfo {author} {\bibfnamefont {A.~J.}\ \bibnamefont
  {Brennan}}, \bibinfo {author} {\bibfnamefont {M.~F.}\ \bibnamefont
  {McDonald}}, \bibinfo {author} {\bibfnamefont {J.}~\bibnamefont {Gramling}},
  \ and\ \bibinfo {author} {\bibfnamefont {T.~D.}\ \bibnamefont {Jacques}},\
  }\bibfield  {title} {\enquote {\bibinfo {title} {{Collide and Conquer:
  Constraints on Simplified Dark Matter Models using Mono-X Collider
  Searches}},}\ }\href {\doibase 10.1007/JHEP05(2016)112} {\bibfield  {journal}
  {\bibinfo  {journal} {JHEP}\ }\textbf {\bibinfo {volume} {05}},\ \bibinfo
  {pages} {112} (\bibinfo {year} {2016})},\ \Eprint
  {http://arxiv.org/abs/1603.01366} {arXiv:1603.01366 [hep-ph]} \BibitemShut
  {NoStop}%
\bibitem [{\citenamefont {Gupta}\ \emph {et~al.}(2015)\citenamefont {Gupta},
  \citenamefont {Primulando},\ and\ \citenamefont {Saraswat}}]{Gupta:2015lfa}%
  \BibitemOpen
  \bibfield  {author} {\bibinfo {author} {\bibfnamefont {Arpit}\ \bibnamefont
  {Gupta}}, \bibinfo {author} {\bibfnamefont {Reinard}\ \bibnamefont
  {Primulando}}, \ and\ \bibinfo {author} {\bibfnamefont {Prashant}\
  \bibnamefont {Saraswat}},\ }\bibfield  {title} {\enquote {\bibinfo {title}
  {{A New Probe of Dark Sector Dynamics at the LHC}},}\ }\href {\doibase
  10.1007/JHEP09(2015)079} {\bibfield  {journal} {\bibinfo  {journal} {JHEP}\
  }\textbf {\bibinfo {volume} {09}},\ \bibinfo {pages} {079} (\bibinfo {year}
  {2015})},\ \Eprint {http://arxiv.org/abs/1504.01385} {arXiv:1504.01385
  [hep-ph]} \BibitemShut {NoStop}%
\bibitem [{\citenamefont {Buschmann}\ \emph {et~al.}(2015)\citenamefont
  {Buschmann}, \citenamefont {Kopp}, \citenamefont {Liu},\ and\ \citenamefont
  {Machado}}]{Buschmann:2015awa}%
  \BibitemOpen
  \bibfield  {author} {\bibinfo {author} {\bibfnamefont {Malte}\ \bibnamefont
  {Buschmann}}, \bibinfo {author} {\bibfnamefont {Joachim}\ \bibnamefont
  {Kopp}}, \bibinfo {author} {\bibfnamefont {Jia}\ \bibnamefont {Liu}}, \ and\
  \bibinfo {author} {\bibfnamefont {Pedro A.~N.}\ \bibnamefont {Machado}},\
  }\bibfield  {title} {\enquote {\bibinfo {title} {{Lepton Jets from Radiating
  Dark Matter}},}\ }\href {\doibase 10.1007/JHEP07(2015)045} {\bibfield
  {journal} {\bibinfo  {journal} {JHEP}\ }\textbf {\bibinfo {volume} {07}},\
  \bibinfo {pages} {045} (\bibinfo {year} {2015})},\ \Eprint
  {http://arxiv.org/abs/1505.07459} {arXiv:1505.07459 [hep-ph]} \BibitemShut
  {NoStop}%
\bibitem [{\citenamefont {Dam}(2016)}]{Dam:2016ebi}%
  \BibitemOpen
  \bibfield  {author} {\bibinfo {author} {\bibfnamefont {Mogens}\ \bibnamefont
  {Dam}},\ }\bibfield  {title} {\enquote {\bibinfo {title} {{Precision
  Electroweak measurements at the FCC-ee}},}\ }\href@noop {} {\  (\bibinfo
  {year} {2016})},\ \Eprint {http://arxiv.org/abs/1601.03849} {arXiv:1601.03849
  [hep-ex]} \BibitemShut {NoStop}%
\bibitem [{\citenamefont {d'Enterria}(2016)}]{dEnterria:2016sca}%
  \BibitemOpen
  \bibfield  {author} {\bibinfo {author} {\bibfnamefont {David}\ \bibnamefont
  {d'Enterria}},\ }\bibfield  {title} {\enquote {\bibinfo {title} {{Physics at
  the FCC-ee}},}\ }in\ \href
  {https://inspirehep.net/record/1421932/files/arXiv:1602.05043.pdf} {\emph
  {\bibinfo {booktitle} {{17th Lomonosov Conference on Elementary Particle
  Physics Moscow, Russia, August 20-26, 2015}}}}\ (\bibinfo {year} {2016})\
  \Eprint {http://arxiv.org/abs/1602.05043} {arXiv:1602.05043 [hep-ex]}
  \BibitemShut {NoStop}%
\bibitem [{\citenamefont {Ruan}(2016)}]{Ruan:2014xxa}%
  \BibitemOpen
  \bibfield  {author} {\bibinfo {author} {\bibfnamefont {Manqi}\ \bibnamefont
  {Ruan}},\ }\bibfield  {title} {\enquote {\bibinfo {title} {{Higgs Measurement
  at $e^+e^-$ Circular Colliders}},}\ }\bibfield  {booktitle} {\emph {\bibinfo
  {booktitle} {{Proceedings, 37th International Conference on High Energy
  Physics (ICHEP 2014): Valencia, Spain, July 2-9, 2014}}},\ }\href {\doibase
  10.1016/j.nuclphysbps.2015.09.132} {\bibfield  {journal} {\bibinfo  {journal}
  {Nucl. Part. Phys. Proc.}\ }\textbf {\bibinfo {volume} {273-275}},\ \bibinfo
  {pages} {857--862} (\bibinfo {year} {2016})},\ \Eprint
  {http://arxiv.org/abs/1411.5606} {arXiv:1411.5606 [hep-ex]} \BibitemShut
  {NoStop}%
\bibitem [{\citenamefont {Alwall}\ \emph {et~al.}(2014)\citenamefont {Alwall},
  \citenamefont {Frederix}, \citenamefont {Frixione}, \citenamefont {Hirschi},
  \citenamefont {Maltoni}, \citenamefont {Mattelaer}, \citenamefont {Shao},
  \citenamefont {Stelzer}, \citenamefont {Torrielli},\ and\ \citenamefont
  {Zaro}}]{Alwall:2014hca}%
  \BibitemOpen
  \bibfield  {author} {\bibinfo {author} {\bibfnamefont {J.}~\bibnamefont
  {Alwall}}, \bibinfo {author} {\bibfnamefont {R.}~\bibnamefont {Frederix}},
  \bibinfo {author} {\bibfnamefont {S.}~\bibnamefont {Frixione}}, \bibinfo
  {author} {\bibfnamefont {V.}~\bibnamefont {Hirschi}}, \bibinfo {author}
  {\bibfnamefont {F.}~\bibnamefont {Maltoni}}, \bibinfo {author} {\bibfnamefont
  {O.}~\bibnamefont {Mattelaer}}, \bibinfo {author} {\bibfnamefont {H.~S.}\
  \bibnamefont {Shao}}, \bibinfo {author} {\bibfnamefont {T.}~\bibnamefont
  {Stelzer}}, \bibinfo {author} {\bibfnamefont {P.}~\bibnamefont {Torrielli}},
  \ and\ \bibinfo {author} {\bibfnamefont {M.}~\bibnamefont {Zaro}},\
  }\bibfield  {title} {\enquote {\bibinfo {title} {{The automated computation
  of tree-level and next-to-leading order differential cross sections, and
  their matching to parton shower simulations}},}\ }\href {\doibase
  10.1007/JHEP07(2014)079} {\bibfield  {journal} {\bibinfo  {journal} {JHEP}\
  }\textbf {\bibinfo {volume} {07}},\ \bibinfo {pages} {079} (\bibinfo {year}
  {2014})},\ \Eprint {http://arxiv.org/abs/1405.0301} {arXiv:1405.0301
  [hep-ph]} \BibitemShut {NoStop}%
\bibitem [{\citenamefont {Sjostrand}\ \emph {et~al.}(2006)\citenamefont
  {Sjostrand}, \citenamefont {Mrenna},\ and\ \citenamefont
  {Skands}}]{Sjostrand:2006za}%
  \BibitemOpen
  \bibfield  {author} {\bibinfo {author} {\bibfnamefont {Torbjorn}\
  \bibnamefont {Sjostrand}}, \bibinfo {author} {\bibfnamefont {Stephen}\
  \bibnamefont {Mrenna}}, \ and\ \bibinfo {author} {\bibfnamefont {Peter~Z.}\
  \bibnamefont {Skands}},\ }\bibfield  {title} {\enquote {\bibinfo {title}
  {{PYTHIA 6.4 Physics and Manual}},}\ }\href {\doibase
  10.1088/1126-6708/2006/05/026} {\bibfield  {journal} {\bibinfo  {journal}
  {JHEP}\ }\textbf {\bibinfo {volume} {05}},\ \bibinfo {pages} {026} (\bibinfo
  {year} {2006})},\ \Eprint {http://arxiv.org/abs/hep-ph/0603175}
  {arXiv:hep-ph/0603175 [hep-ph]} \BibitemShut {NoStop}%
\bibitem [{\citenamefont {de~Favereau}\ \emph {et~al.}(2014)\citenamefont
  {de~Favereau}, \citenamefont {Delaere}, \citenamefont {Demin}, \citenamefont
  {Giammanco}, \citenamefont {Lemaître}, \citenamefont {Mertens},\ and\
  \citenamefont {Selvaggi}}]{deFavereau:2013fsa}%
  \BibitemOpen
  \bibfield  {author} {\bibinfo {author} {\bibfnamefont {J.}~\bibnamefont
  {de~Favereau}}, \bibinfo {author} {\bibfnamefont {C.}~\bibnamefont
  {Delaere}}, \bibinfo {author} {\bibfnamefont {P.}~\bibnamefont {Demin}},
  \bibinfo {author} {\bibfnamefont {A.}~\bibnamefont {Giammanco}}, \bibinfo
  {author} {\bibfnamefont {V.}~\bibnamefont {Lemaître}}, \bibinfo {author}
  {\bibfnamefont {A.}~\bibnamefont {Mertens}}, \ and\ \bibinfo {author}
  {\bibfnamefont {M.}~\bibnamefont {Selvaggi}} (\bibinfo {collaboration}
  {DELPHES 3}),\ }\bibfield  {title} {\enquote {\bibinfo {title} {{DELPHES 3, A
  modular framework for fast simulation of a generic collider experiment}},}\
  }\href {\doibase 10.1007/JHEP02(2014)057} {\bibfield  {journal} {\bibinfo
  {journal} {JHEP}\ }\textbf {\bibinfo {volume} {02}},\ \bibinfo {pages} {057}
  (\bibinfo {year} {2014})},\ \Eprint {http://arxiv.org/abs/1307.6346}
  {arXiv:1307.6346 [hep-ex]} \BibitemShut {NoStop}%
\bibitem [{\citenamefont {Ruan}()}]{DelphesCEPC}%
  \BibitemOpen
  \bibfield  {author} {\bibinfo {author} {\bibfnamefont {Manqi}\ \bibnamefont
  {Ruan}} (\bibinfo {collaboration} {CEPC-SPPC study group}),\ }\bibfield
  {title} {\enquote {\bibinfo {title} {{
  https://indico.cern.ch/event/550509/contributions/2413240/attachments/1395641/2128022/CEPC\_FCC\_WS\_v2.pdf}},}\
  }\href@noop {} {\ }\bibinfo {note}
  {[https://github.com/delphes/delphes/blob/master/cards/delphes\_card\_CEPC.tcl]}\BibitemShut
  {NoStop}%
\bibitem [{\citenamefont {Patrignani}\ \emph {et~al.}(2016)\citenamefont
  {Patrignani} \emph {et~al.}}]{Olive:2016xmw}%
  \BibitemOpen
  \bibfield  {author} {\bibinfo {author} {\bibfnamefont {C.}~\bibnamefont
  {Patrignani}} \emph {et~al.} (\bibinfo {collaboration} {Particle Data
  Group}),\ }\bibfield  {title} {\enquote {\bibinfo {title} {{Review of
  Particle Physics}},}\ }\href {\doibase 10.1088/1674-1137/40/10/100001}
  {\bibfield  {journal} {\bibinfo  {journal} {Chin. Phys.}\ }\textbf {\bibinfo
  {volume} {C40}},\ \bibinfo {pages} {100001} (\bibinfo {year}
  {2016})}\BibitemShut {NoStop}%
\bibitem [{\citenamefont {Aad}\ \emph {et~al.}(2016)\citenamefont {Aad} \emph
  {et~al.}}]{Khachatryan:2016vau}%
  \BibitemOpen
  \bibfield  {author} {\bibinfo {author} {\bibfnamefont {Georges}\ \bibnamefont
  {Aad}} \emph {et~al.} (\bibinfo {collaboration} {ATLAS, CMS}),\ }\bibfield
  {title} {\enquote {\bibinfo {title} {{Measurements of the Higgs boson
  production and decay rates and constraints on its couplings from a combined
  ATLAS and CMS analysis of the LHC pp collision data at $ \sqrt{s}=7 $ and 8
  TeV}},}\ }\href {\doibase 10.1007/JHEP08(2016)045} {\bibfield  {journal}
  {\bibinfo  {journal} {JHEP}\ }\textbf {\bibinfo {volume} {08}},\ \bibinfo
  {pages} {045} (\bibinfo {year} {2016})},\ \Eprint
  {http://arxiv.org/abs/1606.02266} {arXiv:1606.02266 [hep-ex]} \BibitemShut
  {NoStop}%
\end{thebibliography}%

\end{document}